\documentclass[useAMS,usenatbib,usegraphicx,usenatbib]{mn2e}
\usepackage{ifthen}
\usepackage{amssymb}
\usepackage{wasysym}
\usepackage{multirow}
\usepackage[usenames]{color}
\DeclareSymbolFont{boldss}{OT1}{cmss}{bx}{n}
\DeclareSymbolFontAlphabet\mathbsf{boldss}

\def\jref@jnl#1{{\rm#1}}

\def\aj{\jref@jnl{AJ}}                   % Astronomical Journal
\def\araa{\jref@jnl{ARA\&A}}             % Annual Review of Astron and Astrophys
\def\apj{\jref@jnl{ApJ}}                 % Astrophysical Journal
\def\apjl{\jref@jnl{ApJ}}                % Astrophysical Journal, Letters
\def\apjs{\jref@jnl{ApJS}}               % Astrophysical Journal, Supplement
\def\ao{\jref@jnl{Appl.~Opt.}}           % Applied Optics
\def\apss{\jref@jnl{Ap\&SS}}             % Astrophysics and Space Science
\def\aap{\jref@jnl{A\&A}}                % Astronomy and Astrophysics
\def\aapr{\jref@jnl{A\&A~Rev.}}          % Astronomy and Astrophysics Reviews
\def\aaps{\jref@jnl{A\&AS}}              % Astronomy and Astrophysics, Supplement
\def\azh{\jref@jnl{AZh}}                 % Astronomicheskii Zhurnal
\def\baas{\jref@jnl{BAAS}}               % Bulletin of the AAS
\def\jrasc{\jref@jnl{JRASC}}             % Journal of the RAS of Canada
\def\memras{\jref@jnl{MmRAS}}            % Memoirs of the RAS
\def\mnras{\jref@jnl{MNRAS}}             % Monthly Notices of the RAS
\def\pra{\jref@jnl{Phys.~Rev.~A}}        % Physical Review A: General Physics
\def\prb{\jref@jnl{Phys.~Rev.~B}}        % Physical Review B: Solid State
\def\prc{\jref@jnl{Phys.~Rev.~C}}        % Physical Review C
\def\prd{\jref@jnl{Phys.~Rev.~D}}        % Physical Review D
\def\pre{\jref@jnl{Phys.~Rev.~E}}        % Physical Review E
\def\prl{\jref@jnl{Phys.~Rev.~Lett.}}    % Physical Review Letters
\def\pasp{\jref@jnl{PASP}}               % Publications of the ASP
\def\pasj{\jref@jnl{PASJ}}               % Publications of the ASJ
\def\qjras{\jref@jnl{QJRAS}}             % Quarterly Journal of the RAS
\def\skytel{\jref@jnl{S\&T}}             % Sky and Telescope
\def\solphys{\jref@jnl{Sol.~Phys.}}      % Solar Physics
\def\sovast{\jref@jnl{Soviet~Ast.}}      % Soviet Astronomy
\def\ssr{\jref@jnl{Space~Sci.~Rev.}}     % Space Science Reviews
\def\zap{\jref@jnl{ZAp}}                 % Zeitschrift fuer Astrophysik
\def\nat{\jref@jnl{Nature}}              % Nature
\def\iaucirc{\jref@jnl{IAU~Circ.}}       % IAU Cirulars
\def\aplett{\jref@jnl{Astrophys.~Lett.}} % Astrophysics Letters
\def\apspr{\jref@jnl{Astrophys.~Space~Phys.~Res.}}
\def\bain{\jref@jnl{Bull.~Astron.~Inst.~Netherlands}} 
\def\fcp{\jref@jnl{Fund.~Cosmic~Phys.}}  % Fundamental Cosmic Physics
\def\gca{\jref@jnl{Geochim.~Cosmochim.~Acta}}   % Geochimica Cosmochimica Acta
\def\grl{\jref@jnl{Geophys.~Res.~Lett.}} % Geophysics Research Letters
\def\jcp{\jref@jnl{J.~Chem.~Phys.}}      % Journal of Chemical Physics
\def\jgr{\jref@jnl{J.~Geophys.~Res.}}    % Journal of Geophysics Research
\def\jqsrt{\jref@jnl{J.~Quant.~Spec.~Radiat.~Transf.}}
\def\memsai{\jref@jnl{Mem.~Soc.~Astron.~Italiana}}
\def\nphysa{\jref@jnl{Nucl.~Phys.~A}}   % Nuclear Physics A
\def\physrep{\jref@jnl{Phys.~Rep.}}   % Physics Reports
\def\physscr{\jref@jnl{Phys.~Scr}}   % Physica Scripta
\def\planss{\jref@jnl{Planet.~Space~Sci.}}   % Planetary Space Science
\def\procspie{\jref@jnl{Proc.~SPIE}}   % Proceedings of the SPIE

\def\T1{{\rm 1T}}

\def\msol{m_\odot}

\def\keV{{\rm keV}}
\def\yr{{\rm yr}}
\def\pc{{\rm pc}}
\def\kpc{{\rm kpc}}
\def\Mpc{{\rm Mpc}}

\def\Ddt{{{\partial}\over{\partial t}}}

\def\vvec{\mathbf v}
\def\rvec{\mathbf r}
\def\fvec{\mathbf f}

\def\Lcal{\mathcal L}
\def\Mcal{\mathcal M}

\def\Zcal{\mathcal Z}

\def\sersic{S\'{e}rsic }

\newlength{\kinglen}
\newlength{\sersiclen}
\title[Structure, inflow \& central properties of galaxy clusters]{
   Radial structure, inflow and central mass of stationary radiative galaxy clusters}
\author[C.~J.~Saxton \& K.~Wu]
{ Curtis~J.~Saxton$^{1}$ \& Kinwah~Wu$^1$\\
$^1$ Mullard Space Science Laboratory, University College London, 
   Holmbury St Mary, Dorking, Surrey RH5 6NT  \\
}

\date{Received: }

\begin{document}
\twocolumn 

\maketitle

\begin{abstract}
We analyse the radial structure of 
  self-gravitating spheres consisting of
  multiple interpenetrating fluids,
  such as the X-ray emitting gas and the dark halo of a galaxy cluster.
In these {\em dipolytropic} models,
  the adiabatic dark matter sits in equilibrium,
  while the gas develops a gradual, smooth, quasi-stationary cooling flow.
Both affect and respond to the collective gravitational field.
We find that all subsonic, radially continuous, steady solutions
  require a non-zero minimum central point mass.
For megaparsec-sized halos
   with seven to ten effective degrees of freedom ($F_2$),
   the minimum central mass is compatible with
   observations of supermassive black holes.
Smaller gas mass influxes enable smaller central masses
   for wider ranges of $F_2$.
The halo comprises a sharp spike around the central mass,
   embedded within a core of nearly constant density
   (at $10^1 - 10^{2.5}$ kpc scales),
   with outskirts that attenuate and naturally truncate at finite radius
   (several Mpc).
The gas density resembles a broken power law in radius,
   but the temperature dips and peaks within the dark core.
A finite minimum temperature occurs due to gravitational self-warming,
   without cold mass dropout nor needing regulatory heating.
X-ray emission from the intracluster medium
   mimics a $\beta$-model plus bright compact nucleus.
Near-sonic points in the gas flow are bottlenecks
   to the allowed steady solutions;
the outermost are at kpc scales.
These sites may preferentially develop cold mass dropout
   during strong perturbations off equilibrium.
Within the sonic point,
   the profile of gas specific entropy is flatter than $s\propto r^{1/2}$,
   but this is a shallow ramp and not an isentropic core.
When $F_2$ is large, the inner halo spike is only marginally Jeans stable
   in the central parsec,
   suggesting that a large non-linear disturbance
   could trigger local dark collapse onto the central object.
\end{abstract}


\section{Introduction}

Galaxy clusters consist of baryonic and dark matter in the cosmic ratio
  \citep{spergel2007}. 
Black holes and the stars in galaxies and in the intracluster light 
  only constitute a small (10--15\%) fraction of the baryons; 
  and the intracluster X-ray emitting hot gas
  comprises the majority of the baryons 
  \citep{lin2003,gonzalez2007}.
Relaxed clusters are found to contain a round core
  of approximately constant density,
  attenuating into fringes below detection limits
   \citep{lea1973,cavaliere1976}.
There are also clusters with more centrally peaked core. 
They are thought to be systems 
  with short radiative cooling time in comparison with the Hubble time. 
As radiative cooling causes the depletion of pressure support
  near the cluster centre,
  gas inevitably subsides inwards from the cluster outskirts,
  i.e.\ cooling flows    
  \citep[][]{cowie1977,fabian1977,mathews1978}. 

Cooling flows have been linked with accretion onto, 
  and star formation in the dominant galaxy in the cluster,
  and also the fuelling of their galactic nuclei
\citep[e.g.][]{silk1976,sarazin1983,fabian1984a,nulsen1984}. 
Early models of cooling flow invoked a number of simple assumptions.   
In some fluid formulations for the cluster structure, 
  a static global gravitational potential was used, 
  and there was no consideration of gas or halo self-gravity.   
Kinetic and ram-pressures were often not considered explicitly, 
  and this caused a cooling runaway near the cluster centre, 
  leading to a rapid deposition of a great amount of cold gas.
Approximate deprojected cluster X-ray images
  indicated that the gas inflow rates $\dot{m}$ 
  diminishes nearer the cluster centre \citep{stewart1984,thomas1987}.  
This was taken as evidence for widely distributed ``mass dropout''
  -- thermal instability spawning small, underpressured, invisibly cold clumps 
  within a multiphase medium.
Thermal conduction and magnetic fields 
  were argued to be too weak to inhibit this instability and dropout.

The early cooling flow models were challenged
   by various multi-wavelength observations    
   \citep[see reviews by][]{donahue2004,peterson2006}.
First of all, radio and optical imaging
   have not shown the expected accumulations of cold gas, 
   nor the expected bursts of star formation.  
Moreover, X-ray spectroscopic and imaging deprojections of cluster profiles 
  indicates a temperature floor
  typically a factor 3 or 4 below the peak temperature
\citep{kaastra2001,tamura2001,peterson2001,sakelliou2002,johnstone2002,peterson2003}.
In some systems the temperature even appears to increase at the smallest radii
\citep[e.g][]{osullivan2007a}. 
Spectral analyses suggest that the intracluster medium (ICM)
  is likely to be single-phase
\citep{boehringer2001,david2001,molendi2001,matsushita2002}. 
These difficulties prompted the search of possible processes  
  that could suppress the cooling flows, 
  e.g.\  thermal conduction, 
  or non-gravitational heating,
  such as the power injected by active galaxies (AGN). 
\citep[See review by][and references therein.]{peterson2006} 
There are still open questions  
  whether the heating processes can fine-tune to counteract the cooling stably, 
  and whether the heating would distribute appropriately
  across the relevant regions in the cluster 
\citep{fabian1994b,johnstone2002,brighenti2003,conroy2008}.

Meanwhile, theories of halo structure have been overturned several times.
Once, it was assumed that cluster dark matter
   follows the distribution of galaxies,
   in approximately isothermal, flat-cored assemblages
   \citep[e.g.][]{king1966,rood1972,cavaliere1976,cowie1977,fabian1981}.
This view was naturally compatible with the classic signs
   (in the rotation profiles of disc galaxies)
   that galaxian dark matter is more shallowly
   and widely spread than the baryons.
By the 1990s,
   cosmological N-body simulations were becoming fine enough
   to resolve cluster and galaxy halos,
   under the simplifying assumption that dark matter
   acts like a collisionless stellar-dynamical fluid
   (without any short- or long-range gauge fields of its own).
Simulated halos develop sharp power-law central density cusps,
\citep[see e.g.][]{dubinski1991,nfw1996,moore1998,diemand2004,navarro2004,diemand2005}.
The redistribution of cooling, contracting gas
   may steepen the dark cusp further
\citep[e.g.][]{blumenthal1986,gnedin2004,sellwood2005}.

In a circular way, cuspy profiles became an ansatz
   in the fitting cluster observations.
Cuspy profiles have been assumed as templates 
   in composite mass models fitted to gravitational lensing observations.
It has been shown 
   that a cuspy halo can hold a cored X-ray emitting gas distribution
   qualitatively similar to that of traditional cored halo models
   \citep{makino1998}.
Gravitational lensing suggests flat cores in some clusters
   \citep{tyson1998,gavazzi2003,sand2004,sand2008}.
However, on galaxy scales, a considerable weight of evidence
   disfavours the existence dark cusps today
   (or implies that halos are less centrally concentrated than baryons).
These lines of evidence include
velocity fields of dwarf and low surface brightness galaxies
   \citep{flores1994,moore1994,burkert1995,deblok1997,
		weldrake2003,deblok2005,simon2005,kuzio2006};
   and
   kinematics of dwarf spheroidal galaxies
   \citep{lokas2002,gilmore2007}.
   %and the scarcity of dwarf galaxies
   %\citep{klypin1999,moore1999a}.
Current observational data for dwarf spheroidal galaxies
   cannot rule out cuspy profiles from the kinematics alone
   \citep{walker2007},
   though tidal tracers hint circumstantially at gentle cores in specific cases
   \citep{kleyna2003,goerdt2006}.
Lensing analyses of isolated elliptical galaxies suggest cuspy profiles
   near the observed radii
   \citep{read2007},
   while the kinematics of other cases imply flat cores or low dark densities
   \citep{romanowsky2003,douglas2007,forestell2008}.
Possible explanations of the cusp problem may involve
   subtle numerical systematics of N-body methodology,
   extra dark physics,
   or some forms of gaseous, stellar or AGN feedback.

Feedback, if it is responsible for erasing cusps,
   must overturn a substantial fraction of a galaxy's baryonic mass,
   from the deepest zone of its potential,
   without leaving abnormal metallicities and stellar populations.
The implementation of feedback in numerical simulations
   suffers from severe challenges of resolution,
   and considerable arbitrariness or uncertainty
   in recipes for small-scale physics.
Relevant fluid instabilities differ greatly between numerical schemes
   \citep[e.g][]{agertz2006}.
Energy budgets of popular Lagrangian %particle-
   hydrodynamics methods
   are broken by endemic (but rarely mentioned) ``wall heating'' artefacts
   \citep[e.g.][]{noh1987},
   with unknowable consequences in simulated media
   where heating, cooling or thermal instability are important.
A definitive answer to the ``feedback'' question is far off;
   presently it is an almost unfalsifiable proposition.

This paper aims to present a new formalism
  for the structures of relaxed galaxy clusters,
  and to probe the scope of its initial implications
  for cooling flows and dark matter
  (reserving empirical detail and observational fits for future refinements).
We reexamine the classic scenario of inflows in galaxy clusters  
  with a more complete and consistent treatment 
  of the gravitational interaction and energy exchanges  
  in the gas and the dark matter components. 
Also a sensible polytropic equation of state is used for the dark matter, 
  that admits cuspless solutions for some systems 
  and allows the multiple degrees of freedom in the dark matter.
Note that a polytropic halo may arise if
  dark matter has strong self-interactions (SIDM),
  or if the system is formulated properly
  in the framework of Tsallis' statistical thermodynamics 
\citep[cf. the Boltzmann statistical thermodynamics,][]{tsallis1988,plastino1993},
  or in collisionless systems with isotropic velocity distributions.
We illustrate the properties and profiles
   of spherical, spatially continuous, stationary solutions
   relevant to cluster-sized systems.
We quantify certain signature radii of these solutions,
   for the benefit of comparison with simpler models in the literature,
   and to inform future observational tests.
Our solutions indicates that it is inevitable 
   that point-like central masses would emerge 
   in relaxed clusters, groups or pressure-supported galaxies.
For some appropriate regimes of the halo micro-physics,
   the predicted minimum central mass is consistent 
   with those of supermassive black holes in giant galaxies.
We discuss implications for
   the rapid origin of supermassive black holes,
   monolithic condensation of early stellar populations in galaxies,
   and the problem of the central structures of dark halos.

The paper is organised as follows:
\S2 gives the general formulation of
   the multi-component self-graviating systems, 
   and the construction of the gas and dark-matter dipolytropes. 
Next, \S3 describes the valid solutions in the cluster parameter-space, 
   the properties of the solutions, 
   their comparisons with current observations,
   and predictions to be tested by future observations.
In \S4 we discuss our results in the context of galaxy and cluster evolution. 
We conclude in \S5.
The appendices show the derivation of our model's natural units, 
   the normalisation and the rescaling of the model,
   the interpretation of the effective degrees of freedom for the dark matter 
   and comparisons between our model and other standard spherical models 
   for clusters.

\section{Multi-component self-gravitating systems}

\subsection{multi-fluid formulation}

In our model the system has multiple components. 
It is self-gravitating, and the distinct components interact among each other 
   through their shared gravitational potential.
Each component, $i$, has its equation of state, 
   which takes the form 
\begin{equation}
	p_i = \rho_i \sigma_i^2 = s_i \rho_i^{\gamma_i}
\end{equation}
  with partial pressure $p_i$, density $\rho_i$, 
  isotropic velocity dispersion $\sigma_i$ 
  (which corresponds to an isothermal sound speed in the fluid description),
  specific entropy $s_i$ and adiabatic index $\gamma_i$.
The adiabatic index is related to
  an effective number of degrees of freedom $F_i$ via  
\begin{equation}
	\gamma_i = 1 + {2\over{F_i}}
	\ .
\end{equation}  
The mass, momentum and energy conservation equations read 
\begin{equation}
	\Ddt \rho_i+\nabla\cdot\rho_i\vvec_i=0
	\ ,
\label{vector.mass}
\end{equation}
\begin{equation}
	\Ddt \rho_i\vvec_i
	+\nabla\cdot\rho_i\vvec_i\vvec_i+\nabla\rho_i\sigma_i^2
	=\rho_i\fvec_i
	\ ,
\label{vector.momentum}
\end{equation}
\begin{equation}
	\Ddt\epsilon_i+\nabla\cdot(\epsilon_i+\rho_i\sigma_i^2)\vvec_i
	=\rho_i\vvec_i\cdot\fvec_i+\Lcal_i
	\ ,
\label{vector.energy}
\end{equation}
where the energy density is
\begin{equation}
	\epsilon_i\equiv {{\rho_i\sigma_i^2}\over{\gamma_i-1}}
	+{\frac12}\rho_i v_i^2
	={{s_i\rho_i^{\gamma_i}}\over{\gamma_i-1}}
	+{\frac12}\rho_i v_i^2
	\ .
\end{equation}
Equivalently, the energy conservation equation (\ref{vector.energy}) 
  may be expressed in term of the entropy: 
\begin{equation}
	\Ddt s_i+\vvec_i\cdot\nabla s_i
	=(\gamma_i-1)
	\Lcal_i
	\rho_i^{-\gamma_i}
	\ .
\label{vector.entropy}
\end{equation}
The variable $\Lcal_i$ is a volumetric power,  
   which specifies the energy gains and losses.
For a component net loss, say radiative cooling, $\Lcal_i < 0$.
We assume that $\Lcal_i$ is determined 
   by the local thermodynamic and dynamic variables. 
This assumption is justified in the cluster environments 
  as the radiative processes are optically thin. 

The gravitational potential, $\Phi$, satisfies the Poisson equation 
\begin{equation}
\nabla^2\Phi=4\pi G\sum_i\rho_i
\ .
\end{equation}  
The gravitational force is determined
  from the gravitational potential      
  ${\mathbf f} = -\nabla\Phi$ for all $i$, 
  and the force field is the same for all the mass components.   

\subsection{steady spherical solutions}

In this study we consider only stationary spherically symmetric systems.  
Time dependent analysis will be discussed in Saxton et al. (in preparation), 
  and systems with more complicated geometries will be presented elsewhere.  
In a stationary spherically symmetric system, 
  the density, velocity, temperature and gravitational field 
  are functions of radial co-ordinate $r$ only. 
The mass continuity equation (\ref{vector.mass})  
   requires that density and velocity are related by
\begin{equation}
	\dot{m}_i\equiv 4\pi r^2\rho_i v_i 
\ . 
\label{eq.mass.conservation}
\end{equation}
For mass inflow $\dot{m}_i<0$; and for mass outflow $\dot{m}_i>0$. 
A system with a hydrostatic equilibrium has
   $\dot{m}_i=0$ and $v_i=0$ everywhere. 
The entropy equation (\ref{vector.entropy}) becomes  
\begin{equation}
	{{ds_i}\over{dr}}
	=
	(\gamma_i-1){{\Lcal_i}\over{v_i\rho_i^{\gamma_i} }}
	\ , 
\end{equation} 
  and the conservation equations (\ref{vector.mass}) -- (\ref{vector.energy})
  now read 
\begin{equation}
\left[{
\begin{array}{ccc}
v_i&\rho_i&0\\
\sigma_i^2&\rho_iv_i&\rho_i\\
0&\rho_iv_i^2&{{\gamma_i}\over{\gamma_i-1}}\rho_iv_i\\
\end{array}
}\right]
{d\over{dr}}
\left[{
\begin{array}{c}
\rho_i\\v_i\\\sigma_i^2
\end{array}
}\right]
=
\left[{
\begin{array}{c}
\Zcal_{1i}\\\Zcal_{2i}\\\Zcal_{3i}
\end{array}
}\right]
\label{eq.matrix.radial}
\end{equation}
   where the three source/sink terms are 
\begin{equation}
\Zcal_{1i}=-{{2\rho_iv_i}\over{r}}
\ ,
\end{equation}
\begin{equation}
\Zcal_{2i}=\rho_i f
\ ,
\end{equation}
\begin{equation}
\Zcal_{3i}=
\rho_i v_i f +\Lcal_i
\ .
\end{equation} 

The component mass $m_i$ interior to radius $r$ is given by 
\begin{equation}
	{{dm_i}\over{dr}} = 4\pi r^2 \rho_i
	\ , 
\label{eq.mass.gradient} 
\end{equation}
   and is related to the force by 
\begin{equation}
	f=-{{G}\over{r^2}}\sum_i m_i\ 
\ .   
\end{equation}
It follows  that 
\begin{equation}
	{{df}\over{dr}}=
	-{{2f}\over{r}}
	-4\pi G\sum_i\rho_i 
\ . 
\label{eq.dfdr.raw}
\end{equation}  
For a system with a central point-mass
(e.g.\  a supermassive central black hole of a cD galaxy in a cluster)
    $f$ rises asymptotically near the origin.
For systems without a point mass (\S\ref{s.scheme}), $f=0$ at the origin.

Inversion of (\ref{eq.matrix.radial}) gives 
\begin{equation}
	{{d\rho_i}\over{dr}}=
	{1\over{\Delta_i}}
	\left[{
		{{2\rho_iv_i^2}\over{r}}
		+\rho_i f
		-{{(\gamma_i-1)\Lcal_i}\over{v_i}}
	}\right]
	\ ,
\label{general.drhodr}
\end{equation}
\begin{equation}
	{{dv_i}\over{dr}}=
	{1\over{\Delta_i}}
	\left[{
		-{{2\gamma_i\sigma_i^2v_i}\over{r}}
		-v_i f
		+(\gamma_i-1){{\Lcal_i}\over{\rho_i}}
	}\right]
	\ ,
\label{general.dvdr}
\end{equation}
\begin{equation}
	{{d\sigma_i^2}\over{dr}}=
	{{\gamma_i-1}\over{\Delta_i}}
	\left[{
		{{2v_i^2\sigma_i^2}\over{r}}
		+\sigma_i^2 f
		+{{\sigma_i^2-v_i^2}\over{\rho_i v_i}}\Lcal_i
	}\right]
	\ ,
\label{general.dsigmadr}
\end{equation}
  with the sonic factor, $\Delta_i\equiv\gamma_i\sigma_i^2-v_i^2$, 
  where $\Delta_i>0$ corresponds to a subsonic flow.
If the matrix in (\ref{eq.matrix.radial}) is singular, 
   one of the hydrodynamic variables must be eliminated algebraically  
   to reduce the number of equations.
For stationary, spherically symmetric systems, 
   two of $(\rho_i,v_i,\sigma_i^2,p_i,s_i)$
   suffice to describe the stationary solution.

We will distinguish the mass components in the model
   for zero and nonzero $\dot{m}_i$, 
   as the solution to above equations 
   depends on  whether or not the mass component is in a bulk inflow.
In either case the mass conservation equation (\ref{eq.mass.conservation}) 
   will be used to eliminate one of the dynamical variables of each component.

\subsection{stagnant component, $\dot{m}_i=0$}

Usually, a static structure with zero inflow ($\dot{m}_i=0$) is forbidden 
  (see Equations \ref{general.drhodr} and \ref{general.dsigmadr}),     
  if there is a net energy loss ($\Lcal_i \neq 0$) 
  or if the system is not isothermal ($\gamma_i \neq1$). 
An isothermal condition can be established 
  if heat transport is more rapid than other radiative and dynamical processes 
  in the system. 
This is not easily satisfied for the radiative gas components in a cluster. 
However, the situation is different for the dark matter component  
   as dark matter neither radiates nor absorbs light.     
Self-interacting dark matter could behave like a fluid. 
Thus, such a dark halo may have a hydrostatic profile 
  satisfying $\dot{m}_i=0$, $\Lcal_i\equiv 0$, 
  $v_i=0$ and $dv_i/dr=0$ everywhere. 
Moreover, its  structure is completely specified  
  by the density ($\rho_i$)
  or temperature (velocity dispersion) ($\sigma_i^2$) profiles.    

The gradients of the density (\ref{general.drhodr})
  and temperature (\ref{general.dsigmadr})
  are given by  
\begin{equation}
{{d\rho_i}\over{dr}}=
{{\rho_if}\over{\gamma_i\sigma_i^2}}
\ ,
\label{eq.stagnant.dens}
\end{equation} 
and 
\begin{equation}
{{d\sigma_i^2}\over{dr}}=
{{\gamma_i-1}\over{\gamma_i}}f 
\label{eq.stagnant.temp}
\end{equation}
  respectively. 
The latter implies that $\sigma_i^2=-(\gamma_i-1)(\Phi-\Phi_R)/\gamma_i$. 
In the cluster setting, $\Phi_R$ corresponds to the dark halo surface potential.

The specific entropy is uniform in the cluster. 
The effectively polytropic equation of state is appropriate for dark matter
   if it is self-interacting, governed by Tsallis thermostatistics,
   or has a power-law phase-space density (i.e.\ collisionless).
In the latter case, the momentum equation is identifiable 
   as the Jeans equation in the isotropic limit.
(See appendix~\ref{s.freedom}.)
We note that if only one self-gravitating fluid component is present 
   and if there is no central point-mass,
   the object is essentially a Lane-Emden sphere
   \citep{lane1870,emden1907,chandrasekhar1939}.

\subsection{flowing components, $\dot{m}_i\neq 0$}

Fluids with $\dot{m}_i\neq0$ exhibit a central density cusp, 
   $\rho_i\rightarrow\infty$ as $r\rightarrow 0$.
In order to keep the equations numerically tractable,
  we use (\ref{eq.mass.conservation}) to eliminate $\rho_i$.
The equation for the density gradient (\ref{general.drhodr}) is then redundant,
    leaving two relevant gradient equations:
\begin{equation}
	{{dv_i}\over{dr}}=
	{{-v_i}\over{\Delta_i}}
	\left[{
	f+{{2\gamma_i\sigma_i^2}\over{r}}
	-(\gamma_i-1)
	{{4\pi r^2\Lcal_i}\over{\dot{m}_i}}
	}\right]
	\ ,
\label{flowing.dvdr}
\end{equation}
\begin{equation}
	{{d\sigma^2_i}\over{dr}}=
	{{(\gamma_i-1)\sigma_i^2}\over{\Delta_i}}
	\left[{
	f+{{2v_i^2}\over{r}}
	+\left({
	{{\sigma_i^2-v_i^2}\over{\sigma_i^2}}
	}\right)
	{{4\pi r^2\Lcal_i}\over{\dot{m}_i}}
	}\right]
	\  , 
\label{flowing.dsigmadr}
\end{equation}
 and the mass profile equation: 
\begin{equation}
	{{dm_i}\over{dr}}={{\dot{m}}\over{v_i}}
	\ .
\end{equation}

Elimination of $\rho_i$ in terms of $v_i$ and $r$, 
  particularly from the cooling function $\Lcal_i$,
  clarifies the asymptotic behaviour of the differential equations,
  especially in the inner regions. 
The flow velocity $v_i$ may take any value at the origin.
Models in which $v_i\neq0$ at the origin
   describe accretion onto, or winds emerging from, a central object, 
   presumably compact. 
Note that inflowing constituents have some cuspy behaviour near the origin
   (either $\rho_i\rightarrow\infty$, $v_i\rightarrow 0$
   or $v_i\rightarrow\infty$)
   regardless of the occurrence of radiative cooling.
This causes some trouble in numerical integration.
Changes of variables, including a switch of the independent variable, 
   may, however, circumvent the problems.

\subsection{composite system: radiative gas embedded in self-interacting dark matter}

We consider a model cluster, which is a composite system consisting of  
   a cooling gas component with $\dot{m}_1<0$ and $\Lcal_1<0$ 
   and a self-interacting dark matter component
   with $\dot{m}_2=0$ and $\Lcal_2=0$. 
We treat both of them as separate fluid components. 
We omit the stars in galaxies,
   as they are a minor fraction of the cluster baryons.
The dominant radiative loss of the gas component is 
   optically thin thermal bremsstrahlung radiation.   
We omit line cooling, dust, conduction and Compton effects.
Thus, the radiative loss is specified by a cooling function: 
\begin{equation}
	\Lcal_1= -B \rho_1^2 \sigma_1
	=-B\rho_1^{(3+\gamma_1)/2}s_1^{1/2}
\ ,
\label{eq.bremsstrahlung}
\end{equation}
   where the normalisation $B$ depends on the gas composition \citep{rybicki}.
Cooling domination implies $\Lcal_1<0$ everywhere,
    and the specific entropy increases monotonically with $r$.
This also ensures buoyant stability. 

The inner boundary is a gas density cusp, $\rho_1\rightarrow\infty$,
   regardless of whether radiative cooling
   or compressional heating dominates the gas inflow (accretion).
There are two kinds of cusps:  
  the cold cusp ($\sigma_1\rightarrow 0$) and 
  the hot cusp ($\sigma_1\rightarrow\infty$). 
In cold cusps, the gas pressure $p_1$ is finite everywhere.  
The specific entropy vanishes, $s_1\rightarrow 0$, at the cusp. 
For the hot cusps, the entropy decreases smoothly towards the origin.
The flow reaches the origin exactly, $r_*=0$, 
   where the speed and pressure reach infinity.
For fluids with an adiabatic index of $\gamma_i=1+2/F_i$,  
  the asymptotes are 
   $\rho\propto r^{-F_i/2}$,
   $\sigma^2\propto r^{-1}$, and 
   $v\propto r^{(F_i-4)/2}$. 
There may be a positive-mass compact object at the origin, $m_*$.
The flow at a hot cusp is a self-gravitating generalisation 
  of subsonic Bondi accretion \citep{bondi1952}.

A fully general model of multi-fluid self-gravitating objects permits
   supersonic inflows, existence of sonic points and formation of shocks.
However, we focus on quiescent systems in this study.
We consider solutions in which any inflow, if present, 
   is subsonic everywhere in the cluster.

\subsection{central asymptotic behaviours} 

The asymptotic power-law behaviours of the variables in the central region 
  allows us to introduce a new set of variables: 
\begin{eqnarray}
	\beta_{\sigma_i}&\equiv&\sigma_i^2\ r
	\ ,\\
	\beta_{\rho_i}&\equiv&\rho_i\ r^{F_i/2}
	\ ,\\
	\beta_{v_i}&\equiv&v_i\ r^{(4-F_i)/2}
	\ ,
\end{eqnarray} 
   where $i=1$ for the gas and $i=2$ for the dark matter. 
These variables are finite at $r=0$. 
The corresponding equation of state is given by 
\begin{equation}
	\beta_{\sigma_i} = s_i \beta_{\rho_i}^{2/F_i}
	\   .  
\end{equation}  
We define a logarithmic radial coordinate $l\equiv \ln r$. 
In terms of the new variables, the gradient equations are now  
\begin{eqnarray}
	{{d\beta_{v_i}}\over{dl}} 
	&\hspace{-0.3cm}=&\hspace{-0.3cm}
 	 \beta_{v_i}
	  \bigg\{   {{4-F_i}\over{2}}  
	  -  {{1}\over{\gamma_i\beta_{\sigma_i}(1-\Mcal^2) }}
	   \bigg[ \  2\gamma_i \beta_{\sigma_i} 	
	   \cr    	
        & &\hspace*{3.4cm} - G m + {2\over{F_i}}\beta_L r^c \  \bigg]   	 \bigg\} \  , 	
\label{eq.beta.v}
\end{eqnarray}
\begin{eqnarray}
	{{d\beta_{\sigma_i}}\over{dl}}
	&\hspace{-0.3cm}=&\hspace{-0.3cm}
	\beta_{\sigma_i} + %{2\over{F_i}} 
     	{{\gamma_i-1}\over{	\gamma_i(1-\Mcal^2)}}
     \bigg[ \ 	 2\gamma_i\beta_{\sigma_i} \Mcal^2   \cr 
	& & \hspace*{2.4cm}
		-G m	 -(1-\gamma_i \Mcal^2)\beta_L r^c \  \bigg]	 \ ,
\label{eq.beta.sigma}
\end{eqnarray}
\begin{eqnarray}
	{{d\beta_{\rho_i}}\over{dl}}
	&\hspace{-0.3cm}=&\hspace{-0.3cm}
	\beta_{\rho_i}  
	\bigg\{ 	{F_i\over{2}}	+
		{{1}\over{\gamma_i\beta_{\sigma_i} (1-\Mcal^2)}}
		\bigg[ \  2\gamma_i\beta_{\sigma_i} \Mcal^2 \cr  
		& & \hspace*{3.4cm}
		-G m
		+{2\over{F_i}}\beta_L r^c  \ \bigg]
	   \bigg\}     \ ,   
\end{eqnarray}
  where $m = m_1 + m_2$. 
The entropy equation is 
\begin{equation}
	{{ds_i}\over{dl}}
	=
	- s_i\left\{(\gamma_i-1) {{B\beta_{\rho_i}r^c}\over{\beta_{v_i} \sqrt{\beta_{\sigma_i}}}}   \right\}
	\  . 
\end{equation} 
The cooling function  
\begin{equation}
\beta_L =  {{B\beta_{\rho_1} \sqrt{\beta_{\sigma_1}}}\over{\beta_{v_1}}}  \  ; 
\end{equation}  
  and the radial index of cooling term  
\begin{equation}
c \equiv {{7-2F_1}\over{2}} \ . 
\end{equation}  
The Mach number     
 $\Mcal^2 = \beta_{v_1}^2 r^{F_1-3} / \gamma_1 \beta_{\sigma_1}$, 
 and its profile is given by the equation:  
\begin{eqnarray}
	{{d\Mcal^2}\over{dl}}
	&\hspace{-0.3cm}=&\hspace{-0.3cm}
	{{\Mcal^2}\over{1-\Mcal^2}}
	\biggl[
	%-4\left({{\Mcal^2+F}\over{F}}\right)
	-4\left({
		{{\Mcal^2}\over{F_1}} + 1
	}\right)
		+(\gamma_1+1){{Gm}\over{\gamma_1\beta_{\sigma_1} }}
	\nonumber\\
	&&\hspace{2cm}
	-(1+\gamma_1 \Mcal^2)
	{{
		(\gamma_1-1)\beta_L\ r^c
	}\over{
		\gamma_1\beta_{\sigma_1}
	}}
	\biggr]
	\  ,  
\label{eq.dMMdl}
\end{eqnarray}  
  whose solution to the profile equation requires that  
\begin{equation}
\lim_{r\rightarrow 0}\Mcal^2
=\left\{{
\begin{array}{ll}
\infty&\mbox{if $F_1<3$,}
\\
\Mcal_*^2 > 0&\mbox{if $F_1=3$,}
\\
0&\mbox{if $F_1>3$\ .}
\end{array}
}\right.
\end{equation}

Note that the dark matter component does not have radiative cooling.  
This implies     
  $\beta_{v_2} = 0$,
  $d\beta_{v_2}/dl = 0$
  and $ds_2/dl=0$ throughout the system, 
  and the dark-matter structure is determined by
\begin{equation}
	{{d\beta_{\sigma_2}}\over{dl}}
	= \beta_{\sigma_2} - {{\gamma_2-1}\over\gamma_2} G m
	= \beta_{\sigma_2} - {{2Gm}\over{F_2+2}}
	\ ,
\label{eq.beta.sigma2}
\end{equation}
\begin{equation}
	{{d\beta_{\rho_2}}\over{dl}}
	=
	\left({
		{{F_2}\over{2}} - {{Gm}\over{\gamma_2\beta_{\sigma_2} }}
	}\right) \beta_{\rho_2}
	= {{F_2}\over{2s_2}}
		\beta_{\rho_2}^{(F_2-2)/F_2}\ {{d\beta_{\sigma_2}}\over{dl}}
	\ .
\end{equation}

The mass and moment of inertia are given by   
\begin{equation}
	{{dm_i}\over{dl}}
	= 4 \pi \beta_{\rho_i}\ r^{(6-F_i)/2}
\label{eq.beta.mass}
\end{equation}
  and
\begin{equation}
	{{dI_i}\over{dl}}
	= {{8 \pi}\over{3}} \beta_{\rho_i}\ r^{(10-F_i)/2} 
\end{equation} 
   respectively. 
The mass profile would have a steep, cuspy gradient near the origin for $F_i>6$.
The moment of inertia shows a central cusp when $F_i>10$, 
   but this $F_i$ corresponds to systems with infinite mass and radius, 
   which are unphysical and are irrelevant to astrophysical galaxy clusters.       
   
\subsection{numerical calculations}
\label{s.scheme}
   
Some numerical difficulties could arise 
   in solving the structure equations given in the above section when $F_i>6$. 
To overcome these we consider another coordinate 
\begin{equation}
   a\equiv {{2}\over{6-F_2}} r^{(6-F_2)/2} \ ,
\nonumber
\end{equation} 
   instead of $l$ and 
   the transformation  
\begin{equation}
	{{dl}\over{da}} = r^{(F_2-6)/2} = {2\over{(6-F_2)a}}    
\end{equation} 
   for the derivatives. 

We set the boundary conditions at the surface of the dark-matter component, $R$,  
    and the integration will proceeds inwards to the cluster centre. 
At the outer boundary ($r=R$), 
   we specify the total mass ($m(R)=m_1(R)+m_2(R)$),
   the matter inflow rate ($\dot{m}$),
   the specific entropy of dark matter (a constant $s_2>0$),
   the gas temperature ($T_R\equiv\sigma_1^2(R)$)
   and Mach number ($\Mcal_R\equiv \Mcal(R)$). 
The density and temperature of the dark-matter are zero at $r=R$. 
The entropy $s_2$ is non-zero, 
  and it defines the dark-matter density and temperature gradients.

An adaptive-step Runge-Kutta scheme \citep{NRinC} is used in the integration.  
We first integrate a small step radially inwards, 
    using $\sigma_2$ as the independent variable,
    to avoid numerical troubles that could be caused by the steep gradients 
    at the cluster boundary surface. 
We then proceed with the main integration, 
    using the variable $l$ or $a$,
    approaching a reference radius chosen to be $r=10^{-15}U_x$. 
(Here $U_x\equiv B/G$ is the natural unit of distance; appendix~\ref{s.units}.)
We examine how the variables behave near this radius.  
If the gradients become too steep,  
    we would consider alternative variables for the integration.  
There are two types of breakdown that may necessitate a switch.  
We name them as ``cold catastrophe'' and ``supersonic catastrophe'',  
   and will discuss each of them in more detail.  

The ``cold catastrophe'' arises when the cooling is too efficient, 
   causing the temperature to plummet steeply.  
We make use of  the local monotonicity of $s_1$  
   to define a new variable $z\equiv s_1^{1/2}$ for the integration. 
Although there are a steep radial gradients for the variables $\beta$,  
  i.e.\ the corresponding $|d\beta/dl |$ diverge,  
  the derivatives $d\beta/dz$ are still well-behaved and finite. 
Thus, it allows a smooth integration towards the centre, 
   where $z \rightarrow 0$ monotonically. 
Note that if the cooling catastrophe occurs at a non-zero radius
   -- forming a zero-temperature shell --
   we may discard it as it is not a viable steady solution.  
The cold shell lacks pressure support,  
   and material at large radius would fall inwards
   until a more stable configuration emerges.    
Such a system would have a variety of interesting dynamic behaviours, 
   and we will discuss it and related systems in a separate paper. 
 
The ``supersonic catastrophe'' arises 
  when the gas Mach number increases towards unity at a certain radius.
The $(1-\Mcal^2)^{-1}$ factor will diverge and create numerical problems.
As a resolution, we switch to $\Mcal^2$ as the independent variable 
  when the integration proceeds and approaches the sonic horizon ($\Mcal^2\rightarrow 1$). 
Two situations would occur.   
In the first one, the solution has a discontinuity, 
   with causal disconnection between the inner and outer regions. 
This is the shock solution, 
   which is interesting but does not correspond to steady galaxy clusters, 
   the prime interest in this work. 
The second one correspond to a smooth transonic flow, 
   in which the inflow would pass a sonic point \citep{bondi1952}, 
   beyond which the accretion becomes supersonic. 
The steady transonic solutions are valid, 
   but they tend to give lower gas densities than 
   the solutions with subsonic inflows throughout the entire cluster. 
This also implies a greater residual central mass $m_*$.
In this paper we seek to minimise $m_*$ 
   within the set of truly steady solutions,  
   and we prefer the wholly subsonic solutions.

\begin{figure}
\begin{center} 
\begin{tabular}{cc}
\includegraphics[width=8cm]{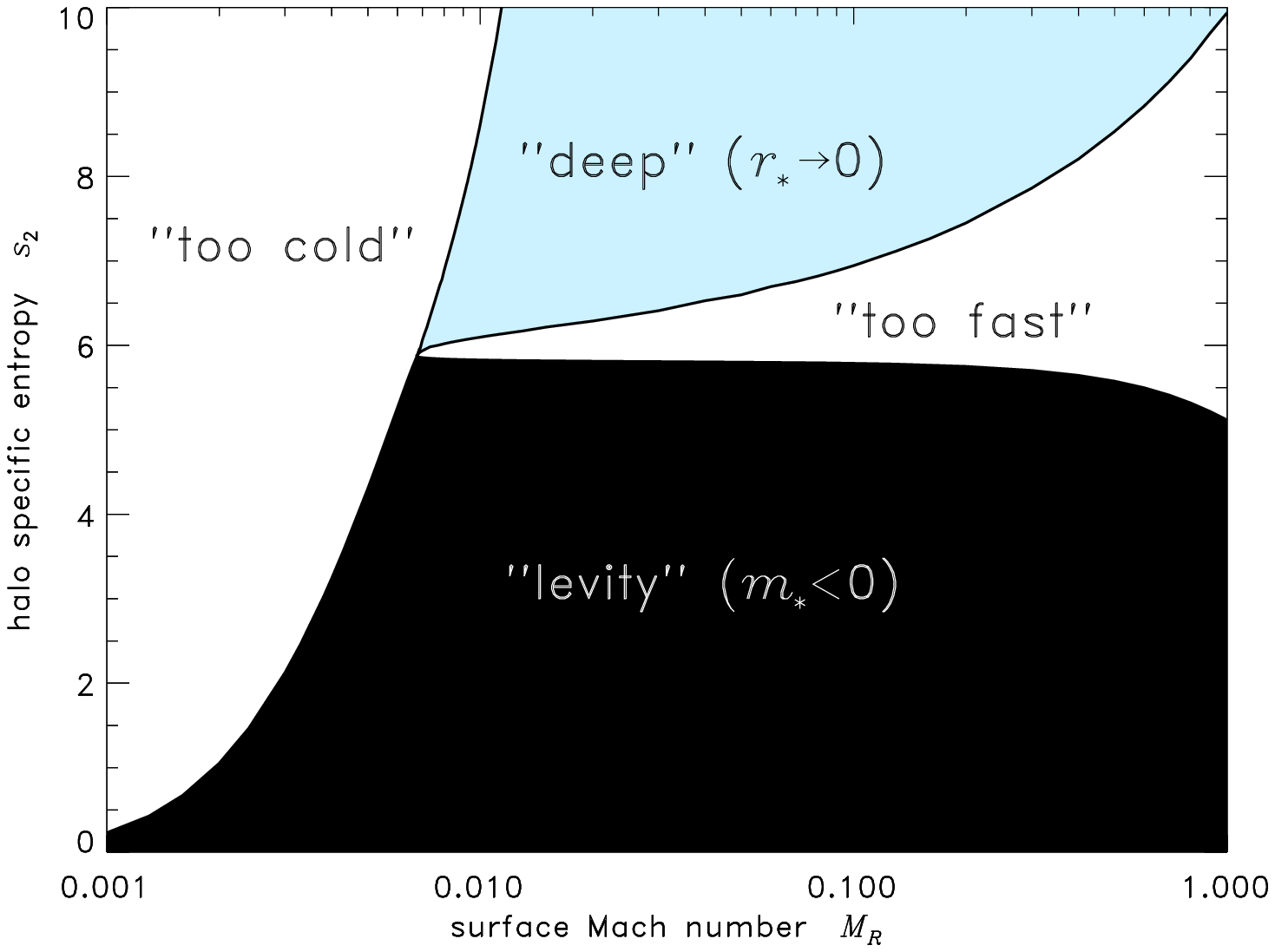}  
\\
\\
\includegraphics[width=8cm]{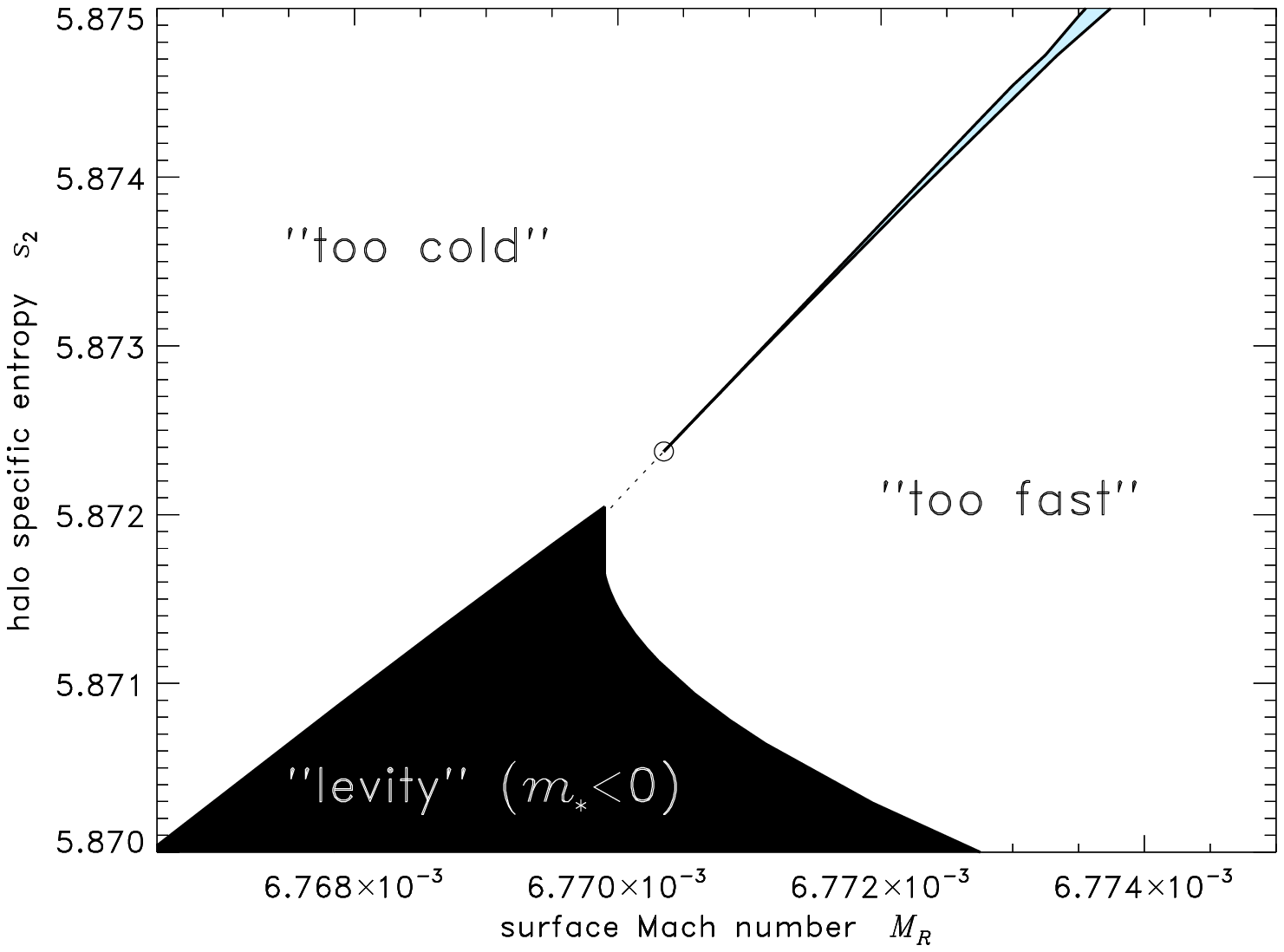}  
\end{tabular}
\end{center} 
\caption{ 
Broad and detailed maps of the key parameter domains of models
with standard mass $m(R)=40U_m\approx3.57\times10^{14}~m_\odot$,
radius $R=4U_x\approx0.983~\Mpc$,
inflow $\dot{m}=10 m_\odot~\yr^{-1}$,
gas surface temperature 1~keV
and degrees of freedom $F_1=F_2=3$.
The axes are the surface gas Mach number
and the dark specific entropy.
In a wedge-shaped domain
{%\color{MidnightBlue}
(``deep'', top panel)}
steady models reach from the dark surface to the origin.
To the right (``too fast'') the profile suffers a supersonic break
at an intermediate radius.
To the left (``too cold'') a cooling catastrophe occurs.
Models in the shaded region (``levity'')
require a negative central mass.
The ``fast'' and ``cold'' borders intersect at a point
(circled) above the $m_*=0$ contour.
Thus $m_*$ has a positive minimum for truly steady models.
}
\label{fig.zones}
\end{figure}

We consider various trial $(\Mcal_R,s_2)$ at $R$ in the integration. 
In each trial we record the radius, $r_*$ 
  where integration stops, and the central, interior mass, $m_*$.
For fixed $(F_1,F_2,R,m(R),\dot{m},T_R)$,
   we map the $(\Mcal_R,s_2)$ plane and divide it into zones 
   according to their physical and numerical characteristics 
  (see Figure~\ref{fig.zones}).
Qualitatively, we have four principal zones. 
Three of them are either unphysical
   or irrelevant to astrophysical galaxy clusters:  
(i) ``too cold'' --- afflicted by a cooling catastrophe at a certain radius; 
(ii) ``too fast'' --- containing a supersonic discontinuity; 
(iii) ``levity'' --- with insufficient pressure support,  
   implying a negative central gravitating mass in compensation. 
The acceptable, physical solutions lie in the wedge-shaped region  
   between the ``too cold'' and ``too fast'' zones. 
These solutions correspond 
   to steady structure and subsonic flow throughout the cluster. 
The $(\Mcal_R,s_2)$ values where
   the ``too cold'' and ``too fast'' boundaries intersect 
   depend on $(F_1,F_2,R,m(R),\dot{m},T_R)$.
The tip of this wedge region of the acceptable solutions 
   does not generally touch the contour where $m_*=0$,
   i.e. the boundary of the ``levity'' zone.  
The central mass $m_*$ of a physical cluster must be positive 
  and exceed some certain value.
All steady, self-gravitating, spherical, cooling multi-component clusters 
   would require a central point mass.
Strictly speaking, a spherical, cooling cluster
   with gas and dark matter composites  
   without a central mass condensation is never steady.  
It will eventually evolve into another configuration on a dynamical timescale.
Readjustment may start with the growth of a central mass,
   with waves of disturbances propagating outwards
   like the ``swallowing waves'' as described in \cite{mathews1971}.
In the later sections, we will present two of the mechanisms 
   that lead to the formation a central condensation 
   in multi-component galaxy clusters. 

\section{Steady inflow solutions}

\subsection{size, mass and compositional families}

\begin{figure}
\begin{center}
\begin{tabular}{ccc}
\includegraphics[width=8.2cm]{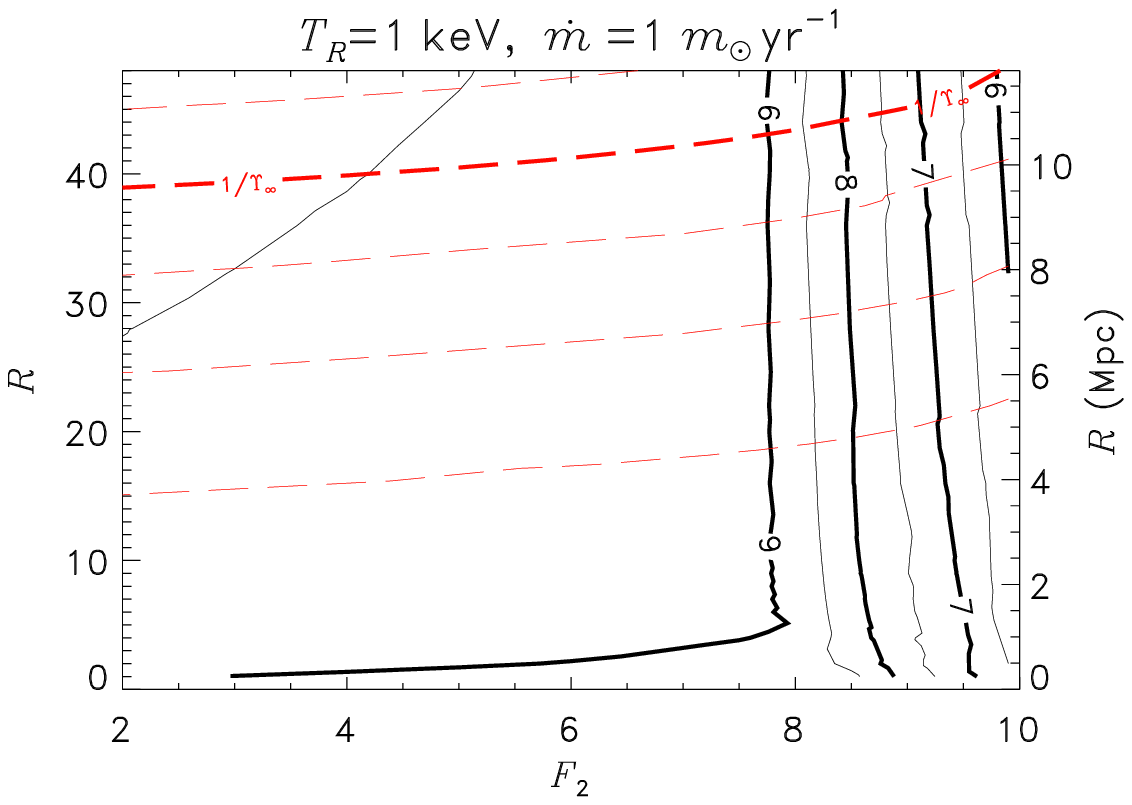}
\\
\includegraphics[width=8.2cm]{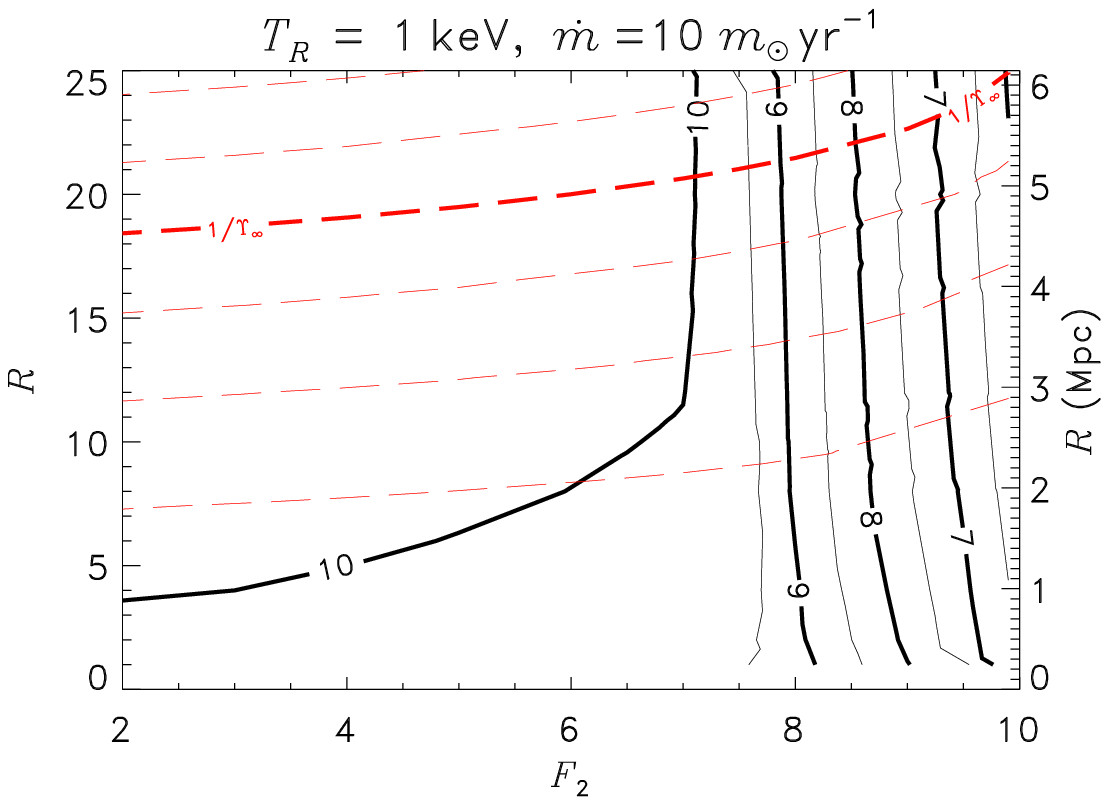}
\\
\includegraphics[width=8.2cm]{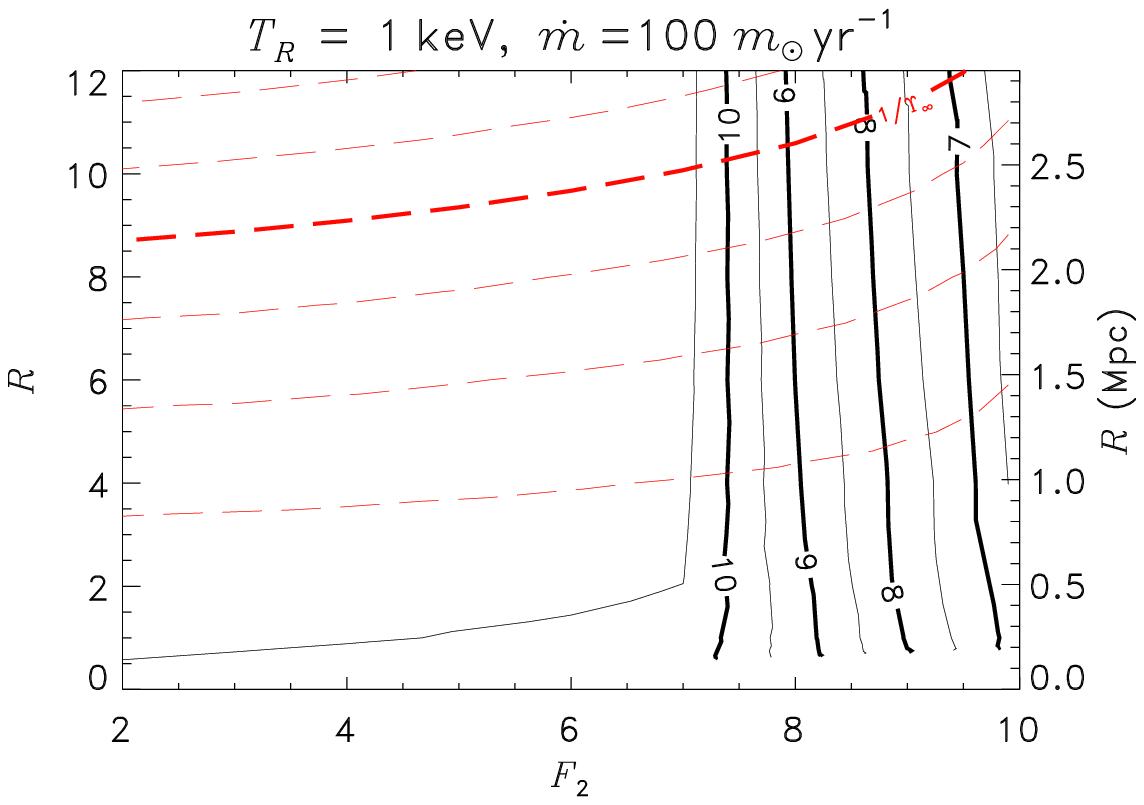}
\end{tabular}
\end{center}
\caption{
For a fixed cluster mass ($m=40$),
surface temperature ($T_R=1$~keV)
and inflow rate,
we vary the outer radius $R$ and dark degrees of freedom $F_2$.
Black/solid contours map the minimal values of
$\log_{10} (m_*/m_\odot)$, the central point mass.
Red/dashed contours show the gas fraction,
$1/\Upsilon$ relative to the cosmic baryon fraction
$({\frac14}, {\frac12}, {\frac34}, 1, {\frac54}\ldots)/\Upsilon_\infty$.
This sequence of panels shows the effect of varying the inflow rate,
with $\dot{m} = 1, 10$ and $100\ m_\odot\ {\mathrm yr}^{-1}$
from top to bottom respectively.
For large $\dot{m}$,
the $\Upsilon$-tracks occur at smaller radii,
and $m_*$ varies more widely along each track.
}
\label{fig.RF-maps}
\end{figure}

We now compare model clusters 
  with different inflow rates ($\dot{m}$), gas surface temperatures ($T_R$)
  and dark-matter degrees of freedom ($F_2$) for a given total mass $m(R)$.  
We choose to fix $m(R)=40U_m\approx3.56\times10^{14}~m_\odot$,  
  unless otherwise specified, 
  although the masses in the cluster solution are rescalable (see Appendix~\ref{s.scaling}).  
We fix $F_1=3$ for the gas.
The values of $\dot{m}$ span the range 
  $1~m_\odot~{\rm yr}^{-1}\le\dot{m}\le 1000~m_\odot~{\rm yr}^{-1}$,  
  inferred from X-ray imaging and spectral observations of cooling core clusters 
\citep[e.g.][]{fabian1981,nulsen1982b,stewart1984,white1994,edge1994}.
For given $(F_1,F_2,R,m,\dot{m},T_R)$,
  we minimise $m_*$ over the $(\Mcal_R,s_2)$ plane.
The contours of $m_*$ and gas fraction ($1/\Upsilon=m_1/m$)
   are plotted in the $(F_2,R)$ plane (Figure~\ref{fig.RF-maps}), 
   with other parameters held constant.

The radius of a minimal-$m_*$ cluster increases with $F_2$ 
  along a $\Upsilon$-track, 
All else being equal,  gas-richer tracks have larger cluster radii.
Each family of solutions shows a similar variation of $R$ with $F_2$ 
  when following a particular $\Upsilon$-track: 
  for cosmic composition, 
  we find that  $R(F_2=9)\approx1.229 R(F_2=2)$.
Also, $m_*$ decreases as $F_2$ increases in a $\Upsilon$-track.
However, for fixed $F_2$, $m_*$ varies with $R$.
As shown in the upper panel of Figure~\ref{fig.RF-maps},  
    a peak $m_*$ appears  for $R\approx1.25$~Mpc,  
    in the models with $F_2\approx8$, $\dot{m}=1\ m_\odot\ {\rm yr}^{-1}$ and $T_R=1~$keV.  
   
The $m_*$ contours behave qualitatively differently
in several distinct regions of the $(F_2,R)$ plane: \\  
%(i) {\em Plunge into dark freedom:}
(i) For $F_2\ga 7$, the values of $m_*$ drop steeply with increasing $F_2$.
In the case of $\dot{m}=1\ m_\odot\ {\rm yr}^{-1}$,
$m_*$ drops by a factor $\sim0.1$
for each increment of $1$ in $F_2$.
The drop is steeper for larger $\dot{m}$.
The cooling and sonic constraints permit a smaller central mass
if the central density profile is steep.
This occurs most easily for halos with more degrees of freedom. \\ 
%Prematurely halting integration at a larger inner radius
%reduces the plunge in $m_*$  
%(Figure~\ref{fig.RF-map-cutoff}).  
%(ii) {\em Gas-poor slope:}
(ii) In another regime,
with small radius $R$,
both the gas fraction and the minimal $m_*$ drop steeply with decreasing $R$,
regardless of $F_2$.
The poverty of gas loosens the ``cold'' and ``fast'' constraints,
enabling smaller $m_*$.
As $R$ shrinks,
  the solution approaches that of the Lane-Emden ideal polytrope 
  (which lacks a central mass)
   or else it becomes a point mass lacking both halo and gaseous envelope. \\ 
%(iii) {\em Plateau:}
(iii) The rest of the $(F_2,R)$ plane is a relatively featureless plateau
(top-left region of Figure~\ref{fig.RF-maps}):
$m_*$ increases only slightly 
  even when there is a large  increase in $R$.    
The $\Upsilon$-tracks,
   however, vary smoothly across the borders from plateau
   to the low-$m_*$ slopes.
This insensitivity of $\Upsilon$ occurs because
   the determination of bulk composition is global,
  whereas $m_*$ is governed by local gas constraints
   acting in local bottlenecks of the inflow at small radii.

The attainable range of $m_*$ values across the $(F_2,R)$ map
depends on the gas inflow rate, $\dot{m}$.
Smaller $\dot{m}$ reduces variation in $m_*$,
with lower values on the plateau.
The three panels of Figure~\ref{fig.RF-maps}
compare families of solutions that differ only in $\dot{m}$.
The $m_*$ contours are completely different for different choices of $\dot{m}$.
For $\dot{m}=1\ m_\odot\ {\rm yr}^{-1}$,
the cosmic-$\Upsilon$ track crosses $\sim4$ orders of magnitude in $m_*$.
For $\dot{m}=10\ m_\odot\ {\rm yr}^{-1}$,
the equivalent track crosses almost $\sim4.5$ orders of magnitude.
For $\dot{m}=100\ m_\odot\ {\rm yr}^{-1}$,
the track crosses $\sim5$ orders of magnitude.
Note that increasing $\dot{m}$
also shifts each $\Upsilon$-track to smaller radii,
i.e. for a given composition, clusters with strong inflows
tend to be more compact.

\begin{figure}
\begin{center}
\begin{tabular}{c}
\includegraphics[width=8.2cm]{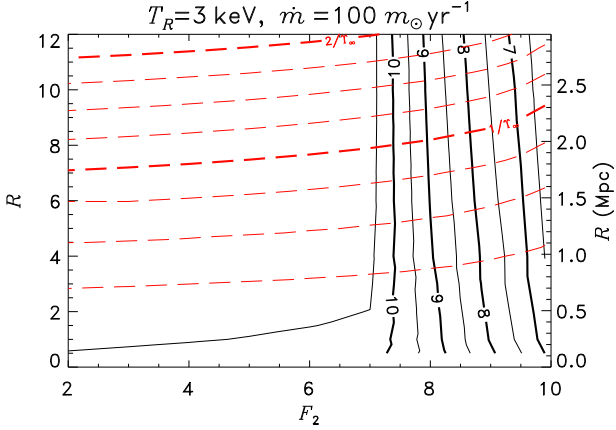}
\end{tabular}
\end{center}
\caption{
Minimal-$m_*$ and $1/\Upsilon$ map
corresponding to the bottom panel
($100\ m_\odot\ \yr^{-1}$)
of Figure~\ref{fig.RF-maps},
but for a warmer (3keV) cluster surface.
The $m_*$ contours are almost unchanged.
The tracks for given $1/\Upsilon$ occur at smaller radii.
}
\label{fig.RF-map-hot}
\end{figure}

Increasing the gas surface temperature $T_R$
(with everything else fixed)
shifts the $\Upsilon$-tracks to smaller radii.
However varying $T_R$ has negligible effect on the $m_*$ contours.
Thus the $m_*=m_*(F_2)$ profile of a given track
shifts to slightly lower masses,
but this is only due to migration of the track
across fixed $m_*$-contours.
(Compare
the lower panel of Figure~\ref{fig.RF-maps}
with Figure~\ref{fig.RF-map-hot}.
Thus, all else being equal,
a hotter cluster is a smaller cluster
but with a similar central mass.

\begin{figure}
\begin{center}
\begin{tabular}{c}
\includegraphics[width=8.2cm]{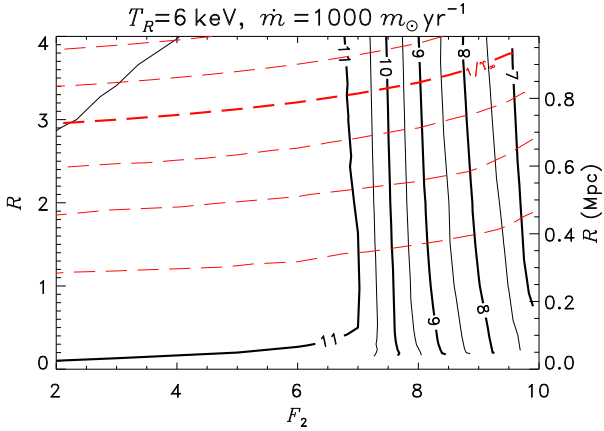}
\end{tabular}
\end{center}
\caption{
Minimal-$m_*$ 
and $\Upsilon$-track map
as in Figure~\ref{fig.RF-maps},
but with hotter gas and more inflow
($T_R=6$~keV,
$\dot{m}=1000~m_\odot~{\rm yr}^{-1}$).
This map includes our most compact solutions
of a given $\Upsilon$.
}
\label{fig.RF-map-compact}
\end{figure}

\subsection{external atmosphere}
\label{s.atmosphere}
\label{s.obc}

As $\dot{m}$ is a constant at all radii,
  all stationary solutions have some tenuous gas
  extending indefinitely far beyond the dark halo.
This atmosphere must lose its identity in the Hubble flow at some point.
In this work we take this cosmic atmosphere or accretion flow for granted.
We will not consider its distribution in detail,
   but briefly discuss the qualitative implications of two scenarios.

We may suppose that infinite atmosphere is a 
   continuation of the cooling, polytropic gas inflow
   but without a dark component.
The density attenuates with distance,
   and bremsstrahlung cooling becomes negligible.
The effects of local self-gravity dwindle.
If no sonic horizon appears,
   then the asymptotic atmospheric structure follows some power-law decline.

Alternatively,  we may choose to interpret the halo surface gas 
  as a cosmic accretion shock \citep[e.g.][]{bagchi2006}.
Its temperature depends on the cluster mass and radius.
In practice we select $m(R)$ and $T_R$.
Then shock conditions constrain the plausible range of radii,
\begin{equation}
	{{F_1}\over{(F_1+1)^2}} {{Gm}\over{T_R}}
	\le R \le
	{{F_1}\over{F_1+2}} {{Gm}\over{T_R}}
	\ ,
\label{eq.shock.limit}
\end{equation}
with the lower and upper limits
corresponding to strong and weak shock extremes respectively.
More exact constraints emerge if we consider the post-shock Mach number
and pre-shock cooling.
Not all families of models
enable the cosmic $\Upsilon$-track
to satisfy (\ref{eq.shock.limit}).
For $\dot{m}=1\ m_\odot\ {\rm yr}^{-1}$
and $T_R=1$~keV,
the radii of $(m_*,\Upsilon)$-optimal solutions are too large,
however a cooler family of models with $T_R=0.4$~keV is satisfactory.
By adjusting $\dot{m}$ and $T_R$ in the opposite direction,
we obtain hotter and more compact cluster models
that also suit a shock interpretation
(e.g. Figures~\ref{fig.RF-map-hot}
and \ref{fig.RF-map-compact}).

For very high $T_R$  the atmosphere would extend infinitely,
  with pressure dropping to some asymptotic value,   
  and temperature rising as a power-law.
This restricts the upper limit of $T_R$ for physical solutions. 
Such phenomena also occur in the adiabatic, hydrostatic clusters  
 \citep{gull1975}, where  
  below some threshold (effectively a minimum $T_R$),
  a gas inflow truncates at finite radius \citep{mathews1978}. 
 
\begin{figure}
\begin{center}
\begin{tabular}{c}
\includegraphics[width=8.2cm]{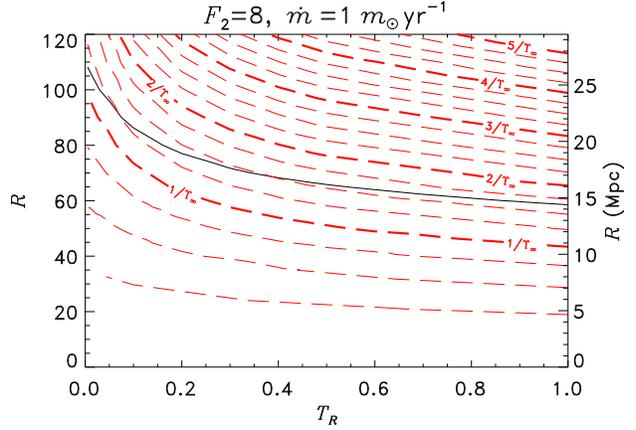}
\end{tabular}
\end{center}
\caption{
Minimal-$m_*$ and $1/\Upsilon$ map
for $F_2=8$,
$\dot{m}=10~m_\odot~\yr^{-1}$
and variable surface temperature $T_R$.
The $\Upsilon$-tracks shift,
but the minimal $m_*$ hardly changes.
}
\label{fig.varyT}
\end{figure}

In our solutions, flow continuity (constant $\dot{m}$)
means that the sub-critical gas atmospheres
can break at an external sonic point ($\Mcal^2=1$) with non-zero density.
It is unclear what external conditions
   should match onto such a supersonic break.
Reducing $T_R$ further
   causes the gas profile to break somewhere inside the halo, $r<R$.
These are not numerically feasible, searchable solutions,
   and we avoid them.
We have coarsely scanned the parameters $(T_R,R)$ for fixed $(F_2,\dot{m})$
and found little qualitative variation in the inner profiles
or minimal-$m_*$ values, aside from the $\Upsilon$-tracks shifting
(see Figure~\ref{fig.varyT}).

\subsection{radial structure of particular clusters}

\subsubsection{general properties and density profiles}

We now examine the internal structures of specific clusters in detail.
Here we discuss only the minimal-$m_*$ models
where the overall composition is cosmic,
$1/\Upsilon\approx0.163$.
Table~\ref{table.models}
   lists the parameters and some global properties of these cluster solutions.
In each model we chose $F_2$ for the halo
   and $(\dot{m},T_R)$ for the gas,
   then tuned the cluser radius $R$ to obtain cosmic composition.
We tabulate signature radii of the models
   defined in Appendix~\ref{s.metrics}:
$R_{I_1}, R_{I_2}, R_I$ are effective core or lever radii,
   weighted by inertial moments of gas, the halo and both combined;
$R_w$ characterises the concentration of gravitational potential energy;
$R_1, R_2, R_3, R_4$ are radii where the total density
   has a radial logarithmic slope of
   $-1, -2, -3$ and $-4$;
$R_o$ is the outermost peak of the rotation curve.
Appendices~\ref{s.gasless}--\ref{s.comparisons}
   compare these signature radii to other spherical models in the literature
   (see Table~\ref{table.halos} for data).

The gas + halo models differ from gasless models
   (Appendix~\ref{s.gasless})
   in several key respects.
For $F_2\ga 7$ the gassy clusters are gravitationally more compact:
   $R_w/R$
   is smaller than for corresponding gasless, non-singular polytropes.
The central mass and dense cusp deepen the potential well significantly
   (especially when $F_2$ is large).

The concentration of gas mass ($R_{I_1}/R$) is rather insensitive to $F_2$.
In all of our minimal-$m_*$ models.
   the gas is less centrally concentrated
   ($0.74 \la R_{I_1}/R \la 0.81$)
   than in a simple $F=3$ polytrope
   ($R_{I}/R\approx0.715$).
The presence of gas affects the halo concentration ($R_{I_2}/R$),
   depending on $F_2$. 
For $F_2=2,3$, the dark mass is slightly more concentrated
   (smaller $R_{I_2}/R$ than for gasless spheres).
For larger $F_2$, gas makes the halo less concentrated
   (larger $R_{I_2}/R$ than for polytropes).
The combined mass distribution has an effectively intermediate concentration:
   either $R_{I_1}<R_{I}<R_{I_2}$ or $R_{I_2}<R_{I}<R_{I_1}$.

The rotation curve peaks farther out than in gasless halos
   (Lane-Emden spheres) of the same $F_2$,
   and moreso for large $F_2$.
Specifically, $R_{\rm o}$ enlarges by $\sim$1\% for $F_2=2$
   but by $\sim$20\% for $F_2=9$ cases.
In all cases we find that $R_{\rm o}>R_3$.
Importantly, this means that the rotation curve doesn't peak
   until outside a radius where the combined density slope
   is steeper than $-3$.
For optimised gassy models with $F_2=2,3$ we find $R_{\rm o}>R_4$;
   however we find $R_3<R_{\rm o}<R_4$
   for $F_2\ge 5$ generally
   (and for $F_2\ge 4$ for the compact models).
In contrast, the gasless halos have $R_2<R_{\rm o}<R_4$ for $F_2=2,3,4,5$.
For real, relaxed galaxy clusters,
   the comparison of measured $R_3$ and $R_4$
   (e.g. from gravitational lensing at the outskirts)
   and of $R_{\rm o}$ (e.g. via member galaxy kinematics)
   could constrain the actual effective value of $F_2$,
   and enable extrapolation of the halo radius $R$.

\begin{figure*}
\begin{center}
\includegraphics[width=13cm]{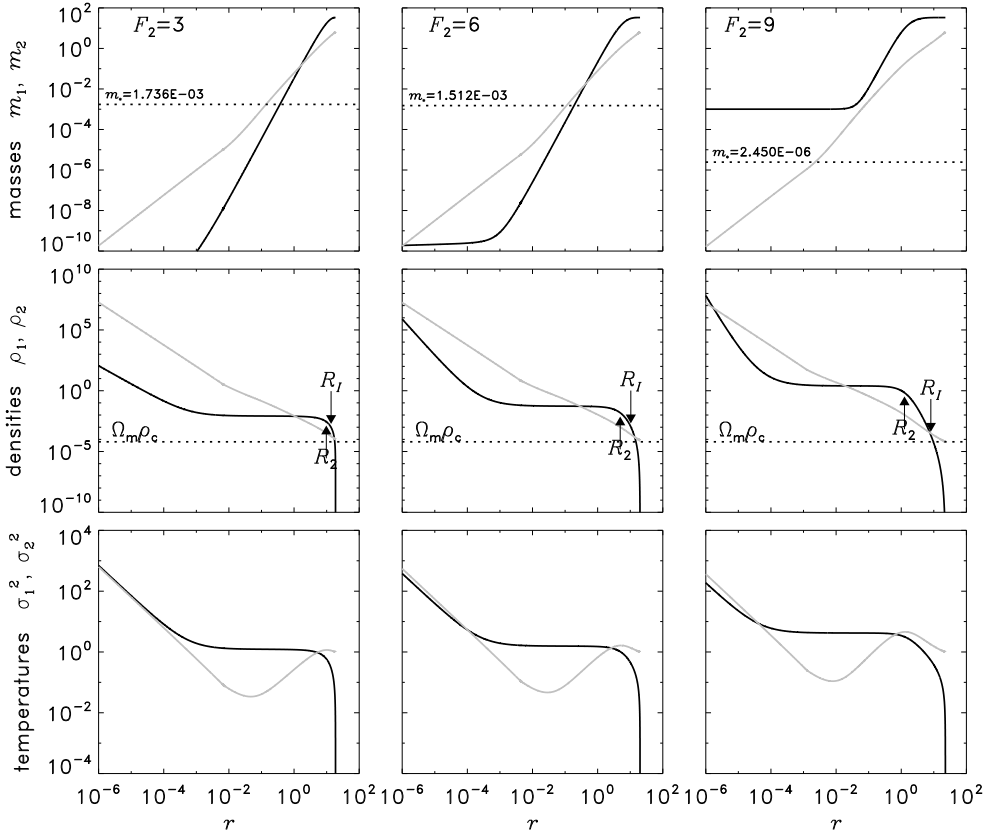}
\end{center}
\caption{
Steady inflow structure solutions for models
with
with the same mass,
$\dot{m}=10~m_\odot~{\rm yr}^{-1}$,
$T_R=1$~keV
but $F_2=3,6,9$,
in columns from left to right respectively.
Gas and dark matter properties are marked in grey and black respectively.
Top panels show
the masses interior to radius $r$,
with the central mass $m_*$ marked by a dotted line.
Middle panels show densities,
compared to the cosmic mean density (dotted line).
The inertial ``core'' radius $R_I$
and the slope-2 radius $R_2$ of the total density
are marked with arrows.
The bottom panels are temperature profiles.
The natural radial unit depends on gravity and bremsstrahlung constants,
$1U_x\equiv B/G\approx0.246$~Mpc.
}
\label{fig.steady}
\end{figure*}

\begin{figure*}
\begin{center}
\includegraphics[width=13cm]{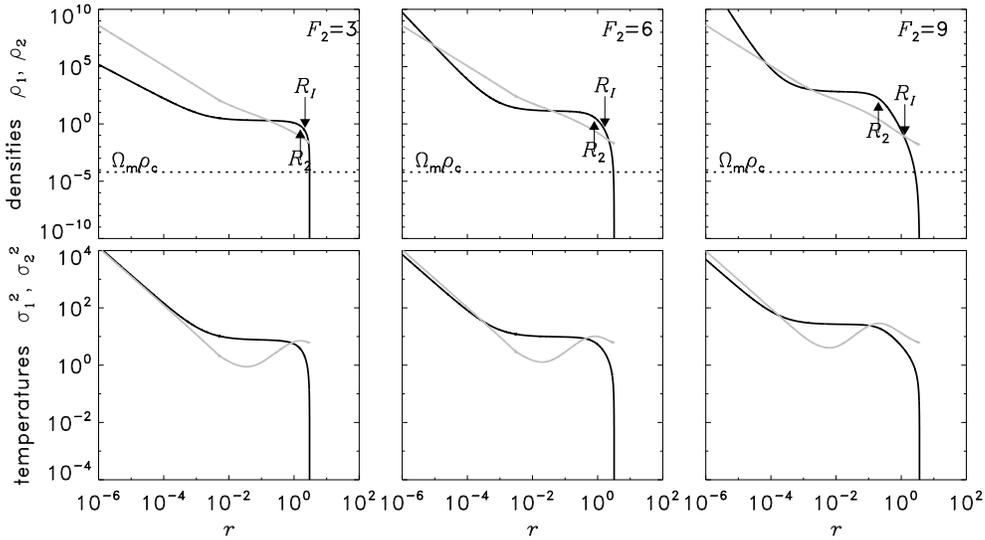}
\end{center}
\caption{
Density (top) and temperature (bottom) profiles
as in Figure~\ref{fig.steady}
but with $\dot{m}=1000~m_\odot~{\rm yr}^{-1}$
and $T_R=6$~keV.
These are the hottest and most compact models.
They have the smallest temperature ratio $T_{\rm max}/T_{\rm min}$
of the $(m_*,\Upsilon)$-optimised models.
}
\label{fig.steady.e}
\end{figure*}

\begin{figure*}
\begin{center}
\includegraphics[width=13cm]{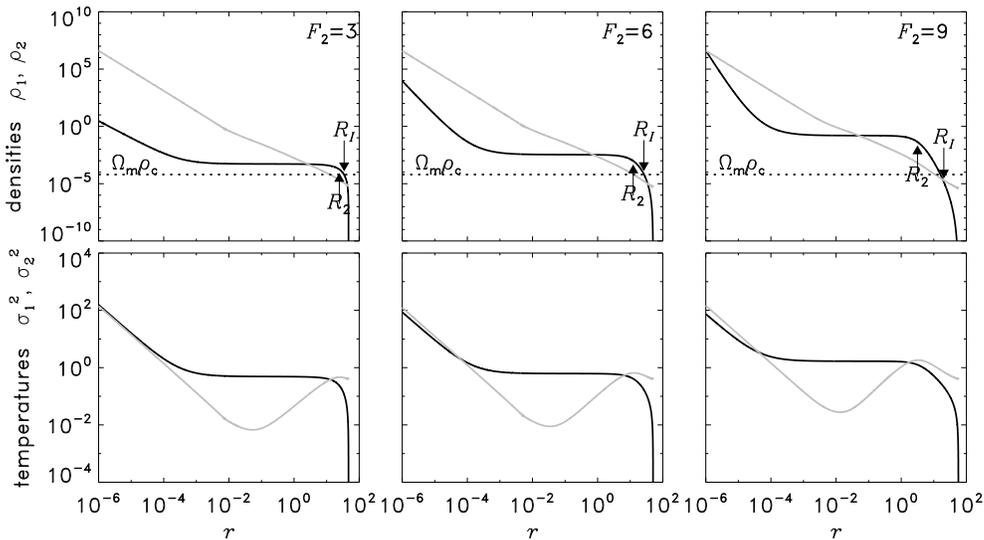}
\end{center}
\caption{
Cluster profiles
as in Figure~\ref{fig.steady.e}
but with $\dot{m}=1~m_\odot~{\rm yr}^{-1}$
and $T_R=0.4$~keV.
These clusters are very radially extended,
and have a large temperature variation ($T_{\rm max}/T_{\rm min}$) in the ICM.
}
\label{fig.steady.g}

\end{figure*}

Figure~\ref{fig.steady}
   illustrates the radial structure of our baseline minimal-$m_*$ solutions
   with $\dot{m}=10\ m_\odot\ {\rm yr}^{-1}$
   and $T_R=1$~keV,
   but differing in $F_2$.
Figure~\ref{fig.steady.e} depicts comparable models
   with a stronger gas inflow, hotter surface and smaller radius.
Figure~\ref{fig.steady.g} shows solutions with
   weak inflow, and a cool surface at large radius.
The gas density profile is monotonic in radius,
   approximating a broken power-law
   with the break appearing at kiloparsec scales,
   and a slightly shallower slope on the outside.
In the outer parts, the index of $\approx -1$
   is consistent with the simplest early models of cooling flows
   \citep[e.g.][]{cowie1977,fabian1977,mathews1978}.
As expected from analysis,
   in the innermost regions
   both the gas and halo have singular density profiles
   approximating a Bondi accretion flow,
   $\rho_i \propto r^{-F_i/2}$ for both gas and dark matter.
The dark cusp is radially smaller than the gas cusp.
Note that the dark cusp emerges for different reasons than
   the cusps of hypothetical collisionless halos in N-body simulations.
The dark cusp emerges as
   a self-consistent, time-independent, hydrostatic response
   to the central mass $m_*$ and the gaseous mass inflow.
In the latter sense it is related to an ``adiabatic contraction'' effect,
   \citep{blumenthal1986}.
It is not a time-dependent relic
   of cosmic structure formation or merger history.

The dark halo density slope varies more than that of gas.
A core of approximately constant dark matter density surrounds the cusp, 
   spanning from ten to hundreds of kiloparsecs' radius.
The halo outskirts
   are a rapid decline to zero density at the surface.
As in gasless models (Appendix~\ref{s.gasless}),
   halos with fewer dark degrees of freedom exhibit a larger core.
The density gradients of the dark matter  in the core are flattest in the cool, puffy solutions
   (e.g. Figure~\ref{fig.steady.g})
   than in the more compact cases, where the slope is appreciably nonzero
   (e.g. Figure~\ref{fig.steady.e}).

Locally, the dark matter density outweighs the gas
   in some but not all layers of each cluster.
Constancy of $\dot{m}$ at the dark matter surface
   means that $\rho_1 > \rho_2$ in a thin surface layer.
However the halo density dominates gas throughout most of the volume,
  as far inwards as the core radius and deeper.
This halo-dominated layer is thicker when $F_2$ is larger:
e.g. reaching in to $r \sim 2$~kpc in the $F_2=9$ model shown in
Figure~\ref{fig.steady}.
For modest $F_2$ ($\la 8$),
   gas dominates the Bondi-like accretion region of the deep interior.
For larger $F_2$, the steep dark matter density cusp dominates over gas
in the innermost regions.
In the $F_2=9$ case shown, we have $\rho_2 > \rho_1$
on sub-parsec radii near the origin,
beneath a gas-rich layer several kiloparsecs thick.
The layers dominated by dark matter density are smaller 
   for the hot, compact solutions (top row, Figure~\ref{fig.steady.e}
   than for cool, extended clusters (Figure~\ref{fig.steady.g}).
In principle, this might become observationally apparent
   in cD galaxy kinematics
   if the gaseous, stellar and dark components could be distinguished perfectly.

The gravitating mass at the outskirts is predominantly dark,
   $m_2(r)>m_1(r)$,
   which follows naturally from the assumption of cosmic baryon fraction.
For models with small $F_2$,
   (e.g. $F_2=3$ in Figure~\ref{fig.steady}),
   the gas mass dominates within $r\la 100$~kpc,
   and the central mass $m_*$ is the dominant component farther in 
   (e.g. $r\la 10$~kpc in this example).
Since halos with larger $F_2$ are more concentrated
   (smaller $R_{I_2}/R$)
   the dark-dominated part of the mass profile is thicker for greater $F_2$
   (middle and right upper panels of Figure~\ref{fig.steady}).
Increasing $F_2$ shrinks the minimal $m_*$,
   so the central object becomes less dynamically significant too.

\begin{figure}
\begin{center}
\begin{tabular}{c}
\includegraphics[width=8.2cm]{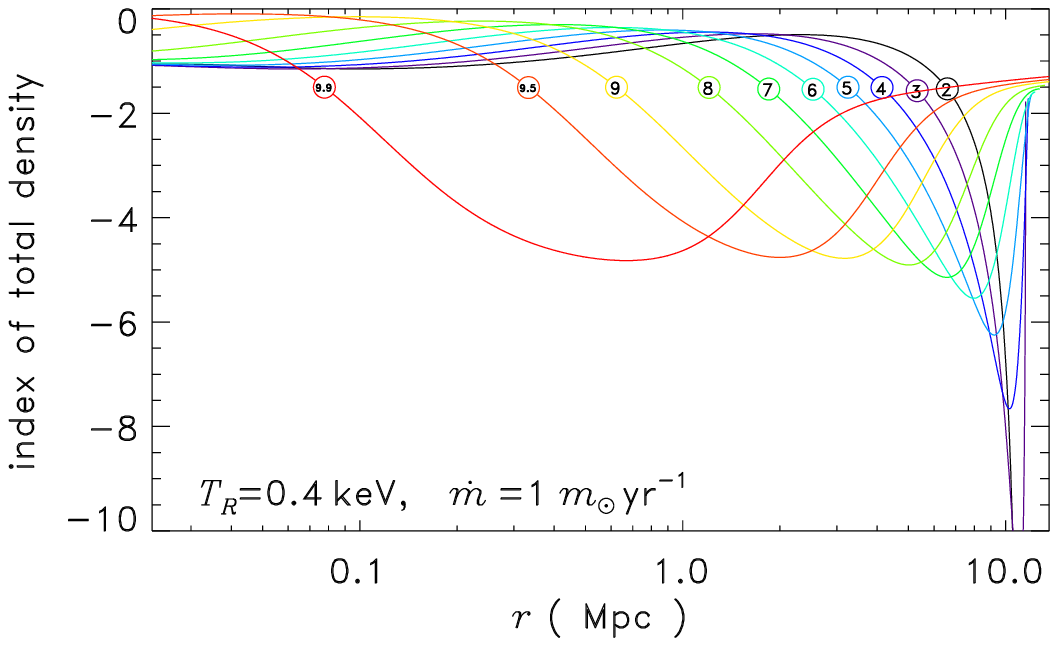}\\
\includegraphics[width=8.2cm]{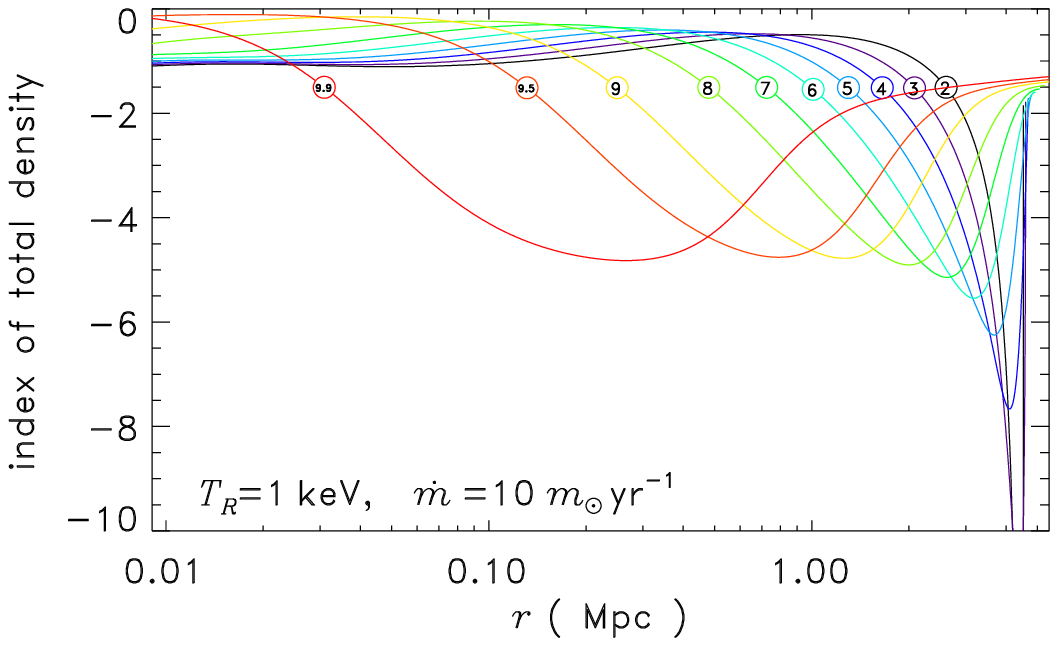}\\
\includegraphics[width=8.2cm]{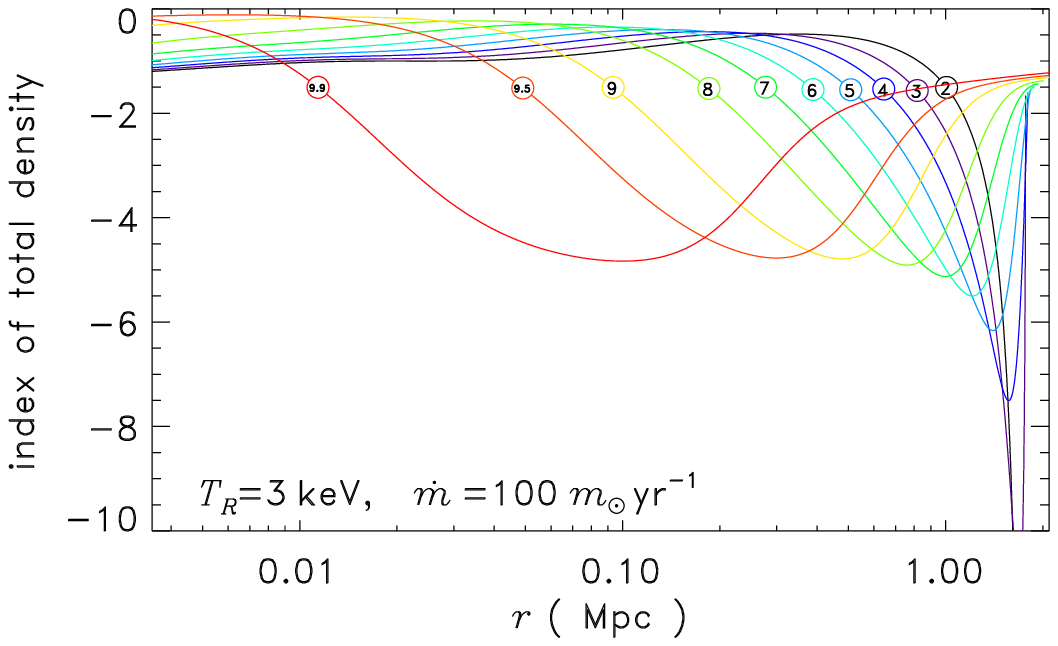}\\
\end{tabular}
\end{center}
\caption{
Profiles of the radial log-index of the total density $\rho_1+\rho_2$,
for $(m_*,\Upsilon)$-optimal solutions with $F_2$ values annotated.
Panels from top to bottom show families 
$(\dot{m}/m_\odot~\yr^{1},T_R/\keV)=(1,0.4)$,
$(10,1)$ and $(100,3)$
respectively.
}
\label{fig.slopes}
\end{figure}

Figure~\ref{fig.slopes}
   shows radial variations of the logarithmic index of the total density
   ($\rho_1+\rho_2$)
   for some of the $(m_*,\Upsilon)$-optimal solutions.
Cases with lower $F_2$ are flatter out to larger radii,
   as the dark core is larger.
The gas inflow tends as
   $\rho_1\propto r^{-1}$ (or steeper) near the centre,
   and this contribution prevents the overall index from reaching zero exactly,
   even deep within the dark core.
The steepest index is $-4$ or lower,
   occurring where the dark fringe drops.
Gas dominates increasingly at larger radii,
   bringing the index up to $\approx-1.3$ near the halo surface.
From panel to panel in Figure~\ref{fig.slopes},
   curves of a given $F_2$ but different $(\dot{m},T_R)$
   look alike except for a radial dilation.
With $\Upsilon$ implicitly fixed,
   the halo parameter $F_2$ controls the proportions of the core
   relative to the outer surface $R$
   (see radii ratios in Table~\ref{table.models}
   and Appendices~\ref{s.metrics},\ref{s.gasless}).
Thus, observing a few signature radii of a real cluster
   could constrain its $F_2$ and $R$.
As a consistency check, satellites orbiting beyond $R$
   ought to exhibit Keplerian motion.
We may disfavour some solutions based on their radii:
   the family of $(1000,6)$ models are smaller than $1~\Mpc$;
   %cannot describe clusters larger than $r\approx2~\Mpc$;
   while the bloated family $(1,0.4)$ exceed $10~\Mpc$
   (too cosmologically large).

Many X-ray, kinematic and gravitational lensing observations of clusters
   find total density indices dropping with $r$
   from tens of kiloparsecs outwards.
X-ray analyses typically assume hydrostasis,
   and probe out to radii of a few hundred kiloparsecs
   (or $\sim$Mpc in rare cases).
Strong gravitational lensing also constrains mass profiles
   out to $10^2$~kpc radii,
   while weak lensing gives statistical evidence at Mpc scales.
In many instances where measured indices happen to range from $-1$ to $-3$,
   this is presented as support for NFW-like profiles
   \citep[][see Appendix~\ref{halo.nfw}]{nfw1996}.
Our reference models show similar indices at comparable radii,
   but with crucial differences in the core (flatter)
   and fringe (steeper, then the halo truncates).
The present paper does not attempt to fit specific clusters,
   but will compare model properties
   to results in observational literature.
Such comparisons are tentative:
   the commonly assumed $\beta$-model density law
   (Appendix~\ref{halo.hubble})
   %\citep[][see Appendix~\ref{halo.hubble}]{cavaliere1976}
   may over-flatten the central gas,
   overestimating the halo density.
Likewise, gravitational lens models involve subtle degeneracies
   \citep[e.g.][and their references]{saha2006,liesenborgs2008}
   that might confuse cores with cusps.

Dipolytrope models can naturally describe
   those clusters observed to have soft cores.
Some X-ray deprojection studies
   have fitted parametric halo models assuming a density cusp,
   and a few appear softer than CDM predictions.
\cite{katayama2004}
   find an index $0.47\pm0.31$ in the cental 100~kpc of A1835.
\cite{ettori2002}
   indicate $\sim0.6$ in A1795.
\cite{ettori2002b} prefer a modified Hubble model (\S\ref{halo.hubble})
   over NFW fits for 10/20 clusters,
   giving $r_{\rm s}$ of a few hundred kpc.
\cite{voigt2006}
   found indices $<1$ in 4/12 of their clusters.
\cite{zhang2006} fit a wide scatter of flattish cusp indices among 13 clusters.
Combining gravitational lensing with stellar kinematics of the cD galaxy,
   \cite{sand2002,sand2004,sand2008}
   find indices $\sim0.6$,
   and $<1$ confidently.
\cite{rzepecki2007} finds an index $\approx0.7$.
These results at radii $<100$~kpc
   are consistent with the shallow regions ($r<R_1$)
   of many curves in Figure~\ref{fig.slopes}.
However the most bloated family
   (upper panel, $\dot{m}=1~\msol~\yr^{-1}$, $T_R=0.4~\keV$)
   flattens through $R_1$ at Mpc scales,
   (implausibly large) disfavouring cases with $F_2\le9.5$.
Among the compact family
   (lower panel)
   the $F_2=9.5$ and $9.9$ curves are too steep in relevant ranges.
The medium cases (middle panel)
   or their homologous relatives (Appendix~\ref{s.scaling})
   are more likely representations of normal cored clusters.

Many gravitational lensing studies treat cluster cores
   as pseudo-isothermal spheres (PIS, Appendix~\ref{halo.pis}),
   which bear comparison to dipolytrope cores.
Given any empirical PIS core size $r_{\rm s}$,
   we can infer $R_2>R_1\approx r_{\rm s}$.
Then one can extrapolate $R$ from Table~\ref{table.models} ratios,
   and this should enclose the observable cluster.
\cite{dahle2003}
   found $r_{\rm s}\approx66~\kpc$ for an ensemble of clusters.
Appraising that this core is too small,
   they rejected fluid-SIDM models of the day.
However in our calculations this core size
   predicts a plausible halo surface radius of $R\ga1.9, 3.8$ or $17~\Mpc$
   (for $F_2=9.0, 9.5$ and $9.9$ respectively).
PIS and NFW fits by \cite{ettori2002}
   imply $R_1\sim0.10$~Mpc and $R_2\sim0.49$~Mpc in A1795.
\cite{diego2005} fit $r_{\rm s}\approx15~\kpc$ for A1689.
This is awkwardly small, favouring higher $F_2\approx10$.
However
   \cite{broadhurst2005a} found a core $\sim200~\kpc$ in the same cluster,
   and a fringe truncating around 2~Mpc (favouring $F_2\approx7$).
\cite{halkola2006} found a similar profile, but with $r_{\rm s}\approx66~\kpc$,
   which constrains ($F_2,R)$ like \cite{dahle2003}.
\cite{limousin2007} 
   fit two halo core elements with $r_{\rm s}\approx99~\kpc$ and $66~\kpc$.
For RX~J1347-1145, a very massive cluster,
   \cite{halkola2008} find a core $r_{\rm s}\approx117~\kpc$
   while \cite{bradac2008b} fit $160~\kpc$.
A core larger than A1689's befits a heavier system,
   assuming universal values of $F_2$ and $R_1/R$.
\cite{rzepecki2007} find flat cores in RCS0224-002,
   with $r_{\rm s}\approx112~\kpc$ and $12~\kpc$.
We interpret the larger measurement as the true halo core,
   and the smaller feature as baryon-induced contraction.
\cite{saha2008} find cuspy profiles for $r>25~\kpc$ in ACO~1703,
   but possibly a density shelf at $\approx100~\kpc$.
They suggest the shelf is meso-structure;
   we suggest an innate core with a partly contracted interior.

Simple collisionless dark matter models
   predict $\rho_2\propto r^{-3}$ asymptotically forever,
   whereas we predict ever steeper indices until finite truncation
   at megaparsec scales.
In future, cluster outskirts will become key observational tests of halo models.
Deeper exposures from newer-generation X-ray observatories are needed,
   and more conclusive gravitational lens models would help.
Already there are lensing hints of outskirts steeper than NFW
   \citep{diego2005,broadhurst2005a,umetsu2008}.
An X-ray deprojection by \cite{nevalainen1999}
   implies a periphery with $\rho_2\propto r^{-4}$.
A much wider cluster X-ray deprojection out to $r\approx1.7~\Mpc$
   \citep{george2008}
   shows that a hydrostatic NFW-based model cannot fit observations,
   because of an excess of mass or a deficit of gas pressure in the fringe.
We interpret this as evidence for a finite polytropic halo.

\begin{table*}
\caption{
Parameters and global properties of the minimal-$m_*$ cluster models.
We fix a fiducial total mass,
$m=40U_m\approx 3.57\times10^{14}m_\odot$,
and seek cosmic composition,
$1/\Upsilon\approx0.163$, inside $R$.
%$1/\Upsilon = 0.16296\pm0.00004$, inside $R$.
From left to right, the columns are:
dark degrees of freedom,
inflow rate ($m_\odot\ {\rm yr}^{-1}$);
surface gas temperature (keV);
outer radius ($U_x\approx0.246~\Mpc$ units);
gas, dark and total concentrations;
gravitational radius ratio;
four of the density slope-radii;
the peak of the rotation curve;
and the minimal central mass
(given in solar units and as a fraction of the cluster mass).
The signature radii ($R_{I_1}, R_{I_2}, R_I, R_w, R_1, R_2, R_3, R_4, R_o$)
are defined in Appendix~\ref{s.metrics}.
}
\begin{center}
$\begin{array}{rrr@{.}lr@{.}lr@{.}lr@{.}lr@{.}lr@{.}lr@{.}lr@{.}lr@{.}lr@{.}lr@{.}lr@{.}lr@{.}lr@{.}lr@{.}lr@{.}lr@{.}lcccccccccccccc}
\multicolumn{1}{c}{F_2}
&\multicolumn{1}{c}{\dot{m}}
&\multicolumn{2}{c}{T_R}
&\multicolumn{2}{c}{R}
&\multicolumn{2}{c}{R_{I_1}/R}
&\multicolumn{2}{c}{R_{I_2}/R}
&\multicolumn{2}{c}{R_{I}/R}
&\multicolumn{2}{c}{R_w/R}
&\multicolumn{2}{c}{R_1/R}
&\multicolumn{2}{c}{R_2/R}
&\multicolumn{2}{c}{R_3/R}
&\multicolumn{2}{c}{R_4/R}
&\multicolumn{2}{c}{R_{\rm o}/R}
&\multicolumn{2}{c}{m_*/m_\odot}
&\multicolumn{2}{c}{m_*/m}
\\
\hline
\\
2
&1	&0&4
	&46&2
	&0&829
	&0&806
	&0&810
	&0&798
&0&470	&0&646&  0&739&  0&797
&0&882
&6&45(9)	&1&81(-5)
\\
3
&1	&0&4
	&47&0
	&0&820
	&0&714
	&0&733
	&0&744
&0&355	&0&522&  0&628&  0&704
&0&761
&3&45(9)	&9&68(-6)
\\
4
&1	&0&4
	&47&8
	&0&812
	&0&625
	&0&659
	&0&686
&0&272	&0&417&  0&525&  0&612
&0&637
&3&35(9)	&9&39(-6)
\\
5
&1	&0&4
	&48&9
	&0&806
	&0&536
	&0&588
	&0&621
&0&207	&0&326&  0&427&  0&519
&0&515
&3&22(9)   &9&03(-6)
\\
6
&1	&0&4
	&50&2
	&0&802
	&0&446
	&0&521
	&0&548
&0&154	&0&246&  0&334&  0&426
&0&400
&3&06(9)  &8&59(-6)
\\
7
&1	&0&4
	&51&8
	&0&801
	&0&355
	&0&458
	&0&00534
&0&108	&0&176&  0&245&  0&329
&0&291
&2&85(9)  &8&00(-6)
\\
8
&1	&0&4
	&53&9
	&0&803
	&0&260
	&0&402
	&1&07(^-5)
&0&0691	&0&113&  0&162&  0&229
&0&190
&3&70(8)  &1&04(-6)
\\
9
&1	&0&4
	&56&8
	&0&813
	&0&155
	&0&358
	&0&00233
&0&0340	&0&0555& 0&0816&  0&123
&0&0950
&1&27(7)  &3&57(-8)
\\
9.5
&1	&0&4
	&59&1
	&0&824
	&0&0927
	&0&343
	&0&0115
&0&0174	&0&0284& 0&0422& 0&0662
&0&0490
&2&66(6)	&7&45(-9)
\\
9.9
&1	&0&4
	&62&5
	&0&842
	&0&0274
	&0&341
	&0&00980
&0&00386 &0&00631&0&00944& 0&0153
&0&0109
&5&71(5)	&1&60(-9)
\\
\\
2
&10	&1&0
	&18&4
	&0&829
	&0&806
&0&809 &0&798
&0&470 &0&647&  0&739&  0&797
&0&882
&1&60(10)  &4&47(-5)
\\
3
&10	&1&0
	&18&7
	&0&820
	&0&714
	&0&733
	&0&744
&0&355 &0&522&  0&628&  0&704
&0&761
&1&55(10)  &4&34(-5)
\\
4
&10	&1&0
	&19&1
	&0&812
	&0&625
	&0&659 &0&686
&0&272 &0&417&  0&525&  0&612
&0&637
&1&49(10)  &4&19(-5)
\\
5
&10	&1&0
	&19&5
	&0&806
	&0&536
	&0&588 &0&621
&0&207 &0&326&  0&427&  0&519
&0&515
&1&43(10)  &4&00(-5)
\\
6
&10	&1&0
	&20&0
	&0&802
	&0&446
	&0&521 &0&499
&0&154 &0&246&  0&334&  0&426
&0&400
&1&35(10)  &3&78(-5)
\\
7
&10	&1&0
	&20&6
	&0&800
	&0&355
&0&458
&2&84(^-5)
&0&108 &0&176&  0&245&  0&329
&0&291
&1&24(10)  &3&48(-5)
\\
8
&10	&1&0
	&21&5
	&0&803
	&0&260
&0&402
&1&53(^-5)
&0&0690 &0&112&  0&161&  0&229
&0&190
&5&82(8)  &1&63(-6)
\\
9
&10	&1&0
	&22&7
	&0&813
	&0&155
&0&357
&0&00293
&0&0340 &0&0555& 0&0815&  0&123
&0&0950
&2&18(7)  &6&13(-8)
\\
9.5
&10	&1&0
	&23&6
	&0&824
	&0&0919
&0&343
&0&0102
&0&0172 &0&0281& 0&0417& 0&0656
&0&00186
&5&22(6)  &1&46(-8)
\\
9.9
&10	&1&0
	&24&9
	&0&841
	&0&0273
&0&341
&0&00986
&0&00385 &0&00629&0&00941& 0&0152
&0&0109
&9&02(5)  &2&53(-9)
%\\
%\\
%2
%&10^2	&1&0
%	&8&70
%	&0&788
%	&0&804
%	&0&801
%	&0&793
%&0&639&  0&732&  0&790
%&0&873
%&7&50(10)  &2&10(-4)
%\\
%3
%&10^2	&1&0
%	&8&87
%	&0&774
%	&0&712
%	&0&723
%	&0&737
%&0&516&  0&622&  0&697
%&0&752
%&7&25(10)  &2&03(-4)
%\\
%4
%&10^2	&1&0
%	&9&09
%	&0&762
%	&0&623
%	&0&647
%	&0&678
%&0&412&  0&519&  0&605
%&0&629
%&6&96(10)  &1&95(-4)
%\\
%5
%&10^2	&1&0
%	&9&35
%	&0&752
%	&0&533
%	&0&574
%	&0&612
%&0&322&  0&423&  0&514
%&0&509
%&6&64(10)  &1&86(-4)
%\\
%6
%&10^2	&1&0
%	&9&67
%	&0&745
%	&0&444
%	&0&505
%	&0&0244
%&0&244&  0&332&  0&423
%&0&396
%&6&26(10)   &1&76(-4)
%\\
%7
%&10^2	&1&0
%	&10&0
%	&0&740
%	&0&354
%	&0&441
%	&1&18(^-7)
%&0&175&  0&245&  0&330
%&0&290
%&5&14(10)  &1&44(-4)
%\\
%8
%&10^2	&1&0
%	&10&6
%	&0&741
%	&0&260
%	&0&382
%	&1&88(^-5)
%&0&113&  0&162&  0&232
%&0&190
%&7&47(8)  &2&10(-6)
%\\
%9
%&10^2	&1&0
%	&11&4
%	&0&751
%	&0&157
%	&0&335
%	&0&00277
%&0&0564& 0&0832&  0&127
%&0&0965
%&2&98(7)  &8&35(-8)
\\
\\
2
&10^2	&3&0
	&7&10
	&0&835
	&0&806
	&0&811
	&0&799
&0&472 &0&648&  0&740&  0&798
&0&883
&7&06(10)  &1&98(-4)
\\
3
&10^2	&3&0
	&7&20
	&0&827
	&0&715
	&0&734
	&0&745
&0&357 &0&523&  0&629&  0&705
&0&763
&6&82(10)  &1&91(-4)
\\
4
&10^2	&3&0
       &7&33
	&0&820
	&0&625
	&0&661
	&0&687
&0&273 &0&418&  0&525&  0&613
&0&638
&6&54(10)  &1&83(-4)
\\
5
&10^2	&3&0
       &7&48
	&0&814
	&0&536
	&0&590
	&0&623
&0&208 &0&326&  0&427&  0&520
&0&516
&6&22(10)  &1&74(-4)
\\
6
&10^2	&3&0
       &7&67
	&0&811
	&0&447
	&0&524
	&0&0376
&0&154 &0&246&  0&334&  0&426
&0&400
&5&83(10)  &1&64(-4)
\\
7
&10^2	&3&0
       &7&90
	&0&810
	&0&355
	&0&461
	&2&12(^-7)
&0&109 &0&176&  0&245&  0&329
&0&291
&4&87(10)  &1&37(-4)
\\
8
&10^2	&3&0
       &8&19
	&0&813
	&0&260
	&0&405
	&2&12(^-5)
&0&0690 &0&112&  0&161&  0&229
&0&190
&8&00(8)  &2&24(-6)
\\
9
&10^2	&3&0
	&8&62
	&0&823
	&0&153
	&0&361
	&0&00287
&0&0335 &0&0548& 0&0806&  0&122
&0&0937
&3&35(7)  &9&40(-8)
\\
9.5
&10^2	&3&0
	&8&95
	&0&833
	&0&0908
	&0&346
	&0&0107
&0&0168 &0&0276& 0&0410& 0&0645
&0&0475
&8&25(6)  &2&31(-8)
\\
9.9
&10^2	&3&0
	&9&43
	&0&850
	&0&0267
	&0&344
	&0&00970
&0&00375 &0&00612&0&00916& 0&0148
&0&0106
&1&48(6)  &4&15(-9)
\\
\\
2
&10^3	&6&0
	&2&95
	&0&828
	&0&805
	&0&809
	&0&798
&0&469 &0&646&  0&739&  0&797
&0&881
&3&19(11)  &8&96(-4)
\\
3
&10^3	&6&0
	&2&99
	&0&818
	&0&714
	&0&732
	&0&744
&0&354 &0&522&  0&628&  0&703
&0&761
&3&07(11)  &8&60(-4)
\\
4
&10^3	&6&0
	&3&05
	&0&810
	&0&624
	&0&658
	&0&685
&0&271 &0&416&  0&524&  0&611
&0&636
&2&93(11)  &8&21(-4)
\\
5
&10^3	&6&0
	&3&12
	&0&804
	&0&535
	&0&587
	&0&616
&0&206 &0&325&  0&426&  0&519
&0&514
&2&77(11)  &7&76(-4)
\\
6
&10^3	&6&0
	&3&20
	&0&799
	&0&446
	&0&520
	&2&55(^-4)
&0&152 &0&246&  0&333&  0&425
&0&399
&2&58(11)  &7&25(-4)
\\
7
&10^3	&6&0
	&3&31
	&0&798
	&0&354
	&0&456
	&1&65(^-8)
&0&107 &0&174&  0&244&  0&329
&0&289
&8&06(10)  &2&26(-4)
\\
8
&10^3	&6&0
	&3&43
	&0&800
	&0&259
	&0&400
	&3&02(^-5)
&0&0685 &0&112&  0&161&  0&229
&0&189
&1&12(9)  &3&14(-6)
\\
9
&10^3	&6&0
	&3&64
	&0&810
	&0&154
	&0&356
	&0&00323
&0&0334 &0&0548& 0&0807&  0&123
	&0&0938
&4&86(7)  &1&36(-7)
\\
9.5
&10^3	&6&0
	&3&79
	&0&821
	&0&0910
	&0&342
	&0&0112
&0&0168 &0&0276& 0&0411& 0&0648
	&0&0475
&1&24(7)  &3&47(-8)
\\
\\
\hline
\end{array}$
\end{center}
\label{table.models}
\end{table*}


\subsubsection{thermal structure}

For dark matter,
   $\sigma_2^2$ decreases monotonically with $r$
   (e.g. dark curves in lower panels of
   Figures~\ref{fig.steady}-\ref{fig.steady.g}).
The temperature peaks (like $\sigma_2^2\propto r^{-1}$)
   in the inner density spike,
   then stays flat over several decades in radius,
   but plummets between the core and the dark surface.
The upturn from core to spike occurs at what is effectively
   a central gravitational sphere of influence,
   $r\sim G m / \sigma_2^2$.
Note that the central point-mass $m_*$ doesn't yet dominate at this radius:
  the intervening gas and halo masses also contribute.
Thus the border of the spike is essentially a sphere of {\em self}-influence,
   where the self-gravity terms become important.

The gas temperature gradient is negative in the cusp.
Accretion power dominates over cooling,
   and the temperature profile evolves a Bondi-like slope,
   $\sigma_1^2\propto r^{-1}$.
Near the central object,
the ratio of dark to gaseous temperatures approaches a constant.
Taking the limit $r\rightarrow 0$ in equations
(\ref{eq.beta.sigma}) and (\ref{eq.beta.sigma2}) 
 gives 
\begin{equation}
	\beta_{\sigma_2*} = {{2 G m_*}\over{F_2 + 2}}
	\hspace{2cm}\mbox{and}
\end{equation}
\begin{equation}
	{{\sigma_{2*}^2}\over{\sigma_{1*}^2}}
	= {{F_1+2}\over{F_2+2}}
	\left({
		1+{{4-F_1}\over{F_1}} \Mcal_*^2
	}\right)
	\ ,
\end{equation}
explicitly involving the central mass and gas Mach number.

Outside this hot, parsec-scale accretion zone,
   the gas thermal structure depends upon the local balance
   of radiative cooling, compression and accretion power.
Gas temperature gradients
   may be either positive or negative,
   depending on which terms dominate equation (\ref{general.dsigmadr}).
Our optimal solutions show a local maximum temperature ($T_{\rm max}$)
   at a radius typical of the dark core,
   and a global temperature minimum
   ($T_{\rm min}$)
   somewhere in the kiloparsec-scale interior.
The region between these extrema,
   where $dT/dr>0$,
   is one reasonable definition of the ``cooling core''.
The peak typically appears on the order of a few times $0.1~\Mpc$,
   as in X-ray observations since the earliest studies of clusters.
This characteristic radial scale $\sim U_x=B/G$
   may be natural to bodies governed by gravity and bremsstrahlung radiation
   (see Appendix~\ref{s.units}).
The temperature peak occurs at smaller radii in the compact solutions
   than in wide clusters
   (compare Figures~\ref{fig.steady.e} to \ref{fig.steady.g})
The temperature dip appears
   at around $r\sim 1 - 10^1~\kpc$,
   with little sensitivity to $\dot{m}$.
The dip radius is outside both the halo density cusp
   and the break-radius in the gas density profile.

Most of the cooling core overlaps a deep layer
   where dark matter is hotter than gas
   ($\sigma_2^2 > \sigma_1^2$).
This layer comprises the dark core (except the central spike).
Here,
   any disturbance from the cluster equilibrium
   is likely to cause waves of adjustment
   that propagate faster via the halo than gas acoustic modes 

In all our solutions, the gas temperature drops off at large radii, 
  as expected in any well bound polytrope.
Cooling is ineffective at the low densities on the fringe,
  and at large $r$ the velocity terms may also vanish from
(\ref{flowing.dsigmadr}).
This leaves the gravity term dominant, which guarantees $dT/dr<0$.
The chosen boundary values of $T_R$ correspond roughly
  to bound or accretion-warmed configurations (\S\ref{s.obc}).
We predict that all isolated clusters have a temperature decline
    at sufficiently large radii.
Many observations agree
\citep[e.g.][]{markevitch1998,finoguenov2001a,degrandi2002,
piffaretti2005,vikhlinin2005,zhang2006,pratt2007}.
In cases where outer temperature profile seems flat
\citep[e.g.][]{allen2001b,kaastra2004,arnaud2005}
we predict that a decline will eventually appear farther out.

The finding of $T_{\rm min}\neq0$ is a highlight of the model.
There exists a non-zero floor temperature for every steady cluster.
Gas does not cool indefinitely,
and we have no need to invoke distributed mass dropout.
It is no surprise that observed cooling cluster cores
lack massive condensations of cold gas and extragalactic stars.
More significantly, the existence of a temperature floor
does not require non-gravitational heating.
It emerges simply from the co-adaptation of gas and halo profiles
in their shared gravitational potential
(whether an active galaxy operates or not).

\begin{figure}
\begin{center}
\begin{tabular}{c}
 \includegraphics[width=8cm]{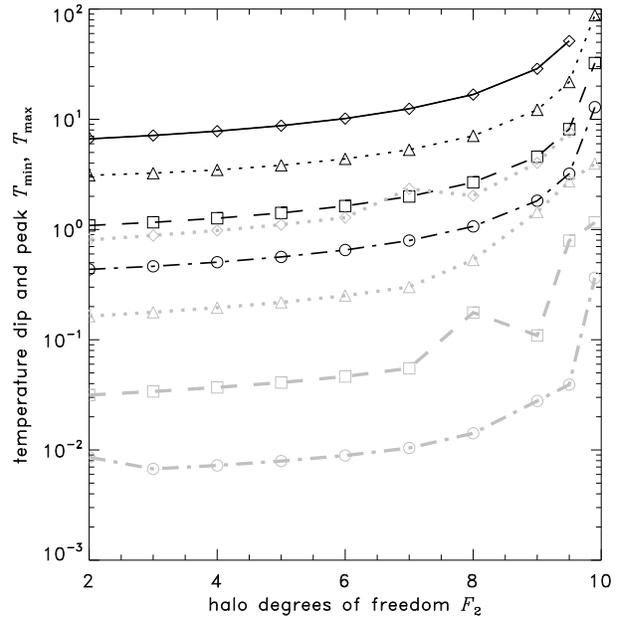}
\end{tabular}
\end{center}
\caption{
$F_2$-dependency of the peak and dip temperatures 
  of the $(m_*,\Upsilon)$-optimal models.
Black lines show the ICM fringe peak temperature.
Grey/cyan lines show the inner dip temperature.
The families of solutions are symbolised by:
$(T_R/{\rm keV},\dot{m}/m_\odot~{\mathrm yr}^{-1})=(6,1000)$
($\Diamond$ solid);
$(3,100)$ ($\triangle$ dot);
$(1,10)$ ($\square$ dash)
and $(0.4,1)$ ($\bigcirc$ dot-dash).
}
\label{fig.TmaxTmin}
\end{figure}

When constrained to cosmic composition and minimal $m_*$,
   both $T_{\rm max}$ and $T_{\rm min}$
   increase with increasing $F_2$
   (Figure~\ref{fig.TmaxTmin}).
However the ratio 
   $T_{\rm max}/T_{\rm min}$
appears less sensitive to $F_2$ than to the other global parameters,
such as $R$.
In the moderate models
   (with $\dot{m}=10\ m_\odot\ \yr^{-1}$ and $T_R=1$)
   the ratio is
   $10\la{T_{\rm max}/T_{\rm min}}\la41$,
   and usually $\approx35$.
For the smaller, strong-inflow models
   (with $\dot{m}=100\ m_\odot\ \yr^{-1}$, $T_R=3$)
   we find
   $8\la{T_{\rm max}/T_{\rm min}}\la22$,
For the most compact and heavily accreting model
   (with $\dot{m}=1000\ m_\odot\ {\rm yr}^{-1}$ and $T_R=6$)
   we find $5.3\la T_{\rm max}/T_{\rm min} \la 8$.
At the opposite extreme,
for the radially largest, coldest series of solutions
with lowest gas influx ($1~m_\odot\ {\rm yr}^{-1}$, $T_R=0.4$),
we find a deep temperature contrast:
$35\la T_{\rm max}/T_{\rm min}\la 81$.
The widest clusters provide the greatest cooling length
   before the inflow reaches the accretion-warmed interior.
Relatively high temperature, compactness and heavy inflow
yields the smallest temperature variation in the model ICM.

X-ray cluster observations reveal $T_{\rm max}/T_{\rm min}\sim3$ to $4$,
   which is softer than the temperature range of our optimal models.
However this difference is reconcilable,
   since $T_{\rm max}/T_{\rm min}$ ratios vary across the solution space.
The minimal-$m_*$ model typically has
   a temperature ratio near the upper extreme.
Neighbouring solutions with greater $\Mcal_R$ or $m_*$
   can have a warmer thermal dip,
   and the soft limit of $T_{\rm max}/T_{\rm min}\approx1$
   is attainable for $m_*$ several times heavier than the minimum.

\begin{figure}
\begin{center}
\begin{tabular}{c}
 \includegraphics[width=8cm]{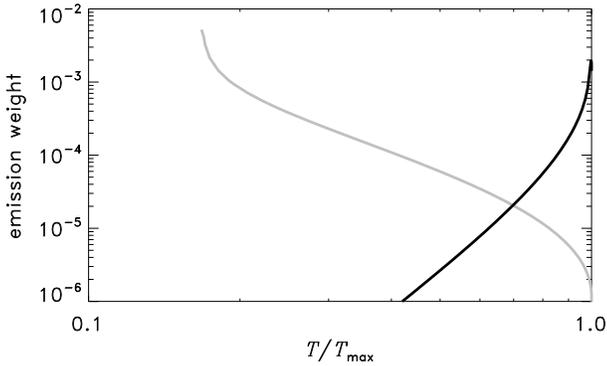}
\end{tabular}
\end{center}
\caption{
Any observational projected annulus 
   includes gas of temperatures ($T$)
   superimposed from a range of 3D shells.
We plot emission weights
   of gas shells crossing two mock-observational annuli:
the grey curve depicts matter in an annulus around the thermal dip
   $r\in[0.9,1.1]R_{\rm min}$;
   the black curve shows an annulus around the thermal peak
   $r\in[0.9,1.1]R_{\rm max}$.
The model has $m_*=2.85\times10^7~\msol$,
   $F_2=9$, $T_R=1~\keV$ and $\dot{m}=10~m_\odot~{\mathrm yr}^{-1}$.
Projection reduces the {\em apparent} peak/dip temperature ratio
   from $T_{\rm max}/T_{\rm min}\approx 5.96$ to $3.96$.
}
\label{fig.emeasure}
\end{figure}

\begin{figure}
\begin{center}
\begin{tabular}{c}
 \includegraphics[width=8cm]{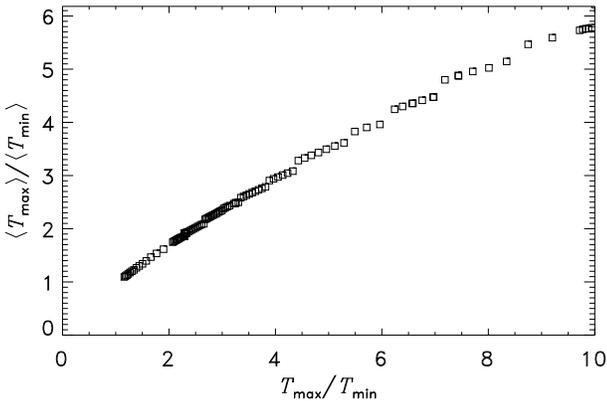}
\end{tabular}
\end{center}
\caption{
Comparison of actual ICM temperature range $T_{\rm max}/T_{\rm min}$
   and the projected ratios
   $\langle T_{\rm max}\rangle/\langle T_{\rm min}\rangle$
   calculated with weights as in Figure~\ref{fig.emeasure},
   with the same $(F_2,\dot{m},T_R)$.
Each point is a model with non-minimal $m_*$ near the ``cold'' border.
}
\label{fig.ToT}
\end{figure}

One may also ask how far the observed
   $T_{\rm max}/T_{\rm min}$ could underestimate actual ratios
   due to finite radial binning and imperfect deprojection.
The thermal minimum is a thin layer at small $r$;
   and an observational annulus superimposes hotter gas from outer shells.
If a shell of density $\rho_1$
   and temperature $\sigma_1$
   overlaps the annulus by area $\delta A$
   then its emission weight is $\propto\rho_1^2\sigma_1(\delta A)$.
Figure~\ref{fig.emeasure} shows the contributions of shells
   to two annuli centred on the temperature dip and peak of a cluster.
For annuli of relative radial thickness $\pm10\%$,
   the integrated, weighted temperature ratio
   $\langle T_{\rm max}\rangle/\langle T_{\rm min}\rangle$
   shrinks by $\la60\%$
   (see Figure~\ref{fig.ToT} for examples).
Thickening the annuli makes little more difference.
Projection effects cannot wholly hide the strongest thermal contrasts.

Ratios of $T_{\rm max}/T_{\rm min}\sim3$
   may occur naturally if clusters have non-minimal $m_*$
   and the gas physics varies from our ideal.
Raising the ICM effective heat capacity, $F_1>3$ 
   (describing micro-scale turbulence or a cosmic-ray component)
   may help.
Conduction and non-gravitational heating may play a role,
   though less influentially than popularly thought.

\begin{figure}
\begin{center} 
\begin{tabular}{cc}
\includegraphics[]{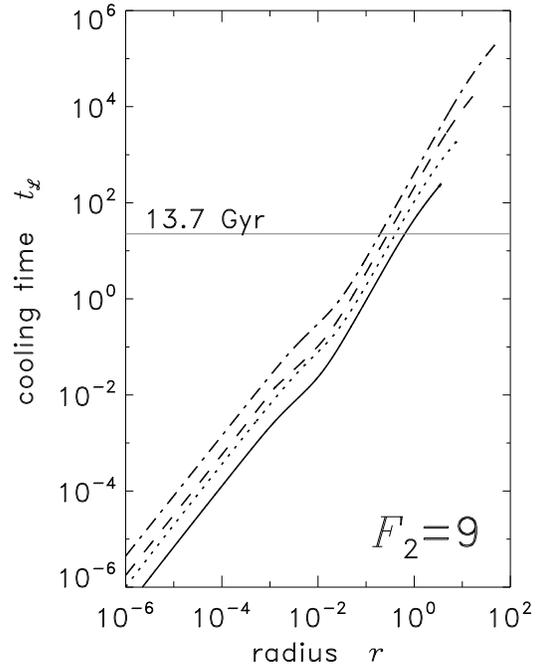}
\end{tabular}
\end{center} 
\caption{ 
Cooling timescale profiles
of $(m_*,\Upsilon)$-optimal cluster solutions,
for $F_2=9$ halos.
Curve patterns correspond to the $(\dot{m},T_R)$ cases in
Figure~\ref{fig.TmaxTmin}.
The grey horizontal line marks a time of 13.7~Gyr,
assuming an ion composition factor $\zeta=1$,
as in Appendix~\ref{s.units}.
We assume the default cluster mass scale $m=40U_m$.
}
\label{fig.tcool}
\end{figure}

\begin{figure}
\begin{center}
\begin{tabular}{c}
 \includegraphics[width=8cm]{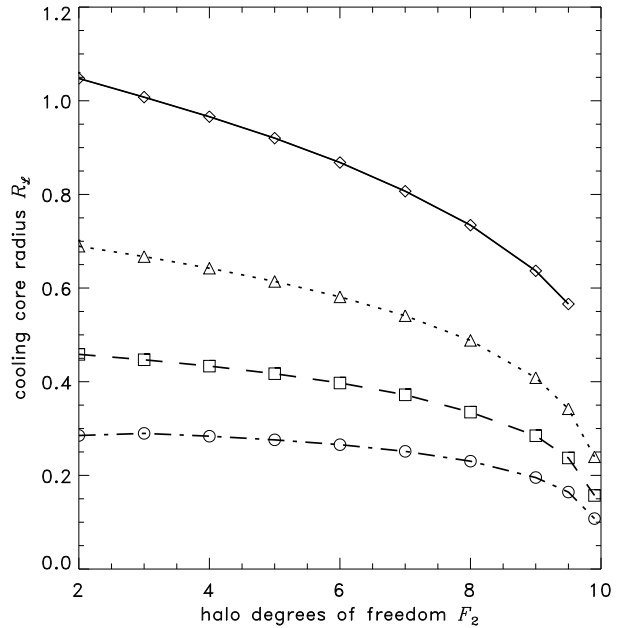}
\end{tabular}
\end{center}
\caption{
Variation with $F_2$ of the radius $R_{_\Lcal}$,
within which cooling is cosmologically relevant,
i.e. the cooling time matches the Hubble time.
We have assumed the cluster mass $m=40 U_m$.
The symbols represent the same families of solutions as in
Figure~\ref{fig.TmaxTmin}.
The hot/compact, large-$\dot{m}$ clusters have the largest $R_{_\Lcal}$.
}
\label{fig.Rcool}
\end{figure}

Bremsstrahlung radiative cooling becomes locally dynamically significant
on timescales of
\begin{equation}
	t_{_\Lcal}={{p_1/(\gamma_1-1)}\over\Lcal}
	={{F_1\sigma_1}\over{2B\rho_1}}
	\ .
\end{equation}
In our optimal models,
  the radial profile of the cooling time is approximately a power-law,
  $t_{_\Lcal} \approx t_{_{\Lcal}*} (r/r_*)^\alpha$
  (see Figure~\ref{fig.tcool}).
The constant $t_{_\Lcal*}$ is approximately the same 
  within families of clusters with the same 
  $(\Upsilon,m,\dot{m},T_R)$ but different $F_2$.
For minimal-$m_*$ solutions,
  the index is fairly consistent from kiloparsec to megaparsec scales,
  $\alpha\approx1.6-1.7$.
This is mildly steeper than X-ray evidence.
For example, \cite{voigt2004} and \cite{bauer2005}
  deproject tens of clusters observed out to moderate redshifts,
  showing $1.3\la\alpha\la1.5$ in strata from 10~kpc to 0.5~Mpc.
Their normalisation also seems comparable to our standard scaling,
  since their curves also cross the Hubble time around $\sim0.1$~Mpc.
This is a fair agreement, especially considering
   the constancy of the fiducial mass that we imposed,
   and the difference between their assumption of hydrostasis
   and the use of the full Euler equation.
{%\color{MidnightBlue}
Improvement might be possible
   if we were to vary the state of the gas $F_1>3$.

}Cooling is cosmologically relevant wherever $t_{_\Lcal}$
is shorter than the Hubble time.
This occurs in some sufficiently dense inner zone of each cluster.
If we define this region's cooling radius,
$R_{_\Lcal}$ where $t_{_\Lcal}<t_{\rm H}$,
then we typically find that $0.2\la R_{_\Lcal}\la 1$
(about 50~kpc to 250~kpc, Figure~\ref{fig.Rcool}).
The ``cool core'' can reasonably be defined as the layer $r<R_{_\Lcal}$
   rather than in terms of thermal gradients.
The cool core is usually smaller than the halo core ($R_{_\Lcal}\la R_{I_2}$).
Along the $(m_*,\Upsilon)$-optimised tracks,
   we find that $R_{_\Lcal}(\Upsilon,F_2)$ shrinks with increasing $F_2$,
   despite the increase of $R(\Upsilon,F_2)$ with $F_2$
   and the near constancy of the gas concentration ($R_{I_1}/R$).

Although formally the cool core is a minor part of the cluster volume,
   cooling controls or affects the exterior gas structure indirectly.
Contraction and subsidence of cooling gas
   reduces pressure support and 
   draws in the effectively adiabatic gas farther out.
\citep[Thus the flow doesn't need breaking around $R_\Lcal$, as in][]{binney1981}.
In steady solutions, the core inflow matches the global rate $\dot{m}$
   of cosmic accretion from outside the halo.
If the central inflow caused by cooling
   doesn't match the global inflow at the outer boundary,
   then a corrective acoustic wave, rarefaction wave or shock
   must propagate outwards into the external cosmic medium.
Pedantically, it would be misleading
   to describe a cooling flow as driven by external pressure;
   the core slumps because cooling undermines local hydrostasis,
   and the outskirts merely follow in sinkage.

\subsubsection{entropy profiles}

The gas entropy profile is potentially an important diagnostic
of the structure and history of a cluster.
For a settled, convectively stable system, $ds_1/dr>0$ everywhere.
Hot bubbles float (and cool clumps sink)
whilst changing volume adiabatically,
until settling at a level with comparable $s_1$.
Spherical adiabatic accretion is expected to yield a power-law slope,
$s_1\propto r^\alpha$.
Central non-gravitational heating may create a radially constant
``entropy floor'' comprising an ``isentropic core.''
For gas that shocks as it accretes into a cuspy halo,
theory and hydrodynamic simulations
predict $\alpha\approx1.1$ and no flat core
\citep{tozzi2001,kay2004,voit2005b}.

\begin{figure}
\begin{center}
\begin{tabular}{c}
  \includegraphics[width=8cm]{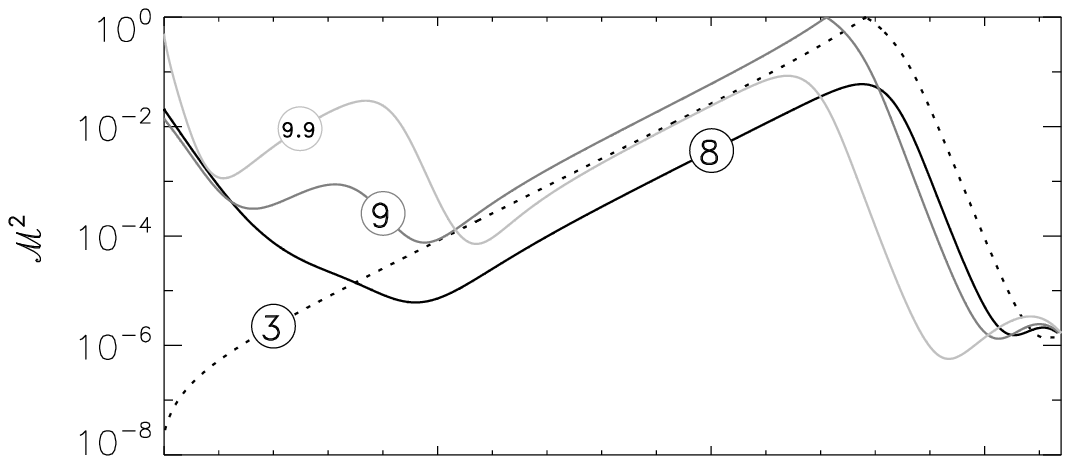}
\\
  \includegraphics[width=8cm]{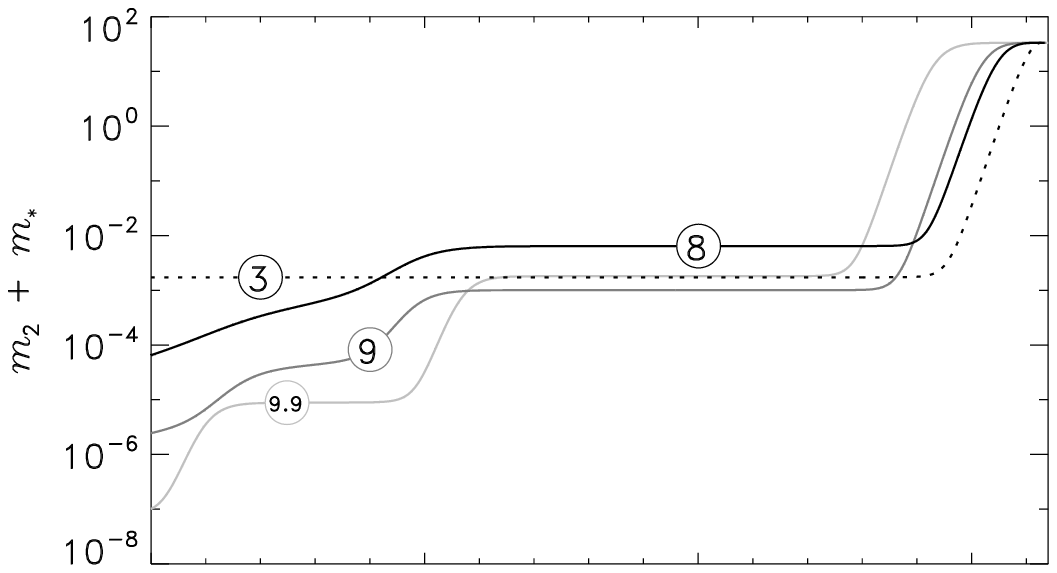}
\\
  \includegraphics[width=8cm]{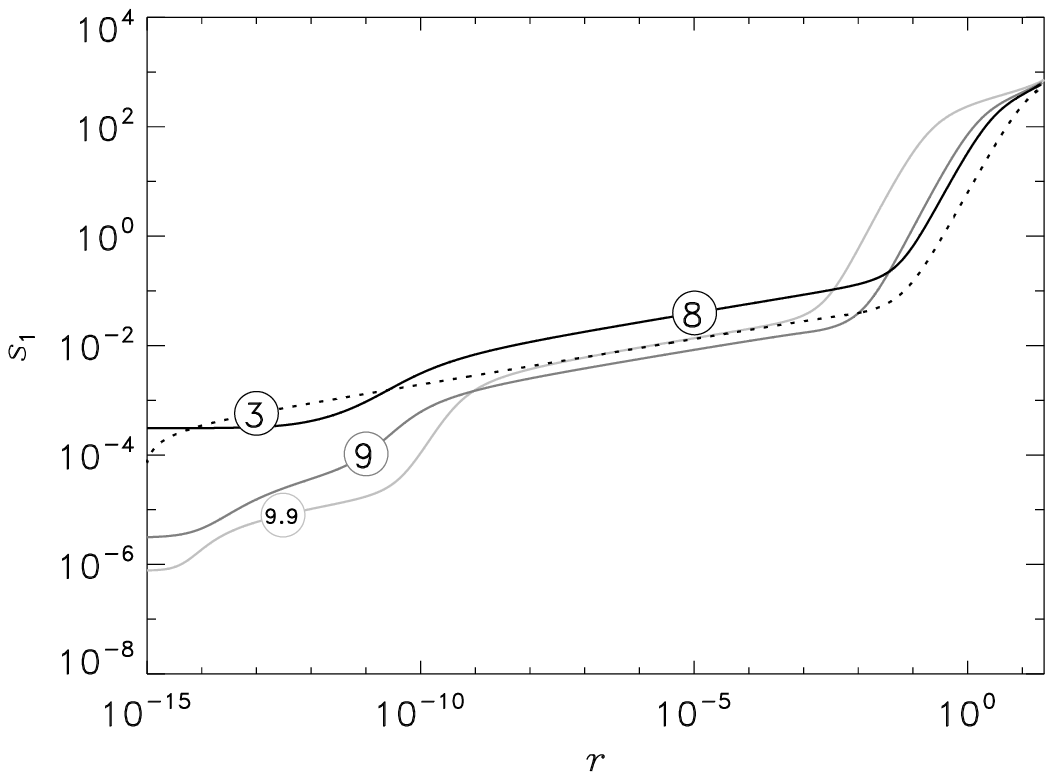}
\end{tabular}
\end{center}
\caption{
Radial profiles reaching the deep interior,
for $(m_*,\Upsilon)$-optimal solutions
where $\dot{m}=10\ m_\odot\ {\rm yr}^{-1}$,
$T_R=1$~keV
and $F_2=3,8,9,9.9$ (annotated).
Panels from top to bottom show Mach number squared,
enclosed non-gaseous mass, and gas entropy.
}
\label{fig.deep}
\end{figure}

Our solutions show different indices in distinct layers.
In the cooling core
   ($10~\kpc\la r \la R_{_\Lcal}$)
   we find $\alpha\approx1.7$,
   but $\alpha\la 1$ in the effectively adiabatic outskirts.
Our model lacks non-gravitational heating,
   so our solutions never develop a constant entropy floor.
Instead, the entropy profiles soften to $\alpha\approx0.2$
   in the hot accreting interior around/within
   the outermost (quasi-) sonic-point
   ($r\sim 10~\kpc$; see e.g. the bottom panel of Figure~\ref{fig.deep}).
This slope persists inwards
   for several orders of magnitude in radius.
The flatness means that cooling is less significant
   in the warm kpc-scale interior than in the cool-core,
   but there still remains a shallow stratification of $s_1$.

The entropy slopes in the ICM are not very different from observation,
   though less agreeably than the indices of the cooling time.
Flat isentropic cores are not observed
\citep{ettori2002,ponman2003,pratt2003,piffaretti2005},
   though there is debate about whether the entropy normalisation
   of smaller systems is affected by to feeback heating or preheating.
The entropy ramp appears at
   $s_1\sim20-140~\keV~{\rm cm}^2$ in observed profiles
   \citep[e.g][]{lloyd-davies2000,david2001}.
It begins at $\sim1~\keV~{\rm cm}^2$ in minimal-$m_*$ models,
   but higher for non-minimal cases
   (e.g. $10~\keV~{\rm cm}^2$ for that in Figure~\ref{fig.emeasure}).
Crudely, we expect the stellar matter potential
   to have a raising effect similar to large $m_*$
   (work in preparation).
Most authors find slopes $\alpha\approx1$ in the cooling core.
\cite{ettori2002,piffaretti2005,pratt2006,zhang2006}
   found $\alpha\approx0.97$,
   $\alpha\approx0.95$,
   $\alpha\approx1.1$
   and $\alpha=0.99\pm0.06$ respectively.
\cite{lemze2007}
   combined X-ray and lensing maps to model the mass of A1689,
   assuming spherical hydrostasis
   and outskirts declining as $\rho_1,\rho_2\propto r^{-3}$.
In the range $10~\kpc\la r \la 1~\Mpc$,
   they exclude any entropy floor,
   and found indices $\alpha=0.82\pm0.02$ when the halo was freely fitted
   (or $1\pm0.2$ when forcing a cuspy halo model).
Thus observed slopes are slightly flatter
   than either simulations or our solutions.
This may be an artefact of the NFW-like profiles
   %(appendix~\ref{halo.nfw})
   assumed in the analysis of data and construction of simulations.
Alternatively, gas physics with $F_1>3$
   %(e.g. turbulence)
   or thermal conduction might improve the match.

Several X-ray observations have seen a softer entropy slope
   in regions inside $r\la 20~\kpc$,
   and the plots show an index $0.1<\alpha\la 0.5$,
   in agreement with the results here.
Such observations are uncommon as they require fine resolution of thin annuli
   at small radii.
\cite{ponman2003} plot entropy profiles of 66 varied objects,
   and the curves seem softer at smaller radii.
Though unclear, this might be the start of an $\alpha\la0.5$ ramp.
\cite{pratt2006}
   observed mild flattening at $r<10~\kpc$.
\cite{david2001}
   deproject the Hydra~A cluster finely in over 30 annuli,
   showing $\alpha\la0.5$ clearly across the innermost four of them.
\cite{russell2007}
   observed a cool-core group with no active AGN,
   and an entropy slope of 0.5 appears within $r<15~\kpc$.
When this kind of entropy break appears and attracts comment,
   AGN heating is conventionally invoked.
The observed shallow region
   has been regarded as merely the outer edge
   of an (unseen) slope-zero isentropic interior.
We argue that this slope is actually a signature of subsonic inflow
   inside the (quasi-) sonic point (\S\ref{s.deep}),
   and not necessarily due to heating.
We predict that the $\alpha\approx0.2$ zone persists inwards to the nucleus.

Nevertheless we caution that the inner entropy slopes may prove hard to test.
Fitting a flat-cored $\beta$-model
%(appendix~\ref{halo.hubble})
could potentially underestimate the central gas density,
since a classic cooling flow has $\rho_1\propto r^{-1}$ at relevant radii.
Temperature variations can hide this slope,
giving an apparently flat X-ray brightness core.
Underestimating gas density could lead to overestimates of the entropy,
exagerrating the flatness of an observed inner $s_1(r)$ profile.

\subsubsection{illusory mass deposition}

Early X-ray imaging studies
used approximate formulae to estimate $\dot{m}$
from their deprojected temperature and luminosity profiles.
In the notation adopted in this work,
the gradient equation for gas temperature (\ref{general.dsigmadr})
was commonly reduced to
\begin{equation}
	\dot{m}(r)={{L(<r)}\over{
		{\frac52}\sigma_1^2 +\Phi(r)-\Phi({\rm inner})}
		}
	\ ,
\label{eq.fake.mdot}
\end{equation}
   where $L$ is the luminosity emitted by annuli interior to $r$.
Kinetic terms are dropped.
For example, \cite{stewart1984}
   plotted $\dot{m}(r)$ dropping towards the centre.
This was taken as evidence for mass dropout throughout the cooling core,
   and was justified in terms of runaway local thermal instability
   in a multiphase medium,
   depending on some initial spectrum of clumpiness
   \citep{nulsen1986}.
Radially varying $\dot{m}$ and multiphase gas
   became standard ingredients of cooling flow fits
   \citep[e.g.][]{thomas1987,johnstone1992}
   and theories
   \citep[e.g.][]{white1987a}.
However the implied deposition products
   (cold gas and stars)
   are not observed in sufficient amounts,
   leading to the ``cooling flow problem.''

\begin{figure}
\begin{center}
\begin{tabular}{c}
\includegraphics[width=8cm]{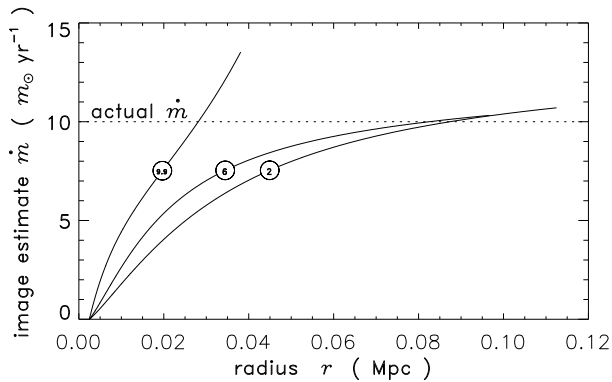}
\end{tabular}
\end{center}
\caption{
Hydrostatic imaging estimates of $\dot{m}$
give a misleading appearance of radial variation.
The true inflow rate (dotted) is constant,
$\dot{m}=10~m_\odot~\yr^{-1}$.
We plot cases with $T_R=1~\keV$,
and varied halo types: $F_2=2, 6, 9.9$ (annotated).
Curves are plotted up to the cooling radius $R_{_\Lcal}$.
}
\label{fig.fakemdot}
\end{figure}

When we apply formula (\ref{eq.fake.mdot})
   to the gas density and temperature solutions,
   it reproduces the true $\dot{m}$ somewhere near $r\la R_\Lcal$,
   but at smaller radii the formula underestimates $\dot{m}$
   (Figure~\ref{fig.fakemdot}).
The fictitious radial variation of ``$\dot{m}$''
   resembles observationally derived curves.
Such profiles are caused by omission of velocity terms.
It is a signature of proximity to the kpc-scale (quasi-) sonic point.
Thus we propose to reconcile centrally depressed ``$\dot{m}$'' profiles
   with the dearth
   of massive cold condensates on cluster scales.
Mass dropout from the keV-temperature cluster medium may be unnecessary,
   and a single-phase ICM may be plausible after all.
We conclude that that gas inflows
   reach the central object when the system is settled,
   or perhaps stall at small, intra-galactic radii
   during any episodes when the structure is driven briefly off stationarity.
We discuss this further in \S\ref{discuss.evolution}.

We have also tested the effect of the hydrostatic approximation
on mass profiles estimated from gas density and temperature observables.
We find that the errors are neglible outside $\ga 10~\kpc$,
and the mass underestimate is only a few percent
at kiloparsec radii.

\subsubsection{Mach number profile \& bottlenecks}
\label{s.deep}

The Mach-number profiles of ($m_*,\Upsilon)$-optimal solutions show
   a variety of features
   almost anywhere from the cluster surface to the smallest calculable radii.
   (e.g. top panel of Figure~\ref{fig.deep}).
There is at least one local maximum in $\Mcal^2$,
   and they are more numerous (per decade in $r$) if $F_2\ga 8$.
In many cases, the flow becomes transonic ($\Mcal^2\rightarrow 1$)
   sharply around one of the maxima.
This is corresponds to the ``sonic point'' of a maximal subsonic solution
   in simpler models of adiabatic, non-gravitating accretion
   \citep{bondi1952}.
A slight variation of the outer boundary conditions can change
   a sonic point into a supersonic break.
Thus these points define bottlenecks that incur the ``too fast'' border,
   (as in Figure~\ref{fig.zones})
   thereby constraining the set of steady solutions.
The radially outermost $\Mcal^2$ maximum
   is often (but not always) the tightest bottleneck.
Throughout our survey of $\dot{m}$ and $T_R$ parameters,
   the outer bottleneck usually occurs somewhere in the radial range
   $10^{-4}\la r\la10^{-2}$
   ($\sim$0.25 - 25~kpc).

Bottlenecks in the gas entropy $s_1$ can also occur,
   where small variations in outer boundary conditions
   can trigger a cooling catastrophe in a specific radial band.
The most susceptible radial layers define the ``too cold'' border.

The cause of bottlenecks appears in the structure of
   the Mach-number equation, (\ref{eq.dMMdl}).
The cooling term $\beta_L r^c$ vanishes at small radii,
   so that the geometric and gravitational terms compete
   to determine the sign of $d\Mcal^2/dl$.
Approaching the origin,
   there is less interior mass,
   and in some solutions the ratio $m/\beta_{\sigma_1}$
   shrinks enough to guarantee that $d\Mcal^2/dl<0$.
As $\Mcal^2$ increases nearer the origin,
   the denominators $(1-\Mcal^2)$ shrink,
   which steepens the gradients of gas-related quantities.
Often this leads to a runaway descent into
   a supersonic or cold catastrophe.
In other conditions there is a narrow escape
   and return to low-$\Mcal$ conditions at deeper radii.
Even in these ``narrow escapes'', the behaviour in the deep interior
   can be dominated by ``see-sawing''
   between positive and negative terms in the ODEs.
In the set of surviving solutions, the multiple ``narrow escapes''
   appear as ripples or steps on the density profile.

For larger $F_2$, the central mass gradients are steeper,
   and prone to yield low $m/\beta_{\sigma_1}$ values,
   which triggers more bottlenecks and see-sawing in layers near the origin.
These tend to become more numerous and restrictive as $F_2$ increases.
As the middle panel of Figure~\ref{fig.deep} shows,
   when $F_2$ is big
   the halo structure in the nuclear region
   is a concentric set of uniform cores with steep fringes.
Undulations appear in the gas entropy profile
   as departures from the typical $s_1$ slope,
   coinciding with the deep-core mass shells
   (bottom panel of Figure~\ref{fig.deep}).

\subsection{Jeans stability}
\label{s.jeans}

The outer surface of a finite single-fluid polytrope
occurs at a radius comparable to the local Jeans radius,
\begin{equation}
	r_{_{\rm J}} = \sqrt{ {\pi \gamma \sigma^2}\over{4 G \rho} }   \ .
\end{equation}
As equilibrium structures governed by the balance of self-gravity
and pressure,
they are necessarily stable against gravitational collapse.
Thus these bodies are Jeans masses, albeit with non-uniform interiors.

\begin{figure*}
\begin{center} 
\includegraphics[height=4cm]{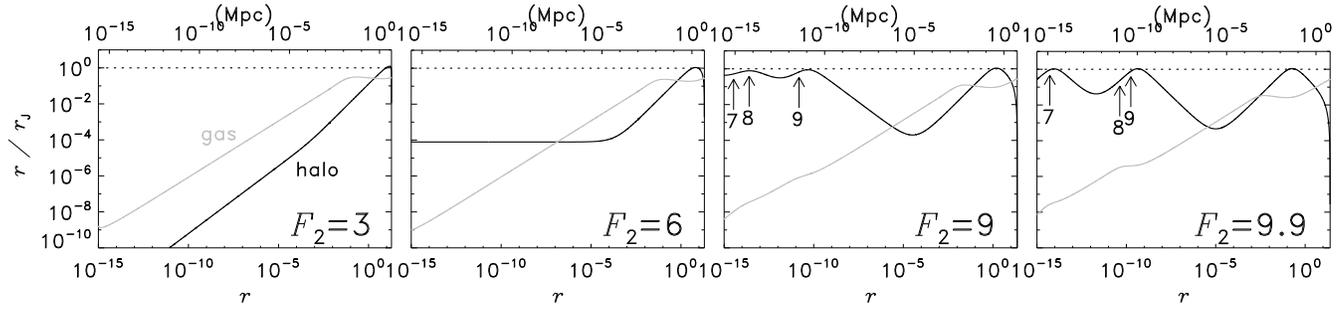}
\end{center} 
\caption{ 
The ratio of the radius to the local Jeans radius,
for $(m_*,\Upsilon)$-optimal solutions
with
$\dot{m}=10\ m_\odot\ {\rm yr}^{-1}$,
$T_R=1$~keV and $F_2$ dark degrees of freedom.
Halo and gas profiles are marked in black and grey respectively.
A dotted line marks the instability threshold, $r=r_{_{\rm J}}$.
In the $F_2>6$ cases the arrows mark where the marginally stable inner halo
encloses a dark mass of $10^7, 10^8, 10^9m_\odot$.
A sufficiently large perturbation might make this mass collapsible.
}
\label{fig.jeans}
\end{figure*}

The gravitational stability of a two-component system 
  is not as immediately obvious,
  with or without complications of inflow and cooling.
The effective sound speeds and densities of the constituents differ at every location,
  and so their Jeans radii differ as well,
  $r_{_{\rm J1}}\neq r_{_{\rm J2}}$.
It is possible for the dark matter to be locally Jeans-unstable
   while the gas is Jeans-stable, or vice versa.
Figure~\ref{fig.jeans}
shows the ratio of the radial coordinate to the local Jeans scale,
throughout a set of reference models with cosmic composition.
As expected, $r$ approaches $r_{\rm J}$
   in the fringe, for both gas and dark matter.
Descending from megaparsec- to parsec-scale radii,
   each becomes more Jeans-stable.
   Indefinitely nearer to the origin,
   the gas becomes ever more stable.

However the central, sub-parsec gravitational stability of the halo
is more complex and contingent.
For low $F_2$, the dark halo is Jeans-stable at all radii.
In the $F_2=6$ case,
the halo sits at a nearly constant degree of Jeans-stability
in the sub-parsec interior.
For $F_2>6$ halo stability lessens nearer the origin.
For $F_2>8$ the halo hovers near marginal Jeans stability
for several decades in radius near the central mass.
The upturn towards this condition begins at radii as large as 10~pc,
and is sharper for larger $F_2$.
In steady solutions, the ratio $r/r_{\rm J}$ within the cusp never exceeds
its maximum value near the dark surface.
Thus the dipolytropic cluster models are formally gravitationally stable,
and we cannot infer a spontaneous collapse of the dark cusp, 
  without the onset of other instabilities. 
In \S\ref{discuss.evolution}
we will discuss the possibility of externally stimulated collapse.
%All our two-fluid cluster models are gravitationally stable,
%though the stability is marginal in the deep nuclei of large-$F_2$ cases.

\subsection{Virial scaling}
\label{s.virial}

Our formulation deals with the stationary, relaxed conditions
to which spherical clusters tend ultimately.
By construction, we omit time-dependent, externally driven evolution
and our results are indifferent to cosmological history.
Collisionless self-gravitating systems are expected collapse to a typical mean density
that is some multiple of the cosmic critical density,
implying a ``virial mass'' within a ``virial radius.''
The model clusters in this work are not collisionless idealisations,
but their constituents may have been
out of acoustic contact before assembly,
which is effectively similar.
Thus virial scaling may have some back-of-the-envelope relevance
in observational comparisons.

Masswise, our solutions rescale freely,
and the virial radius is not an emergent scale.
In order to mimic cosmological virial scaling relations,
we can rescale the mass, temperature and other profiles
such that $R_{\rm v}$ matches one of the signature scales.
In Table~\ref{table.virial}
we show how three choices of virial radius
($R_{\rm v}=R_3$, $R_{\rm v}=R_I$, $R_{\rm v}=R$)
affect the normalisation of the $(m_*,\Upsilon)$-optimal models.
We show the mass within the virial radius,
and the rescaled inflow rate.
Our radially extended solutions become giant clusters.
For them, the choice of $R_{\rm v}=R$ is clearly excessive.
The most compact solutions rescale to the mass of a group
or giant elliptical galaxy.
For the compact cases with high $F_2$,
the density profile is steep almost everywhere,
$R_3$ is small and thus the choice $R_{\rm v}=R_3$
implies a puny galaxy mass and negligible inflow.
If any simple prescription for $R_{\rm v}$
applies to all the model cluster families here,
then $R_{\rm v}\approx R_I$ seems like the most realistic choice.

Virial scale selection is not necessarily the best choice
for linking our solutions to a cosmological scenario.
We could alternatively scale each cluster
so that the acoustic crossing time 
is some fraction of the Hubble time.
However this range of choices is too wide to explore in the present work.
For the purposes of calculating X-ray spectra
(in \S\ref{s.xrays} below)
we shall adopt the simple choice of $R_{\rm v}=R_I$.

\begin{table}
\caption{
Virial masses and inflow rates
for some of our $(m_*,\Upsilon)$-optimal solutions,
under mass rescaling with three choices of virial radius:
$R_{\rm v}=R_3$, $R_{\rm v}=R_I$ and $R_{\rm v}=R$.
Masses are expressed in terms of $\log_{10}(m_\odot)$
and inflows in terms of $\log_{10}(m_\odot\ \yr^{-1})$.
}
\begin{center}
$\begin{array}{lrrr@{.}lr@{.}lr@{.}lr@{.}lr@{.}lr@{.}l}
\multicolumn{1}{c}{ }
&\multicolumn{2}{c}{\mbox{standard}}
&\multicolumn{4}{c}{R_{\rm v}=R_3}
&\multicolumn{4}{c}{R_{\rm v}=R_I}
&\multicolumn{4}{c}{R_{\rm v}=R}
\\
&\multicolumn{2}{c}{\overbrace{~~~~~~~~~~~~}}
&\multicolumn{4}{c}{\overbrace{~~~~~~~~~~~~~~~~}}
&\multicolumn{4}{c}{\overbrace{~~~~~~~~~~~~~~~~}}
&\multicolumn{4}{c}{\overbrace{~~~~~~~~~~~~~~~~}}
\\
\multicolumn{1}{c}{F_2}
&\multicolumn{1}{c}{\dot{m}}
&\multicolumn{1}{c}{T_R}
&\multicolumn{2}{c}{m_{\rm v}}
&\multicolumn{2}{c}{\dot{m}}
&\multicolumn{2}{c}{m_{\rm v}}
&\multicolumn{2}{c}{\dot{m}}
&\multicolumn{2}{c}{m_{\rm v}}
&\multicolumn{2}{c}{\dot{m}}
\\
\hline
\\
2.0 & 0 &0.40 &16&5 &3&17 &16&6 &3&26 &16&9 &3&55	\\
3.0 & 0 &0.40 &16&3 &2&94 &16&5 &3&10 &16&9 &3&58	\\
4.0 & 0 &0.40 &16&1 &2&67 &16&4 &2&94 &17&0 &3&62	\\
5.0 & 0 &0.40 &15&9 &2&35 &16&3 &2&76 &17&0 &3&66	\\
6.0 & 0 &0.40 &15&6 &1&96 &16&2 &2&56 &17&0 &3&71	\\
7.0 & 0 &0.40 &15&2 &1&47 &16&1 &2&36 &17&1 &3&77	\\
8.0 & 0 &0.40 &14&7 &0&777 &15&9 &2&17 &17&1 &3&85	\\
9.0 & 0 &0.40 &13&9 &^-0&400 &15&8 &2&03 &17&2 &3&96	\\
9.5 & 0 &0.40 &13&1 &^-1&58 &15&8 &2&03 &17&2 &4&03	\\
9.9 & 0 &0.40 &11&2 &^-4&36 &15&9 &2&13 &17&3 &4&14	\\
\\
2.0 & 1 & 1.0 &15&3 &2&37 &15&4 &2&46 &15&7 &2&76	\\
3.0 & 1 & 1.0 &15&1 &2&14 &15&3 &2&31 &15&7 &2&79	\\
4.0 & 1 & 1.0 &14&9 &1&87 &15&2 &2&14 &15&8 &2&82	\\
5.0 & 1 & 1.0 &14&7 &1&55 &15&1 &1&96 &15&8 &2&86	\\
6.0 & 1 & 1.0 &14&4 &1&16 &15&0 &1&76 &15&8 &2&92	\\
7.0 & 1 & 1.0 &14&0 &0&670 &14&9 &1&56 &15&9 &2&98	\\
8.0 & 1 & 1.0 &13&5 &^-0&0227 &14&7 &1&37 &15&9 &3&05	\\
9.0 & 1 & 1.0 &12&7 &^-1&20 &14&7 &1&24 &16&0 &3&16	\\
9.5 & 1 & 1.0 &11&9 &^-2&40 &14&6 &1&23 &16&0 &3&24	\\
9.9 & 1 & 1.0 &10&0 &^-5&17 &14&7 &1&33 &16&1 &3&35	\\
\\
2.0 & 2 & 3.0 &14&1 &1&51 &14&2 &1&60 &14&5 &1&89	\\
3.0 & 2 & 3.0 &13&9 &1&27 &14&1 &1&44 &14&5 &1&92	\\
4.0 & 2 & 3.0 &13&7 &1&00 &14&0 &1&27 &14&5 &1&95	\\
5.0 & 2 & 3.0 &13&4 &0&682 &13&9 &1&09 &14&5 &1&99	\\
6.0 & 2 & 3.0 &13&2 &0&293 &13&7 &0&899 &14&6 &2&04	\\
7.0 & 2 & 3.0 &12&8 &-0&206 &13&6 &0&698 &14&6 &2&10	\\
8.0 & 2 & 3.0 &12&3 &-0&906 &13&5 &0&504 &14&7 &2&17	\\
9.0 & 2 & 3.0 &11&5 &-2&11 &13&4 &0&366 &14&7 &2&27	\\
9.5 & 2 & 3.0 &10&6 &-3&32 &13&4 &0&360 &14&8 &2&34	\\
9.9 & 2 & 3.0 &8&74 &-6&12 &13&5 &0&449 &14&8 &2&45	\\
\\
2.0 & 3 & 6.0 &12&9 &0&790 &13&1 &0&878 &13&3 &1&17	\\
3.0 & 3 & 6.0 &12&7 &0&553 &12&9 &0&720 &13&4 &1&20	\\
4.0 & 3 & 6.0 &12&5 &0&285 &12&8 &0&554 &13&4 &1&24	\\
5.0 & 3 & 6.0 &12&3 &^-0&0312 &12&7 &0&375 &13&4 &1&29	\\
6.0 & 3 & 6.0 &12&0 &^-0&419 &12&6 &0&181 &13&4 &1&34	\\
7.0 & 3 & 6.0 &11&6 &^-0&918 &12&5 &^-0&0231 &13&5 &1&40	\\
8.0 & 3 & 6.0 &11&2 &^-1&61 &12&3 &^-0&221 &13&5 &1&47	\\
9.0 & 3 & 6.0 &10&3 &^-2&79 &12&3 &^-0&346 &13&6 &1&58	\\
9.5 & 3 & 6.0 &9&50 &^-4&00 &12&3 &^-0&348 &13&7 &1&66	\\
\\
\hline
\end{array}$
\end{center}
\label{table.virial}
\end{table}

\begin{figure}
\begin{center}
\includegraphics[width=8.2cm]{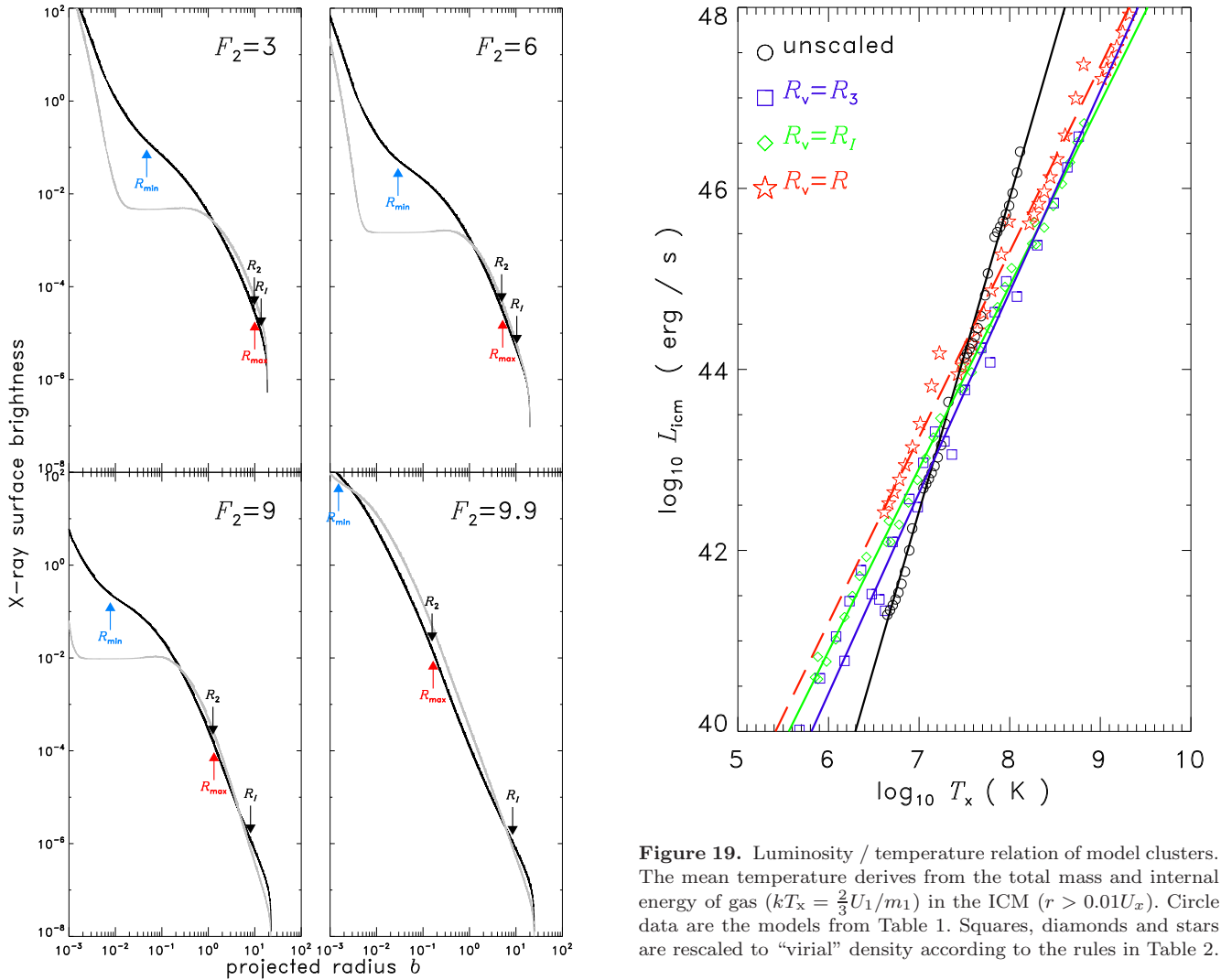}
\end{center}
\caption{
Brightness profiles of X-ray continuum
projected from $(m_*,\Upsilon)$-optimal solutions with
$\dot{m}=10~m_\odot~{\rm yr}^{-1}$ and $T_R=1$~keV,
but various $F_2$.
Black lines  show $0.1 - 2.4$~keV emission;
grey shows the $2 - 10$~keV band.
We have rescaled the mass and temperature to a virial selection of
$R_{\rm v}=R_I$.
The X-ray core is smaller than the halo core,
and both shrink as $F_2$ rises.
}
\label{fig.bright}
\end{figure}

\begin{figure}
\begin{center}
\includegraphics[width=8cm]{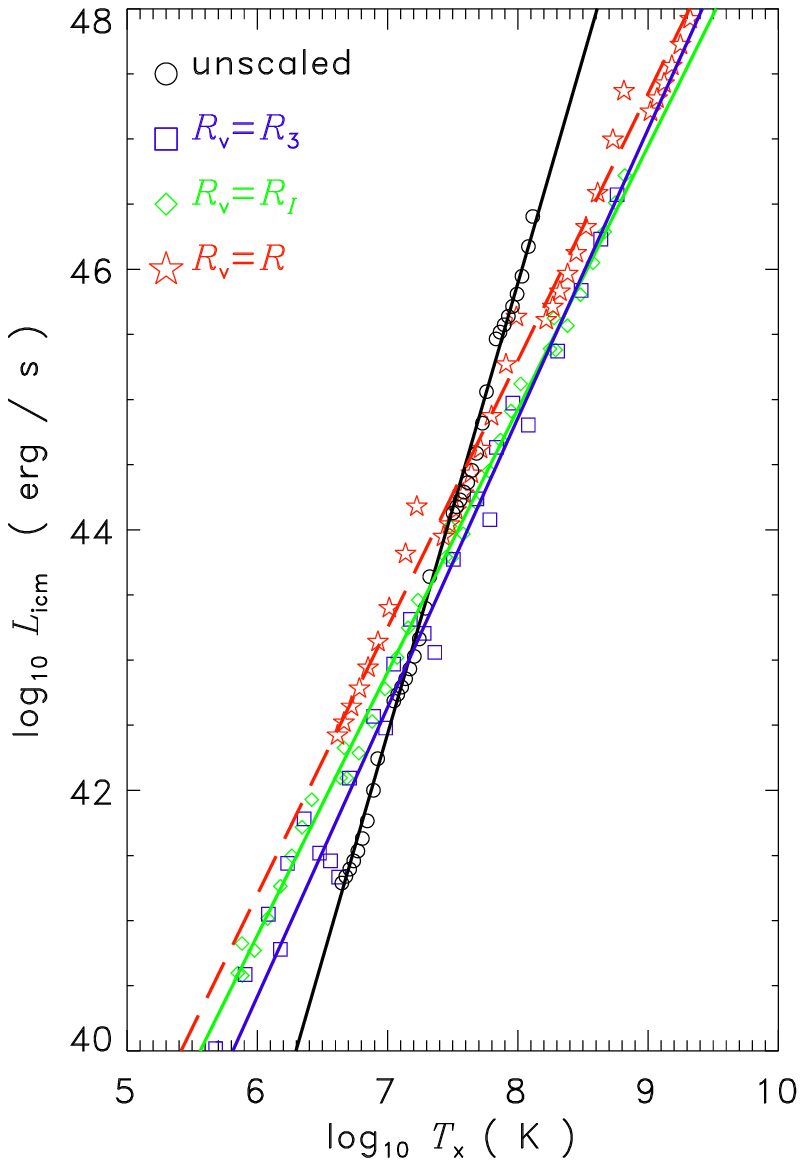}
\end{center}
\caption{
Luminosity / temperature relation of model clusters.
The mean temperature derives from the total mass
and internal energy of gas ($kT_{\rm x}={\frac23}U_1/m_1$)
in the ICM ($r>0.01U_x$).
Circle data are the models from Table~\ref{table.models}.
Squares, diamonds and stars are rescaled to ``virial'' density
according to the rules in Table~\ref{table.virial}.
}
\label{fig.LxTx}
\end{figure}

\subsection{X-ray brightness profiles \& continuum spectra}
\label{s.xrays}

For photons of energy $h\nu$,
the emission power per unit volume
due to thermal bremsstrahlung is
\begin{equation}
	j_\nu = {{B\rho_1^2}\over\sigma_1} e^{-h\nu/\sigma_1^2}
	\ .
\end{equation}
We project the spherical structure of each cluster solution
by integrating this emissivity along sightlines 
to produce simulated X-ray surface brightness maps.
Figure~\ref{fig.bright}
shows the brightness profiles of our baseline models
in $0.1 - 2.4$~keV and $2 - 10$~keV bands.
The dense central spike appears
as a bright, steep spot at radial scales $r\la 10^{-2}\ U_x \approx 3\ $kpc.
This is comparable to the size of a cD galaxy,
a tiny fraction of the cluster volume.
The soft X-ray profile declines at all radii,
but is shallower between the bright spot and $R_2$.
The hard X-ray profile is more clearly structured:
   steep in the central spot;
   flat in some core ranging from $1~\kpc\la r\la 0.1R_2$;
   and steep again in the outskirts.
It is noteworthy that the apparent X-ray core
is smaller than the halo core
and smaller than the radius where gas temperature peaks.
The X-ray core contains the radius of the temperature dip.
The core is smaller when $F_2$ is larger
   (and seems almost to vanish in the $F_2=9.9$ solutions).

Observed X-ray brightness profiles of galaxy clusters
are empirically fitted with a ``$\beta$-model'' atmosphere
within a presumed isothermal distribution of dark matter or galaxies
\citep[][see Appendix~\ref{halo.hubble}]{cavaliere1976}.
In such fits,
the surface brightness varies with projected radius $b$
according to
\begin{equation}
	S(b) = S_0 [ 1 + (b/b_{\rm c})^2 ]^{-3\beta+1/2}
	\ .
\label{eq.beta.brightness}
\end{equation}
The index $\beta$ fits the outer slope,
   while the parameter $b_{\rm c}$ measures the core radius.
The observed range is $0.4\la\beta\la1.4$
   and typically $0.1~\Mpc\la b_{\rm c}\la 0.5~\Mpc$
   \citep[e.g.][]{jones1984,neumann1999,ettori1999,mohr1999}.
Variant models exist to fit more centrally peaked clusters, 
   by combining cusps and/or $\beta$ components of different sizes
   \citep{xu1998,pratt2002}.
The $\beta$-model was originally derived for isothermal atmospheres,
   but it is still applied to clusters
   that are now known to have radial temperature variations.

Excepting the central spike,
   the $\beta$-model resembles the synthetic X-ray profiles presented here.
In the $\dot{m}=10~m_\odot~\yr^{-1}$ series,
   we find core sizes $b_{\rm c}\la 0.3$~Mpc,
   shrinking slightly with rising $F_2$.
However the $(m_*,\Upsilon)$-optimal model for $F_2=9.9$
   forms no X-ray core at all,
   which is empirically unfavourable.
The fringe brightness declines with similar slopes in all bands:
   $\beta\approx1.0\pm0.2$ around $R_2\la r\la R_I$.
These results sit within the observed range.

We note that $\beta$-like profiles
   are not a distinguishing feature of our formulation,
   nor cored halos as a class.
Even in the entirely cuspy profiles predicted for collisionless dark matter,
   the gas settles as a cored atmosphere,
   with shallow central gradients,
   fitting some kind of $\beta$-model
   \citep{nfw1996,eke1998}.
\cite{makino1998} have shown that $\beta$-like profiles are
   broadly natural to isothermal hydrostatic clusters,
   though cuspy halos yield smaller $b_{\rm c}$.

Like the projected light distributions,
the three-dimensional luminosity profile $L=L(r)$
   has a flat core, a declining fringe and a bright nuclear spot.
The central X-ray spike fits inside radii $r\la10^{-2}$,
   so we choose this as a cutoff defining
   the luminosity of the intracluster medium.
At standard scaling, $m(R)=40U_m$,
   the ICM luminosity increases with $\dot{m}$ and with $F_2$.
We find $L_{\rm icm}\ga 2\times10^{41}~{\rm erg}~{\rm s}^{-1}$
   in the wide/cool family of clusters
   ($\dot{m}=1~m_\odot~{\rm yr}^{-1},T_R=0.4$~keV).
For the medium-sized solutions ($\dot{m}=10~m_\odot~{\rm yr}^{-1},T_R=1$~keV),
we find $L_{\rm icm}\ga 5\times10^{42}~{\rm erg}~{\rm s}^{-1}$.
In the more compact solutions
   we have $L_{\rm icm}\ga 1.4\times10^{44}~{\rm erg}~{\rm s}^{-1}$
   (for $\dot{m}=100~m_\odot~{\rm yr}^{-1},T_R=3$~keV);
and $L_{\rm icm}\ga 2\times10^{45}~{\rm erg}~{\rm s}^{-1}$
   for the most compact set ($\dot{m}=1000~m_\odot~{\rm yr}^{-1},T_R=6$~keV).

The nuclear spot is more luminous,
   but this depends on where we truncate the model.
If we had undertaken a relativistic formulation (\S\ref{discuss.F6}),
   then structures at the classical inner radius $r_*=0$
   would shift out to an event horizon at $r_\bullet=2Gm_*/c^2$.
If we truncate $L(r)$ at $r\sim 5r_\bullet$
   (suiting accretion onto a black hole)
   then we find typical nuclear luminosities around
   $L_*\approx 5\times10^{44}
	(\dot{m}/1~m_\odot~{\rm yr}^{-1})~{\rm erg}~{\rm s}^{-1}$,
   which is plausible as the total accretion power
   of an active galactic nucleus.
Realistically, we would expect a large share of this power
   to emerge in forms other than observable bremsstrahlung radiation.
Extra physics and sub-parsec AGN anatomy may reduce
   the emergent luminosity
   to a small fraction of the accretion power
   (see \S\ref{discuss.evolution}--\S\ref{discuss.spin}).
In particular,
   %the complications of non-spherical geometry,
   Compton scattering and dense obscuring interstellar media
   must alter or reprocess the escaping radiation.
Thus,
   the central density and accretion power
   in the present, undetailed model
   may be consistent
   with radiatively inefficient black hole feeding or with AGN,
   which is not such a dire outcome as early reviews assumed.
(Indeed the bright spot is {\em hot} and therefore qualitatively different
   from the {\em cold catastrophe} accumulations that early works predicted.)

All of the ICM and nuclear luminosities given above
   have assumed the standard normalisation,
   with total cluster mass $m(R)=40 U_m \approx 3.57\times10^{14}~m_\odot$.
If we rescale the masses by a factor $X$
   then the luminosity changes by a factor $X^{5/2}$
   (see Appendix~\ref{s.scaling}).
For instance, if we lighten the compact, megaparsec-scale,
   $\dot{m}=1000~m_\odot~{\rm yr}^{-1}$ solutions
   by a factor $X=0.01$
   to represent an isolated giant elliptical protogalaxy,
   then the inflow shrinks to $\dot{m}=1~m_\odot~{\rm yr}^{-1}$,
   with nuclear luminosity
   $L_* \approx 5\times10^{42}~{\rm erg}~{\rm s}^{-1}$
   and ICM luminosity
   $L_{\rm icm}\approx 2\times10^{40} - 2\times10^{41}~{\rm erg}~{\rm s}^{-1}$.
The central mass limit rescales as well:
   the central black hole must weigh at least
   $1.2\times10^5~m_\odot \la m_* \la 3\times10^9~m_\odot$
   (depending on $F_2$).

Luminosity and mean ICM temperature correlate in our basic models,
   $L_{\rm icm}\propto T_{\rm x}^\alpha$ (see Figure~\ref{fig.LxTx}).
For raw models set via inflow condition (\ref{eq.shock.limit}),
   the slope is $\alpha=3.46$,
   which is steeper than observed
   \citep[$\sim2.6$ to 3.0, e.g.][]{markevitch1998,arnaud1999,novicki2002,ikebe2002,lumb2004}.
Clusters rescaled to virial density (\S\ref{s.virial})
   show $\alpha=2.22, 2.02$ and 2.05
   (for $R_{\rm v}=R_3, R_I, R$ respectively)
   resembling the $\alpha=2$
   prediction from gravitational collapse theory
   \citep{kaiser1986}.
Nature's way of breaking the mass homology
   seems intermediate:
   the cause may involve
   the boundary condition,
   heating \citep{ponman1999,loewenstein2000}
   or other aspects of gas physics (\S\ref{discuss.gas}).
   
\begin{figure*}
\begin{center}
 \includegraphics[width=14cm]{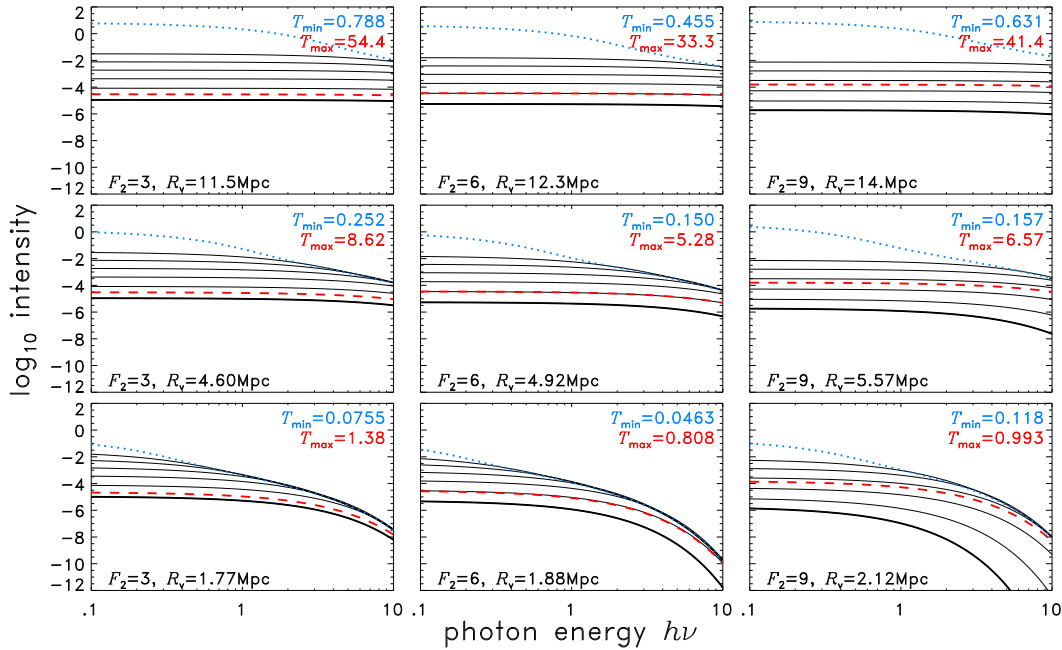}
\end{center}
\caption{
Projected X-ray spectra (continuum only) from the models
with fiducial parameters
$(\dot{m}/m_\odot~\yr^{-1},T_R/\keV)=(1,0.4)$,
$(10,1)$ and $(100,3)$
(from top to bottom rows respectively).
Here, however, the masses and temperatures are rescaled to virial conditions,
choosing $R_{\rm v}=R_I$, as in Table~\ref{table.virial}.
The halo freedom is $F_2=3,6,9$ in the left, middle and right columns.
In each panel,
the blue (dotted) spectrum reaches the temperature dip radius;
red (dashed) reaches the temperature peak;
and the black spectra are projected at fractions of the virial radius,
$b=\{{\frac1{32}},{\frac1{16}},{\frac18},{\frac14},{\frac12},1\}~R_{\rm v}$.
The bold curve is at $b=R_{\rm v}$.
Rescaled values of the peak and dip temperatures (keV)
are noted in the top-right corner of each panel.
}
\label{fig.spectra}
\end{figure*}

We have calculated synthetic X-ray spectra at several projected radii
   in the bloated, medium and small cluster solutions
   (see Figure~\ref{fig.spectra}).
As with the projected brightness profiles,
   each of these plots assumes the virial mass scaling $R_{\rm v}=R_I$.
This makes the large clusters hotter and more massive
   than under the original normalisation ($m(R)=40U_m$)
   becoming a massive supercluster with temperatures of tens of keV
   around the virial radius, which is several megaparsecs.
This rescaling does not resemble any known realistic object;
   the choice of $R_{\rm v}=R_I$ appears unsuited to this family of solutions.
Clusters of medium radius (middle row) show temperatures of several keV
   near the virial radius, which is near or outside the temperature peak.
At the projected radius of the temperature dip,
   the continuum curves in a clearly multi-temperatured way.

The results are similar for the smaller clusters (bottom row)
   but the peak temperature is $\approx1~\keV$
   and the continua are more obviously curved
   in the 0.1--10~keV band that we display.
However line cooling is significant compared to bremsstrahlung
   in low-mass systems such as these,
   which may alter the profiles appreciably
   (\S\ref{discuss.lines}).
Further calculations are needed, with a more detailed cooling function,
   to model galaxy/group spectra including line and edge features.
However line cooling would break the homology of the present solutions
   (Appendix~\ref{s.scaling}),
   requiring a more expensive exploration of parameter-space.
We defer this topic for future investigation.

\subsection{projected mass and gravitational lensing profiles}

\begin{figure*}
\begin{center} 
\begin{tabular}{cc}
 \includegraphics[width=12cm]{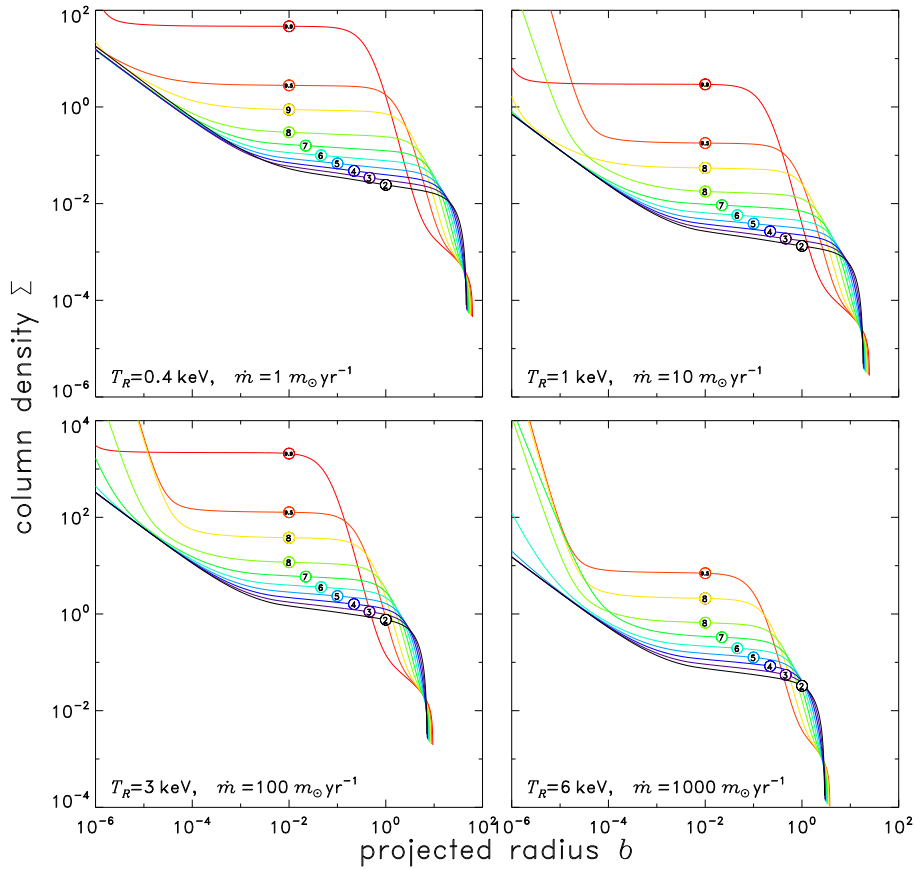}
\end{tabular}
\end{center} 
\caption{ 
Projected profiles of the total projected density,
$\Sigma(b)$,
of $(m_*,\Upsilon)$=optimal cluster solutions
with different $F_2$ values (annotated circles).
We scale each model to the default total mass, $m(R)=40U_m$.
}
\label{fig.lens}
\end{figure*}

To help predict and assess the gravitational lensing signatures 
   of our cluster models,
   we calculate mass maps
   by integrating $\rho_1 + \rho_2$ along lines of sight
   at various projected radii $b$.
Figure~\ref{fig.lens}
   presents the projected profiles
   of total mass column density, $\Sigma=\Sigma(b)$,
   for several represenative $(m_*,\Upsilon)$-optimised models.
The radial gradients are steep on scales from
   several hundreds of kpc to several Mpc
   (beyond the effective core radius $R_I$).
The distribution flattens on scales between tens to hundreds of kiloparsecs.
(This is the mass core.)
In the deepest inner regions,
   within tens of parsecs of the origin,
   the central spike dominates
   and the flat core steepens again.
The border between core and spike typically occurs around
   $10^{-4} U_x - 10^{-3} U_x$,
   regardless of whether the cluster is cool/bloated, moderate or hot/compact.
The radius of the projected mass spike is comparable to, or smaller than,
   the radius where the X-ray brightness spike begins.
The spike in $\Sigma$ exists in theory but may be unobservable in practice:
   dominated by the stellar mass of a cD galaxy.

As was apparent in three-dimensional density profiles,
   the projected $\Sigma$ core is smaller and denser when $F_2$ is greater.
The outer radius $R$ varies more slowly with $F_2$,
   so the mean slope of the outskirts is shallower if $F_2$ is high.
The spike profile depends on $F_2$ also:
   for $F_2<6$ the gas density dominates at $\sim 10$~pc scales,
   and the profile has a logarithmic slope of $\sim-1$.
For $F_2\ge 6$ the spike is more often dominated by dark mass,
   and $\Sigma$ shows a slope steeper than $-1$.
These predominantly dark spikes are more prevalent and more radially extended
   in the compact/hot cluster models.
In the coolest, most diffuse models
  (top left of Figure~\ref{fig.lens})
  all the spikes are gas-dominated down to $10^{-6} U_x$.

We predict that gravitational lensing measurements
that probe images on Mpc scales
will see steep mass slopes of the halo fringe.
Medium-separation lensing systems
(probing just within the core radius)
may reveal shallower gradients of the dark core.
However $\Sigma$ indices are unlikely to reach zero in the core,
   due to the $\rho_1\propto r^{-1}$ gas contribution,
   and the projection of outer layers.
If the innermost regions can ever be discerned
   through the light of the cD galaxy
   then they may appear spiky.

\section{Discussion}

\subsection{implications in galaxy evolution}
\label{discuss.evolution}

As we report in \S\ref{s.deep},
the bottlenecks that select the valid stationary solutions
occur on two different radial scales.
The outer critical region occurs at kiloparsec radii.
It may not be a coincidence
   that this is the characteristic extent of the stellar matter
   in an elliptical galaxy.
If a primitive, initially starless proto-cluster or proto-group 
   were perturbed from a stationary state,
   then the cooling catastrophe will tend to emerge 
   within $\la3$~kpc from the centre.
\cite{nulsen1984} and \cite{fabian1984}
   anticipated this size from back-of-the-envelope arguments;
   our radially complete calculations substantiate it.
During a strong enough disturbance,
   excess cold gas may drop out as star formation,
   until a new stationary configuration settles.
In short, bumping or shaking a proto-galactic globule
   may spawn a spheroidal galaxy monolithically.
However the bottleneck radii limit the extent
   of stellar condensation,
   influencing the high-mass cutoff of galaxies
   \citep[perhaps alongside AGN or conduction effects,
e.g.][]{silk1998,begelman2005,silk2005,best2006,fabian2002b}.

The inner critical regions occur on sub-parsec scales,
   comparable to the sizes of galactic nuclei.
Here again, transient disturbances towards overcooling
   might form a local concentrations of stars.
However in high-$F_2$ cases,
   the inner halo sits on the brink of Jeans instability
   (\S\ref{s.jeans}),
   and the upturn towards this threshold begins as far out as $\sim10~\pc$.
In these conditions, a non-linear perturbation 
   may trigger collective gravitational collapse by the dark matter
   (until a new stationary solution emerges).
Supermassive black holes may be the natural product.
The spikes of high-$F_2$ halos
   may collapse themselves into holes
   whenever anything rattles the central galaxy appreciably.
The collapsible spike mass is comparable to real SMBH masses.

Earlier works have proposed mechanisms for black holes
   to feed and grow from self-interacting dark matter,
   evading Eddington limits
   and emitting little directly luminous evidence
   \citep{soltan1982}.
\cite{ostriker2000} and \cite{hennawi2002}
   considered dark Bondi accretion onto stellar black hole seeds,
   %in a pre-existing cusp,
   followed by slower, diffusive loss-cone refilling.
Their feeding scenario requires SIDM interactions to be weak enough
   to provide an initial NFW-like cusp.
\cite{munyaneza2005}
   show that a SMBH can grow at an asymptotic (exceeding exponential) rate
   if the dark halo has a degenerate fermion core.
\cite{balberg2002a}
   found that gravothermal collapse in conductive, weakly-SIDM
   can spawn $10^6~m_\odot$ black holes directly
   (and these require gas accretion to reach $10^9~m_\odot$).
Like the latter mechanism,
   our scenario requires no seed,
   and its preconditions are self-organised by the gas/halo interplay.
Like the fermion-feeding model,
   we are free to consider interactivity
   at a strength that precludes NFW cuspiness in any epoch.
The mechanism is a differentiated Jeans collapse:
   when stimulated, the hole feeds from a single dark gulp,
   and this intake is only limited by the mass of the dark spike.

Stationarity of the whole cluster demands the development of a central mass.
Note however that the $m_*$ limits
   express only the {\em minimal} object.
This lower limit does not predict
   the correlation between supermassive black holes
   and stellar bulge properties.
Such relations probably involve
   extra regulatory processes operating on galaxy scales.
The depth of the galaxy potential
   and the kick velocities of black hole mergers
   may influence the upper limits
\citep[e.g.][]{haiman2004,madau2004,gonzalez2007a,schnittman2007,campanelli2007a,campanelli2007b,bogdanovic2007,volonteri2007}.
In our scenario, massive black holes
   can form rapidly and darkly.
If a merging galaxy is apt to expel its SMBH
   then we would expect a replacement to condense
   when the scene settles to a resurgent gas inflow and dark spike.

During the tranquil periods between mergers,
there are two plausible fates for the gas inflow
after it penetrates below kiloparsec radii.
It may suffer cold catastrophe
(if the configuration is near the critical border)
depositing cold gas and new stars centrally.
This essentially shrinks the ``cooling flow problem''
down to parsec scales,
which is arguably an improvement over the old $0.1~\Mpc$-sized problem.
More likely, the inflow feeds the central black hole directly.

The existence of powerful quasars in the era $z>6$
   \citep{willot2003,walter2003,barth2003}
   requires that black holes grew rapidly.
Our model not only allows this,
   but {\em requires} it to happen
   before a galaxy or cluster achieves stationarity.
Whether large $\dot{m}$ inflows are sustainable into the modern era
   is another matter.
The accumulated stellar mass distribution
   may eventually alter the central potential enough
   to change the allowed domain of steady feeding solutions
   (\S\ref{discuss.stars}).
If high $\dot{m}$ inflows can persist,
   the implied AGN activity needs explaining.
Much of the time, an AGN may accrete in a radiatively inefficient mode:
many complicating factors could reduce the settled AGN luminosities
to a small fraction of the actual accretion power.
Opacity (\S\ref{discuss.F6})
   and thermal conduction (\S\ref{discuss.conduction})
%and obstructive circumnuclear media
   may smudge or dim the nuclear bright-spot.
An opaque, super-Eddington inflow
   entrains and swallows much of its own luminosity
   \citep{begelman1978,begelman1979}.
The non-luminous power of jets and bulk outflows from sub-parsec regions
   may hide much of the accretion power
   \citep[e.g.][]{dimatteo2003,allen2006}.

If cD galaxies form from massive cooling inflows,
   then they may differ qualitatively
   from elliptical galaxies in the field.
We expect an association between cooling cores
   and the presence of a cD galaxy.
A cluster merger might displace an old cD galaxy,
   and temporarily disrupt the cooling core.
A new cD galaxy would eventually sprout
   in the middle of the resuming cooling flow.

\subsection{extra baryonic and radiative physics}
\label{discuss.gas}

Here we briefly discuss several baryonic phenomena
that could potentially modify our quantitative results.

\subsubsection{line cooling}
\label{discuss.lines}

The presence of line cooling 
  would break the mass/temperature homology of 
  the solutions assuming bremsstrahlung cooling alone.
Additional parameters are therefore needed in the hydrodynamic formulation, 
  thus increasing the dimensionality of the problem
  and  the complexity and computational cost 
  in searching for the structure solutions.

Line emission must affect cooler systems
   (galaxies and groups) more than rich clusters.
The extra cooling may tighten the constraints of the ``cold'' border,
perhaps raising the lower limits on $m_*/m$ for ligher objects.
This may also alter the compactness of clusters
of given mass, comosition and $\dot{m}$:
i.e. the $\Upsilon$-tracks may shift in the $(F_2,R)$ plane.

Hydrogen and helium cooling must have been significant
   for primordial ``mini-halos''
   -- hypothetical, metal-free bodies of gas and dark matter
   that were lighter than modern groups or galaxies.
If their radii are smaller than the most compact models in this paper,
   then mini-halos develop proportionally larger central masses.
The contribution of line cooling above bremsstrahlung
   may raise the minimal level of $m_*/m$ further.
This supports the notion that massive black holes
   condensed directly from primordial envelopes,
   and that condensation may have been
   as fast as free-fall or the onset of cooling catastrophe.

\subsubsection{thermal conduction}
\label{discuss.conduction}

The role of thermal conduction in the intracluster medium
   has been contentious for decades
\citep[e.g.][]{mathews1978,binney1981,nulsen1982b,tucker1983,stewart1984,
friaca1986,bertschinger1986b,boehringer1989,tribble1989,
suginohara1998,narayan2001,loeb2002,voigt2002,zakamska2003,kim2003,voigt2004,
conroy2008}.
Locally tangled magnetic fields may hinder conduction  
   by inhibiting diffusion transverse to field lines.
However, fields that have been aligned radially by a bulk inflow
   might rather promote radial heat conduction
\citep{bregman1988,soker1990}.
Unsuppressed conduction could warm the coldest layers
   and lessen radial temperature variations:
   raising the dip temperature and reducing the $T_{\rm max}/T_{\rm min}$ ratio.

In the conventional models, 
  cluster outskirts have been seen as a plausible reservoir
   for conductively heating the cool core.
This was a natural proposition when solutions were sought
   in the ``too cold'' domain,
   and the whole core was thought to suffer
   a multi-phase, distributed cooling catastrophe.
We emphasise that heat conduction {\em outwards} from the hot nucleus
   may also be relevant.
These regions are ideal for conduction,
   since the densities, temperatures and temperature gradients are all high.
Conductive cooling of the nucleus
   must flatten the central temperature and density profiles,
   dimming the central luminous spot.
Some of the central accretion power conducts outwards,
   ultimately to emerge as ICM luminosity elsewere.
Conductive heating may loosen the constraint
   of the ``too cold'' border, enabling smaller $m_*$ values.
We expect this quantitatively, but it requires proof from extended analyses.

\subsubsection{thermal instability}
\label{discuss.clumps}

The relevance of thermal instability to the phase structure of cool cores
   has also been debated since the early theories of cooling flows
   \citep{fabian1977,mathews1978}.
In uniform, isobaric conditions,
   overdense clumps may overcool radiatively in a runaway manner,
   and condense as a cold phase in thin, hot surroundings
   \citep{field1965}.
However, dense blobs may rain towards the gravitational centre,
   resulting in ablation and warming that counteracts instability.
Blobs that fall deep and fast enough may reheat via shocks
   \citep{cowie1980,nulsen1984}.
Buoyancy may inhibit instability from the start,
   by shifting cool or warm blobs to strata of matching entropy.
Thermal conduction hinders thermal instability
   by warming nascent blobs.
Local magnetic fields that enwrap blobs might help isolate them,
   aiding thermal instability and mass dropout.
Fields that thread blobs and their surroundings
   might bind the phases to co-move
   \citep{nulsen1986,thomas1987}.

Studies of thermal instability in interstellar shocks and stellar accretion
suggest that local, homogeneous, isobaric analyses
are incomplete or too simple for many applications.
The shape of the cooling function affects instability;
if line-cooling complicates the cooling law
   then gas is stable at some temperature.
Macroscopic flow geometry can be influential:
   in general the Eulerian operator $(\partial_t+\vvec\cdot\nabla)$
   couples temporal evolution with motion and gradients.
\citep[See e.g.][and references therein.]{chevalier1982,
bertschinger1986a,saxton1999}
A better treatment of thermal instability in clusters
   requires analysis of global structure as well as local physics
   \citep[e.g.][]{malagoli1987,balbus1989}. 
{%\color{MidnightBlue}
By retaining temporal terms in the governing equations,
   %our model is open to stability analysis:
   our model is ready for stability analyses
   that relate regional thermal (in)stability
   to deformations in the halo and gas profiles
   (in preparation).

However the} strongest original argument for thermal instability
   fades in our 
{%\color{MidnightBlue}
   present}
   results.
A full treatment of the conservation laws and flow velocity
   seems to dispel an old illusion of radially varying $\dot{m}$.
Constant-$\dot{m}$ models resemble observations as well,
and so we need not invoke multiphase effects,
nor mass dropout across the core.
If mass drops out anywhere (for reasons beyond our model)
then it happens deep inside the cD galaxy,
and we have at least reduced the ``cooling flow problem''
to a sub-kiloparsec AGN problem
   \citep[e.g.][]{tabor1993}.
Without dropout, the cooling flow
   leads into hot spherical accretion feeding the nucleus.

Having a warm $T_{\rm min}$, our solutions are consistent with
   the paucity of cold gas and star formation measured at long wavelengths
\citep[see review by][]{donahue2004},
   and also with the X-ray spectroscopic evidence
   that disfavours widespread mass-dropout to very low temperatures
\citep{kaastra2001,tamura2001,peterson2001,peterson2003}.
Observations both inside and outside the cooling radius
   are consistent with single-phase flows but spatially varying temperature
\citep[e.g.][]{boehringer2001,molendi2001,david2001,matsushita2002,ettori2002}.
Thus we justify omitting micro-scale thermal instability
   and retain a smooth, single-phased flow model.

\subsubsection{stellar material}
\label{discuss.stars}

Our present analyses omit the effects of collisionless, stellar mass.
This is justifiable on cluster scales,
   where stars are effectively passive tracers,
   and gas comprises most of the baryons.
Our formulation also applies directly to primordial structures,
   if protogalaxies detached from the background before stars became abundant.

Stellar mass dominates within the effective radii of modern elliptical galaxies,
   such as the brightest central galaxies grown in cluster cores.
A stellar mass distribution
   %\citep[see Appendix~\ref{halo.sersic};][]{prugniel1997}
   (see Appendix~\ref{halo.sersic})
   may alter our solutions within the central kiloparsecs,
   by deepening the potential there.
The extra accretion warming may loosen the ``cold'' constraints on $m_*$.
It may also soften the ICM temperature dip.
We defer the evaluation of gas/dark dipolytropes
   in elliptical galaxy potentials for future study.

If the central galaxy is quiesent then 
   direct stellar interactions with the inflow are negligible.
It was recognised early
   that stellar mass loss in modern elliptical galaxies
   is weak compared to a cooling cluster's inflow
   \citep[e.g][]{nulsen1984},
   and likely to be smothered in terms of mass, momentum and energy.
In our solutions, the central gas is near local virial temperature,
   already similar to the stellar velocity dispersion.
Drag from stellar motions may stir gas locally,
   but the cross-section and covering factor of stars
   is too small for much global effect.
Thus the main effect of stars is gravitational
   and limited to the centre (unless a starburst erupts).
Starbursts may affect the early life of a cD galaxy,
   heating nearby gas as an AGN might.

\subsubsection{heating by active galaxies}
\label{discuss.feedback}

We omit AGN interactions from our model,
   in order to focus on the undriven, natural tendency of clusters.
Feedback effects require extra semi-empirical terms,
   with a diabolically tempting number of adjustable parameters.
If jet activity or other phenomena
   drive the cluster medium to convection or turbulence,
   with eddy kinetic energies comparable to the gaseous internal energy,
   then the gas gains extra effective degrees of freedom, $F_1>3$.
This may steepen gas density profiles,
   and shift the ``cold'' and ``fast'' constraints on $m_*$.

Some authors propose that AGN output
   (such as the mechanical power of radio lobes)
   can suppress or even regulate cooling flows
\cite[e.g.][]{tabor1993,binney1995,churazov2001,brueggen2001,quilis2001,
ruszkowski2002,kaiser2003,ruszkowski2004}.
There seems enough raw available power for low and medium-mass clusters
   \citep{birzan2004,allen2006},
   but the sufficiency of AGN warming in massive cooling flows is debatable.
For AGN to be effective heaters,
   radio lobes must mix with the ICM,
   or the gas must be viscous enough to dissipate disturbances.
The distribution of AGN power is questionable in some systems:
   strong jets may cut themselves channels out of the core,
   dumping their power ineffectually in the outskirts
   \citep[e.g.][]{best2006,vernaleo2006}.
In some clusters, ICM metallicity gradients
   imply that mixing has not been thorough.
Some cool cored clusters lack nuclear activity altogether.

Because of their low entropy, cooling flows may simply sink and slip
   around the sides least disturbed by AGN channels.
If so then active cooling clusters may still resemble our solutions overall.
However if the AGN blocks the inflow near the centre
   then the model needs modification there:
   an additional distributed heating function in $\Lcal$,
   and a local drop in $\dot{m}$.
Starvation on sub-kiloparsec scales may dim the luminous nucleus.
The outskirts (beyond the reach of radio bubbling)
   should match the standard profiles obtained here.

However our adaptive model improves or solves
   key aspects of the ``cooling flow problem''
   without resort to extrinsic heating.
Considering a responsive halo,
   and all relevant momentum and self-gravity terms,
   we find that steady clusters always develop a floor temperature
   due to purely gravitational self-warming.
This level is high enough to explain the rarity of cold condensates.
Consequently, we do not need non-gravitational heating
   to dominate on megaparsec scales.
We dispel the need for finely tuned, two-way {\em feedback} between
   the cooling flow and the heating processes.
Stellar and AGN heating are demoted to an incidental, intermittent role.
Heating need not be persistent nor stable.

\subsubsection{relativity, opacity and cosmic rays}
\label{discuss.F6}

In a relativistic formulation,
   the inner boundary would shift from $r_*=0$
   to the Schwarzschild radius of the central mass.
The bremsstrahlung cooling function acquires relativistic corrections.
Gravitational redshift dims the inner boundary.
Inflow velocities will be naturally subluminal.
As the gas becomes relativistic in the nuclear regions,
   it gains more effective degrees of freedom,
   $F_1\rightarrow 6$.
This implies a higher combined heat capacity
   than a normal $F_1=3$ medium.
The region where the dark spike verges on Jeans-instability may enlarge.
Otherwise the issues of pressure support,
   the subsonic constraint
   and the avoidance of cooling catastrophes
   remain qualitatively unchanged in a relativistic framework.

If gas in the central regions becomes Compton-thick
   or otherwise dominated by radiation pressure 
   then this will also result in $F_1\rightarrow6$.
This will steepen the gas density spike
   and may promote collapse of the dark spike.
The opaque inner inflow must be radiatively inefficient,
   enlarging the domain of effectively adiabatic Bondi-like behaviour.
We cannot presently say in which direction the $m_*$ limits change.
Sub-parsec AGN anatomy may complicate the issue.

If cosmic rays contribute significant pressure to the intercluster medium,
   then they would deserve incorporation as a third fluid in our model.
Extra source terms enter the momentum and energy equations
   to express cosmic ray diffusion
   and heat exchanges with the coterminous thermal gas.
As a relativistic plasma,
   cosmic rays have $F_3\approx6$ degrees of freedom.
The consquences for the innermost structures
   and $m_*$ limits may follow cases of a halo with $F_2\geq6$.

\subsubsection{angular momentum}
\label{discuss.spin}

If some of the cluster gas possesses significant angular momentum
   then the inflow could deviate from radial streaming at small radii.
Some fraction would accrete onto a small disc,
   of a size determined by rotational support
   \citep[e.g.][]{nulsen1984}.
This may alter the minimal-$m_*$ limits slightly,
   and the steady solutions might require
   some minimal disc surface density profile.
If the disc is viscous then it is only a temporary residence for inflowing gas;
   it ultimately feeds the central mass.
A  disc that accretes enough mass
   could self-gravitate and develop density waves, or fragment.
This is a recipe for forming spiral galaxies.

However discs (and filaments) are only viable
   as long as they avoid contact with similar bodies.
Collisions and stirring by asymmetric substructures 
   can restore the spherical approximation in the long run,
   and on the scales relevant to our model.
Turbulence or convection in the inner regions
   could easily destroy or preempt a disc,
   easing angular momentrum efflux and quasi-spherical mass influx.
Empiricially, our spherical approximation is valid
   as far as relaxed clusters and central galaxies are actually roundish.

A spherical model
   cannot address the topic of halo substructure directly.
Collisionless dark matter cosmogonies overpredict the abundance
   of satellites at galaxy and group scales by at least an order of magnitude
   \citep{klypin1999,moore1999a,moore2000,donghia2004}.
Several authors have debated whether a (weakly) SIDM halo
   can be lumpy and aspherical enough
   as a gravitational lens and host of satellite galaxies
   \citep[e.g.][]{meneghetti2001,moore2000,gnedin2001,
	natarajan2002,furlanetto2002}.
Dwarf halos might ablate at dark bow-shocks,
   or (we suggest) might persist as self-bound, dark eddies
   that roll as they orbit in a turbulent background.
Within a cluster, each galaxy perturbs the cluster profile locally,
   and each subhalo centre probably develops a miniature analogue
   of the spherical inflow solutions.

\subsection{nature of dark matter}
\label{discuss.dark}

In detail, our solutions depend on the assumption of
   a polytropic equation of state for the dark matter.
This condition is the emergent equilibrium
   if Tsallis' thermostatistics govern the halo,
   even if dark matter lacks non-gravitational interactions entirely.
If dark matter is a Bose-Einsten condensate,
   then it is effectively polytropic in the classical limit
   \citep{sin1994,goodman2000}.
If it is a degenerate fermion gas
   then it has a more complex equation of state,
   obtainable by integrating a local distribution function
   \citep{munyaneza2005,munyaneza2006},
   or else a polynomial approximation
   \citep{nakajima2007}.
If dark matter is collisionless but ruled by long-range dark forces
%(e.g. if dark particles are charged in relation to extra gauge fields
%for which luminous matter is neutral)
   then a more complicated treatment becomes necessary,
   analogous to collisionless plasma physics.
If dark matter feels strong enough local self-interactions,
   then it is analogous to an adiabatic ideal gas,
   and a polytropic equation is expected.
%SIDM may be essentially polytropic in the strong regime,
   %approximating an adiabatic ideal gas.
%This is not merely a ``fluid approximation'',
%but is an explicit dark fluid hypothesis, following e.g. Peebles (2000).

The possibility of local dark self-interactivity
\citep{spergel2000,goodman2000,peebles2000}
is theoretically and observationally attractive,
but not yet exhaustively tested.
This scenario explains the cored halo profiles evident in many galaxies,
   %evident in low-surface brightness and dwarf spheroidal galaxies,
   %\citep{flores1994,moore1994,gilmore2007},
   and may yield more realistic substructure than collisionless cosmogonies.
   %and suggest more realistic distributions of satellites
   %than the overpredictions by collisionless cosmogonies
   %\citep{klypin1999,moore1999a,moore2000,donghia2004}.
Early numerical studies of weakly self-interacting halos
   mimicked SIDM using particle codes with Monte Carlo scattering,
   which bred realistic cores
   but found a gravothermal catastrophe
   that could eventually degrade the cores into isothermal cusps 
\citep{moore2000,yoshida2000a,yoshida2000b,burkert2000,kochanek2000,dave2001}.
However later analyses \citep{balberg2002b,ahn2005}
   considered more general initial conditions,
   delaying collapse beyond the Hubble time.
Polytropic halos (as in this work) can describe
   a more strongly interacting fluid regime,
   where the mean free path is short enough
   that conduction and gravothermal effects vanish.
    
Despite the indications from galaxy scales,
   the fashionable preference is to defer and displace
   the faults of CDM substructure
   onto ``baryon feedback'',
   which is beset with long-term challenges
   in numerical methods and in theory.
The evidence on cluster scales is still ambiguous enough to allow this.
   Many observers {\em assume} cuspy profiles in their data fits.
   (Fully non-parametric modelling remains rare.)
For relaxed clusters,
   some X-ray deprojections show soft-cored halo profiles,
   \citep[e.g.][]{nevalainen1999,ettori2002b,katayama2004,voigt2006,zhang2005,zhang2006}
   while others seem compatible with cusps
   \citep[e.g.][]{pointecouteau2005,vikhlinin2006}.
In some cases the total mass profile appears cuspy,
   but not to the extent expected of a collisionless halo
   affected by gas
   \citep[e.g][]{zappacosta2006}.
Gravitational lensing analyses also give mixed signs:
   some prefer or allow soft cores
   \citep{tyson1998,dahle2003,gavazzi2003,diego2005,sand2002,sand2004,sand2008,
	halkola2006,halkola2008,rzepecki2007}
   while others prefer cusps
   \citep{broadhurst2005a,sharon2005,saha2006,limousin2007}.
% SOFT: tyson1998, gavazzi2003, sand2002, sand2004, sand2008
% AMBIG: dahle2003,diego2005
% CUSP: sharon2005 (1.5 slope?), limousin2007,saha2006
% gavazzi2003: CUSP OK, until fifth image implies CORE

In more violent circumstances,
   cluster mergers have been treated as probes of dark interactivity.
One gravitational lens ``bullet cluster''
   was claimed as a merger of collisionlessly interpenetrating halos,
   separating from shocked gas
   \citep{clowe2006}.
The mass, speed, timing and rarity
   of the hypothetical merger have been questioned,
   and well-tuned simulations devised in reply
   \citep{hayashi2006,farrar2007,zhao2007,springel2007,milosavljevic2007,angus2008,nusser2008}.
Subsequently, an anti-bullet cluster has appeared,
   where the dark matter is the more dissipative constituent
   (settled in the middle) while galaxies fly on the periphery
   \citep{mahdavi2007}.
Another lensing cluster is reported
   with an encircling ring or shell of dark matter
   \citep{jee2007}.
Taken together,
   these special cases tell an inconsistent story about dark physics.
However, particular projected morphologies admit multiple interpretations:
   for instance the ``bullet'' subcluster velocity vector can be reversed,
   and the line-of-sight shapes and displacements are unknown.
This paper cannot aim to disentangle
   all the latent assumptions in the dark matter merger problem,
   but clearly some alternative gestalts are needed.
Varieties of SIDM remain among the promising candidates.
%\citep[The version of][even appears to allow reversible shocks.]{peebles2000}

In the absence of a central mass or gas inflow,
   a polytropic halo can have a constant-density core
   (Appendix~\ref{s.gasless}),
   compatible with galaxian evidence.
However for galaxy clusters with inflowing gas,
   we do not obtain simple cored profiles
   like those assumed commonly.
The presence of inflow requires a central mass for stationarity,
   and a polytropic cluster halo grows a density spike {\em within} its core.
The spiky halos are effectively a ``contraction''
   induced by the central mass and gas inflow.
For large $F_2$ or large $\dot{m}$,
   the core radius may be small enough
   to give a misleading appearance of a NFW-like cusp.
Smallness of some observed cluster cores
   \citep[e.g.][]{dahle2003,katgert2004,limousin2007}
   is not evidence against SIDM.
Rather, it will help to constrain the dark freedom $F_2$
   and the inflow history of baryons.
Cluster cores as small as a few tens of kiloparsecs
   are possible if $F_2$ is large.
This fits the concordance favouring $8\la F_2\la10$,
   which minimises the central mass $m_*$ (this work)
   and agrees with \cite{nunez2006},
   who effectively find $F_2\approx9.6$
   by fitting galaxy rotation curves.

Given any alternative closed set of equations
   for the dark dynamics and statics,
   one can repeat the formulation of this work,
   to find obtain another set of differential equations
   coupling the gas and halo.
The interplay of these constituents in their shared potential
   must always lead to the exclusion of some domains
   due to cooling catastrophe or acoustic breaks in the gas.
However it must be proven, for each scenario,
   whether nonzero $m_*$ is required (as in the model dipolytropes here)
   and whether structures with cosmic baryon fraction can exist.

\section{Conclusions}

We have presented a self-consistent 2-component model for galaxy clusters,
   bound by a non-static gravitational potential
   that emerges naturally from the solutions
   along with the co-adapted halo and gas profiles.
Applying this formulation to clusters of plausible total mass and composition,
   we reconcile some of the observational difficulties
   involving gas inflows due to cooling.
Furthermore,
   we find that stationary solutions of the cluster structure
   invariably require (or develop) a non-zero central mass.

We have analysed the distribution of cooling gas
   in a responsive spherical halo of megaparsec scale.
Mass, momentum and energy continuity are imposed.
Bremsstrahlung radiative cooling is allowed to become dynamically significant.
All constituents participate gravitationally.
Realistic models emerge when  dark matter has a polytropic equation of state,
   which is justified in terms of
   the equilibria of Tsallis thermostatistics,
   adiabatic dark self-interactions
   or Bose-Einstein condensation.

We find that there exist steady, continuous solutions
   spanning all radii inside the halo.
The joint constraints of cooling and acoustic continuity
   set the minimal central mass.
The minimal $m_*$ varies with $\dot{m}$ and $F_2$
   but only weakly with the gas surface temperature $T_R$.
The cluster's total gas fraction is linked with $\dot{m}$, $R$ and $T_R$.
The masses, $\dot{m}$, densities, temperatures, velocities
   of any particular solution
   can rescale to yield another valid solution with the same radial dimensions.
%The essence of each family of solutions can be characterised
   %by intrinsic radial scales.

A cosmic baryon fraction and observed SMBH masses
   are consistent with the halo's effective microscopic degrees of freedom
   being in the interval $7\la F_2<10$.
The lower limits on $m_*$ are laxer if $\dot{m}^2/m^3$ is smaller.
For cosmic composition,
   the fiducual total cluster mass
   and $\dot{m}\ge 1~m_\odot~{\rm yr}^{-1}$,
   we always find $m_*\ga5\times10^5~m_\odot$
   (or $m_*/m\ga 2\times10^{-9}$).
Smaller central masses are impossible in steady clusters,
   unless extra physics dominate.
To enable $m_*$ as small as $10^6$--$10^7~m_\odot$ in a cluster,
   we need $9\la F_2<10$.
This agrees with galaxy rotation models of \cite{nunez2006}
   which imply $F_2\approx9.6$.

The halo density develops a spike around the central mass,
   surrounded by a flat core attenuating to a fringe on megaparsec scales.
These layers are less distinct when $F_2$ is larger.
This varied structure remains apparent
   in projections of the total column density.
For $F_2$ high enough to enable plausible $m_*$,
   we find halo cores with $10~\kpc\la R_1\la300$~kpc.
Observationally, there are reported core fits with
   $15~\kpc\la R_1\la200~\kpc$
   \citep[e.g.][]{dahle2003,diego2005,rzepecki2007,halkola2008}
   which is similar to what we obtain.
We predict that mass profiles steepen beyond NFW in the dim fringe.
Given observable scale radii such as $R_2$,
   one could predict the outer radius where a halo naturally truncates.

Our solutions belong outside the inevitably overcooling regime
   where classic cooling flows were constructed.
They naturally provide a non-zero floor temperature,
   obviating the need for (unobserved) mass dropout and cold condensation.
The entropy, density and temperature profiles
   broadly resemble observed clusters,
   suggesting that varying the gas parameters
   may enable detailed fits in future.
We find a shallow gas entropy ramp at radii inside the sonic point,
   rather than a flat isentropic core.
In projection, the intracluster medium
   resembles the classic $\beta$-model X-ray profile in the outskirts,
   plus a peaked cooling/warming core.
The central luminous spot is comparable to AGN power,
   though opacity, conduction and detailed AGN anatomy
   could probably soften and spread the emission,
   and lower the radiative efficiency considerably.
In the optimal-$m_*$ solutions,
   the ratio of peak/dip temperatures is a factor 5--40.
This reduces the need for AGN self-regulation
   (as distinct from incidental heating).
The inclusion of thermal conduction and $F_1>3$
   might improve the profiles and lessen the AGN role further.

By construction, the classes of solutions that we obtained 
   comprise the steady configurations
   of bremsstrahlung-cooling Mpc-scale spheres.
If our broad physical assumptions hold,
   and if structural asymmetries are subdominant,
   then these solutions represent end-points of cluster relaxation.
In the configuration-space of spherical clusters,
   the  solutions are fixed points.
If feedback or non-gravitational heating processes
   become globally, persistently important,
   then they drive the evolving cluster state
   in a forced orbit around those natural points.

A temporary disturbance of the system may cause
   a local cooling catastrophe or acoustic disconnection,
   and the structure must somehow adjust
   until reaching another steady state in neighbouring configuration-space.
Our analysis does not outline a particular evolutionary path,
   but the profiles offer clues.
The critical bottlenecks for gas continuity
   occur at radial scales typical of elliptical galaxies.
This may be a preferred layer for cold gas dropout and star formation,
   during any transient, externally driven detour into cooling catastrophe.
Stability analyses are needed to determine whether
   our scenario implies regulation or runaway monolithic collapse.
For $F_2>6$,
   the innermost halo is only marginally Jeans-stable,
   implying that large-amplitude disturbances could trigger
   a local gravitational collapse of dark matter
   (without involving the gas directly).
This mechanism for dark growth of SMBH
   may turn out to be an important process
   (besides baryonic feeding and gravitational ejection)
   influencing SMBH demographics.

\section*{Acknowledgments} 

We thank Ignacio Ferreras for
   comprehensive, stimulating and helpful discussions.
We thank the referee, Michael Loewenstein,
   for abundantly constructive criticism.
We acknowledge the use of numerical routines 
   provided by the Gnu Scientific Library project.
 
%%\bibliographystyle{mn}
%\bibliographystyle{mn2e}
%\bibliography{jour,blob}

\appendix

\section{natural units}
\label{s.units}

If the gas has approximately solar composition then
the bremsstrahlung constant has a value
$B=B_\odot\approx 5.06\times10^{16}~{\rm g}^{-1}~{\rm cm}^{4}~{\rm s}^{-2}$,
calculated according to \cite{rybicki},
(for a Gaunt factor $g_{\rm B}=1.25$)
as in \cite{saxton2005}
with the abundance tables of \cite{anders1989}.
We parameterise the composition dependency of $B$
relative to the solar value
as a correction factor, $\zeta\equiv B/B_\odot$.
Values of $\zeta$
depend on the abundance-weighted ionic mass ($\bar{m}$),
charge ($\bar{Z}$) and charge-squared ($\overline{Z^2}$),
\begin{equation}
	\zeta \propto
		{{\overline{Z^2} }\over{\bar{Z}} }
		\left({
			{\bar{Z}}\over{
				\bar{m}/m_{\rm e} + \bar{Z}
			}
		}\right)^{3/2}
		\sqrt{ {\bar{Z}}\over{1+\bar{Z}} }
		\ g_{\rm B}
		\ .
\end{equation}
For astrophysical plasmas, $\zeta$ is close to unity:
for the same $g_{\rm B}$ in a pure H plasma, $\zeta\approx0.979$;
for a $9:1$ mix of H and He, $\zeta\approx0.925$.

If we define a system of units 
such that $B=1$
and the gravitational constant
$G=6.6732\times10^{-8}~{\rm g}^{-1}~{\rm cm}^3~{\rm s}^{-2}=1$ also,
then the unit of length is
\begin{equation}
	U_x\equiv B/G
	=7.58\times10^{23}~\zeta~{\rm cm}
	=0.246~\zeta~{\rm Mpc}
	\ ,
\end{equation}
It may be significant that this scale,
which is natural to any object governed by
self-gravity and optically-thin bremsstrahlung,
is typical of the observed X-ray core radii of galaxy clusters
\citep[e.g.][]{jones1984}.

If we adopt a velocity scale where
$\sigma^2=1$ corresponds to a temperature of $1~{\rm keV}$,
then the unit of velocity is
\begin{equation}
	U_v=3.95\times10^7~{\rm cm}~{\rm s}^{-1}
	\ ,
\end{equation}
which implies a time unit
\begin{equation}
	U_t\equiv{{U_x}/{U_v}}
	=1.92\times10^{16}~\zeta~{\rm s}
	=0.608~\zeta~{\rm Gyr}
	\ .
\end{equation}
The age of the universe
\citep{spergel2003}
is presently thought to be
$\sim 22.5/\zeta$.
The unit of mass arises from
\begin{equation}
	U_m \equiv  U_x^3 U_t^{-2} G^{-1}
	=1.77\times10^{46}~\zeta~{\rm g}
	=8.91\times10^{12}~\zeta~m_\odot
	\ ,
\end{equation}
and the unit of density is
\begin{equation}
	U_\rho\equiv{{U_m}{U_x^{-3}}}
	=4.07\times10^{-26}~\zeta^{-2}~{\rm g}~{\rm cm}^{-3}
	\ .
\end{equation}
The critical density for the universe today is
$\rho_{\rm c}=2.33\times10^{-4}\ \zeta^2\ U_\rho$,
%2.3283405e-4 = 9.4687321e-30 g cm^-3
and the mean matter density is 
$\Omega_{\rm m}\rho_{\rm c}\approx 6.23\times10^{-5}\ \zeta^2\ U_\rho$.
%6.28652e-05
The unit of particle number density depends on plasma composition
in a more complicated way,
\begin{equation}
U_n\equiv{{U_\rho}\over{\bar{m}+\bar{Z}m_{\rm e}}}
\ .
\end{equation}
For solar composition we have
$U_n\approx 3.96\times10^{-2}~{\rm cm}^{-3}$.
The units of mass accretion and power are
\begin{equation}
U_{\dot{m}}\equiv{{U_m}/{U_t}}
=1.47\times10^4 M_\odot~{\rm yr}^{-1}
\ ,
\end{equation}
\begin{equation}
U_L\equiv{{U_m U_v^2}/{U_t}}
=1.44\times10^{45} {\rm erg}\ {\rm s}^{-1}
=3.77\times10^{11}\ L_\odot
\ .
\end{equation}
and neither depends on composition, $\zeta$.

\section{scaling relations}
\label{s.scaling}

Given one steady cluster model,
it is possible to construct a set of equivalent models
that differ only by uniform multiplicative rescaling of the physical variables.
Let us define the transformation factors as
\begin{eqnarray}
r&\hspace{-0.25cm}\rightarrow&\hspace{-0.25cm}X_r\ r,
\nonumber\\
m&\hspace{-0.25cm}\rightarrow&\hspace{-0.25cm}X_m\ m,
\nonumber\\
\dot{m}&\hspace{-0.25cm}\rightarrow&\hspace{-0.25cm}X_{\dot{m}}\ \dot{m},
\nonumber\\
\sigma^2&\hspace{-0.25cm}\rightarrow&\hspace{-0.25cm}X_\sigma\ \sigma^2,
\nonumber\\
s&\hspace{-0.25cm}\rightarrow&\hspace{-0.25cm}X_s\ s,
\nonumber\\
v&\hspace{-0.25cm}\rightarrow&\hspace{-0.25cm}X_v\ v\ \ \ \ \mbox{and}
\nonumber\\
\rho&\hspace{-0.25cm}\rightarrow&\hspace{-0.25cm}X_\rho\ \rho.
\end{eqnarray}
Mass conservation
(\ref{eq.mass.conservation})
implies a constraint
\begin{equation}
	X_{\dot{m}} = X_r^2 X_\rho X_v\ .
\end{equation}
The equations of the mass profile,
such as
(\ref{eq.mass.gradient}) or
(\ref{eq.beta.mass}),
require that
\begin{equation}
	X_m = X_\rho X_r^3\ .
\label{eq.Xm.3}
\end{equation}
Mach numbers must be left unchanged during the transformation,
and thus
\begin{equation}
	X_v^2 = X_\sigma\ .
\end{equation}
In each of the gas equations,
say (\ref{eq.beta.v}),
all of the additive terms
must rescale by the same product.
After some evaluation, this implies that
\begin{equation}
	X_m = X_\sigma X_r = X_\rho X_r^2\ .
\label{eq.Xm.2}
\end{equation}
Satisfaction of (\ref{eq.Xm.3}) and (\ref{eq.Xm.2})
implies that the spatial dimensions cannot vary,
\begin{equation}
	X_r = 1\ .
\end{equation}
Therefore, any valid similarity transformation
parameterised by a scale $X$,
implies the following scaling factors
for the key physical variables:
\begin{eqnarray}
	X_m&\hspace{-0.25cm}=&\hspace{-0.25cm}X_\rho = X_\sigma = X
	\ ,\nonumber \\
	X_v&\hspace{-0.25cm}=&\hspace{-0.25cm}X^{1/2}
	\ ,\nonumber \\
	X_{\dot{m}}&\hspace{-0.25cm}=&\hspace{-0.25cm}X^{3/2}
	\ ,\nonumber \\
	X_s&\hspace{-0.25cm}=&\hspace{-0.25cm}X^{(F-2)/F}
	\ .
\label{eq.scaling}
\end{eqnarray}
Luminosities and surface brightnesses scale as $X_L=X_{\dot{m}}X_v^2=X^{5/2}$,
so the X-ray luminosity scales as $m^{5/2}$ within any family of solutions.
The relations (\ref{eq.scaling})
imply the existence of two invariant length scales,
associated with the mass inflow and temperatures:
\begin{equation}
R_{\dot{m}} \equiv \left.{{\sqrt{\gamma_1}\sigma_1 m}/{\dot{m}}}\right|_R
%R_{\dot{m}} \equiv \left.{{\sqrt{\gamma_1}\sigma_1 m}\over{\dot{m}}}\right|_R
\ ,
\end{equation}
\begin{equation}
R_\sigma \equiv \left.{{G m}/{\gamma_1\sigma_1^2}}\right|_R
%R_\sigma \equiv \left.{{G m}\over{\gamma_1\sigma_1^2}}\right|_R
\ ,
\end{equation}
The latter is algebraically equivalent to the sonic radius in
simple, adiabatic Bondi accretion,
although cooling and self-gravity mean that
our models needn't develop a sonic point nor $\Mcal^2$ extremum
at this radius.
Together, the parameters ($F_1,F_2,R,R_{\dot{m}},R_\sigma)$
uniquely denote a set of homologous cluster models.

\section{EFFECTIVE DEGREES OF FREEDOM}
\label{s.freedom}

The key property of an ideal fluid is $F$,
the effecitve degrees of freedom.
Thus $F$ is a qualitatively decisive parameter of the halo models.
In ordinary space,
free particles have three translational degrees of freedom.
If the particles are individual, point-like entities lacking substructure
then $F=3$.
However many physically motivated scenarios 
entail $F>3$ or non-integer values.

If individual particles can rotate, twist or distort
then there are additional internal microscopic degrees of freedom
(e.g. $F=5$ for a diatomic gas).
Highly relativistic or radiation-dominated fluids have $F=6$
(e.g. cosmic ray contributions to ICM pressure).
Larger integer values of $F$ could also occur
if the particles experience higher spatial dimensions,
e.g. if their de~Broglie wavelength is smaller than
the scale of compact hidden dimensions.
If a fluid includes subspecies that do not fully interact,
then the effective $F$ is larger than for single species.

Some alternative scenarios involve fewer degrees of freedom.
If the fluid is a classical Bose-Einsten condensate
\citep[as in][]{goodman2000,arbey2003,boehmer2007,lee2008}
then $F=2$,
for an equation of state $p\propto\rho^2$.
A case $F=1$ could describe constrained particles,
   analogous to beads on an abacus.
An incompressible fluid corresponds to $F=0$.
Isobaric conditions can be described by $F=-2$.

If non-local physical interactions are important,
and the medium is described by Tsallis' statistics
\citep{tsallis1988},
then $F$ is effectively some non-integer,
$F=(3q-1)/(q-1)$ for some constant $q$
\citep[][]{plastino1993,hansen2005,nunez2006,zavala2006}.
In a gravitational context,
this includes and entails
the ephemeral constraints and interactions
present on all intermediate levels
between the small scale of two-body scattering
and the large scale of the global potential.
% ephemeral subsystems, quasi-bound interactions
Other meso-scale physics,
such as the energy associated with turbulent eddies,
can also provide larger, non-integer values of $F$.
If highly efficient heat transport processes operate
then the fluid approaches isothermality
%($\sigma^2$ spatially constant)
and $F\rightarrow\infty$.

\section{STANDARD GASLESS HALO MODELS}

\subsection{comparative measures of a halo}
\label{s.metrics}

To found our treatment of two-fluid cluster models,
   we will here review the intrinsic properties
   of polytropic halos (without gas),
   and contrast them with other halo models in the literature.
In order to compare theoretical and semi-empirical halo models
   with each other and with observations,
   it is necessary to define some global physical measurements.
Every real halo ought to have a finite outer radius, $R$,
   but the invisibility of dark matter means that $R$
   isn't directly determinable.
In practice, the cluster's baryons are only visible
   out to certain detection limits (e.g. to a limiting X-ray flux)
   and this extent sets lower bounds on $R$.
We prefer to characterise the models using spatial measurements,
   (which can be compared to the true outer radius)
   since these are invariant under mass rescaling 
   (Appendix~\ref{s.scaling}).

For non-singular and cored halo models,
   a ``King radius'' is defined in terms of the central conditions,
   \citep[e.g.][p.228]{binney1987}
\begin{equation}
	%R_{\rm K}=\left.{ \sqrt{ {9\sigma^2}\over{4\pi G\rho} }}\right|_{r=0}
	R_{\rm K}=\left.{ \sqrt{ {9\sigma^2}/{4\pi G\rho} }}\right|_{r=0}
	\ ,
\end{equation}
This scale typifies the extent of the flat density core in many models.
The cluster rescaling (appendix~\ref{s.scaling}) leaves $R_{\rm K}$ invariant.
However $R_{\rm K}$ is undefined for cuspy halos
   or those with a central point mass.
Thus we require alternative measures of core size
   and overall halo concentration.

Firstly, let us define 
   a radius that contains a majority of the mass,
   or that typifies the central concentration.
%Since our polytropic halo models lack an explicit parameter
   %for the core radius,
We refer to a sphere's total mass $m$ and moment of inertia,
\begin{equation}
	I={{8\pi}\over{3}}\int_0^\infty \rho r^4\ dr
	\ .
\end{equation}
   which give a mass-weighted lever radius,
\begin{equation}
	R_I\equiv \sqrt{
		5I/2m
		%{{5I}\over{2m}}
	}\ ,
\end{equation}
   which is scaled such that $R_I=R$ for a uniform sphere.
The radius $R_I$ is applicable to
   models where the core is not explicitly parameterised.
Since density decreases monotonically in $r$,
   the inner layers dominate $R_I$.
When the the mass is centrally peaked
   or the core is small compared to the fringe,
   the ratio $R_I/R$ is small.
Though the true surface may be invisible
   below some flux or density thresholds,
   truncated observational estimates of $R_I$ or $R_I/R$
   might still approximate the global values acceptably.

We define another radial scale measuring the cluster's self-gravity.
The gravitational potential energy of a spheroid,
\begin{equation}
	W = -4 \pi G \int_0^\infty m \rho r\ dr
	= 2 \pi \int_0^\infty \Phi \rho r^2\ dr
	\ ,
\end{equation}
is finite for realistic models.
This leads to the definition of a gravitational radius,
\begin{equation}
	%R_w = -{{G m^2}\over{W}}
	R_w = -G m^2/W
	\ ,
\end{equation}
\citep[see][pp.68-69.]{binney1987},
which can be regarded as the size of energetically typical orbits
in the halo.

Halo structure can be diagnosed by observable kinematic tracers,
   such as velocities of gas clouds or stars in circular orbits in a galaxy,
   or the motions of galaxies within a cluster.
The circular orbital velocity of test particles
   peaks at some radius $R_{\rm o}$ if
\begin{eqnarray}
	\left.{ ( 4\pi\rho r^3 - m ) }\right|_{r=R_{\rm o}}
	&\hspace{-0.3cm}=&\hspace{-0.3cm}0\ ,
\hspace{1cm}\mbox{and}
\nonumber\\
	\left.{ {{d\ln\rho}\over{d\ln r}} }\right|_{r=R_{\rm o}}
	&\hspace{-0.3cm}\le&\hspace{-0.3cm}-2
	\ .
\label{eq.rot.peak}
\end{eqnarray}

We denote a sequence of signature radii
   where the logarithmic density slope passes specific values:
\begin{equation}
	\left.{
		{{d\ln\rho}\over{d\ln r}}
	}\right|_{r=\left\{{R_1, R_2, R_3, R_4, \ldots}\right\}}
	= \left\{{-1, -2, -3, -4, \ldots}\right\} \ .
\end{equation}
For instance, $\rho\propto r^{-3}$ at the slope-3 radius $R_3$,
   and so on.
These slope radii may be multi-valued,
   if the density profile undulates
   (i.e. exhibiting concentric, alternating steep and flat layers).
If the halo is radially finite ($R<\infty$)
   then all slope radii are finite too.
Infinite models may lack some of the slope-radii.
If the rotation curve peaks anywhere,
   then (\ref{eq.rot.peak}) implies that $R_{\rm o}\ge R_2$.
The total mass cannot be finite without $R_3$ existing.
A finite moment of inertia and $R_I$ requires finite $R_5$.
Finite $W$ and $R_w$ require $R_{2.5}$ and $R_3$ to occur
   at least once in the outskirts.

The literature on cosmological simulations conventionally defines
a ``virial radius'', $R_{\rm v}$,
enclosing a mean density that is some multiple, $\delta_{\rm c}$,
of the cosmic critical density.
This overdensity is in the range $100\la\delta_{\rm c}\la200$,
depending on the cosmological model.
As in \cite{bryan1998},
we use
$\delta_{\rm c}=18\pi^2+82(\Omega_{\rm m}-1)-39(\Omega_{\rm m}-1)^2$,
corresponding to idealised spherical collapse,
and $\Omega_{\rm m}=0.27$
such that
\begin{equation}
	{{3 m_{\rm v}}\over{4\pi R_{\rm v}^3}}
		\approx 97.01\rho_{\rm c}
		\approx 0.02259\ U_\rho
	\ .
\label{eq.virial}
\end{equation}
The virial radius conveniently measures
   idealised, radially infinite models,
   or numerical simulacra which are unresolved in their fringes.
The drawback of $R_{\rm v}$ is that it loses
   information about the outskirts.
It also fails to characterise compact objects
   where $R_{\rm v}$ encloses the entire mass
   (e.g.\  the most compact $(m_*,\Upsilon)$-optimal models with $R<1~\Mpc$).
Since $R_{\rm v}$ is defined relative to an absolute density,
   it does {\em not} transform neatly
   under mass and temperature rescaling
   (Appendix~\ref{s.scaling}).
On the other hand,
   it is always possible to rescale the cluster masses
   so that $R_{\rm v}=R$ or some other signature radius
   (\S\ref{s.virial}).
The same is true for King models 
   or any model with at least one free scale.

X-ray imagery and gravitational lensing studies
constrain the column densities of gas and dark matter,
as projected onto the plane of the sky.
Thus it is useful to calculate comparative
   two-dimensional projected properties.
An effective half-light radius, $R_{\rm e}$,
   is conventionally defined by a line-of-sight cylinder
   that encircles half the emission (or projected mass).
If the total mass and halo radius were known,
   then we can also define a radius $R_\Sigma$
   for the image contour with mean brightness (or column density),
   $\bar\Sigma = m/\pi R^2$.
% define R_h where the brightness is half the central value
% define R_m containing half the 3D mass

\subsection{gasless polytropic halo}
\label{s.gasless}

Here, for the sake of clarifying our main results,
   we review the intrinsic behaviour of 
   finite polytropic dark halos {\em without} gas.
We will examine the influence of a central point-mass.
The upper blocks of Table~\ref{table.halos}
   characterise a set of gasless polytropic halos,
   of different degrees of freedom,
   with and without a central mass.

\begin{figure}
\begin{center} 
\includegraphics[width=8cm]{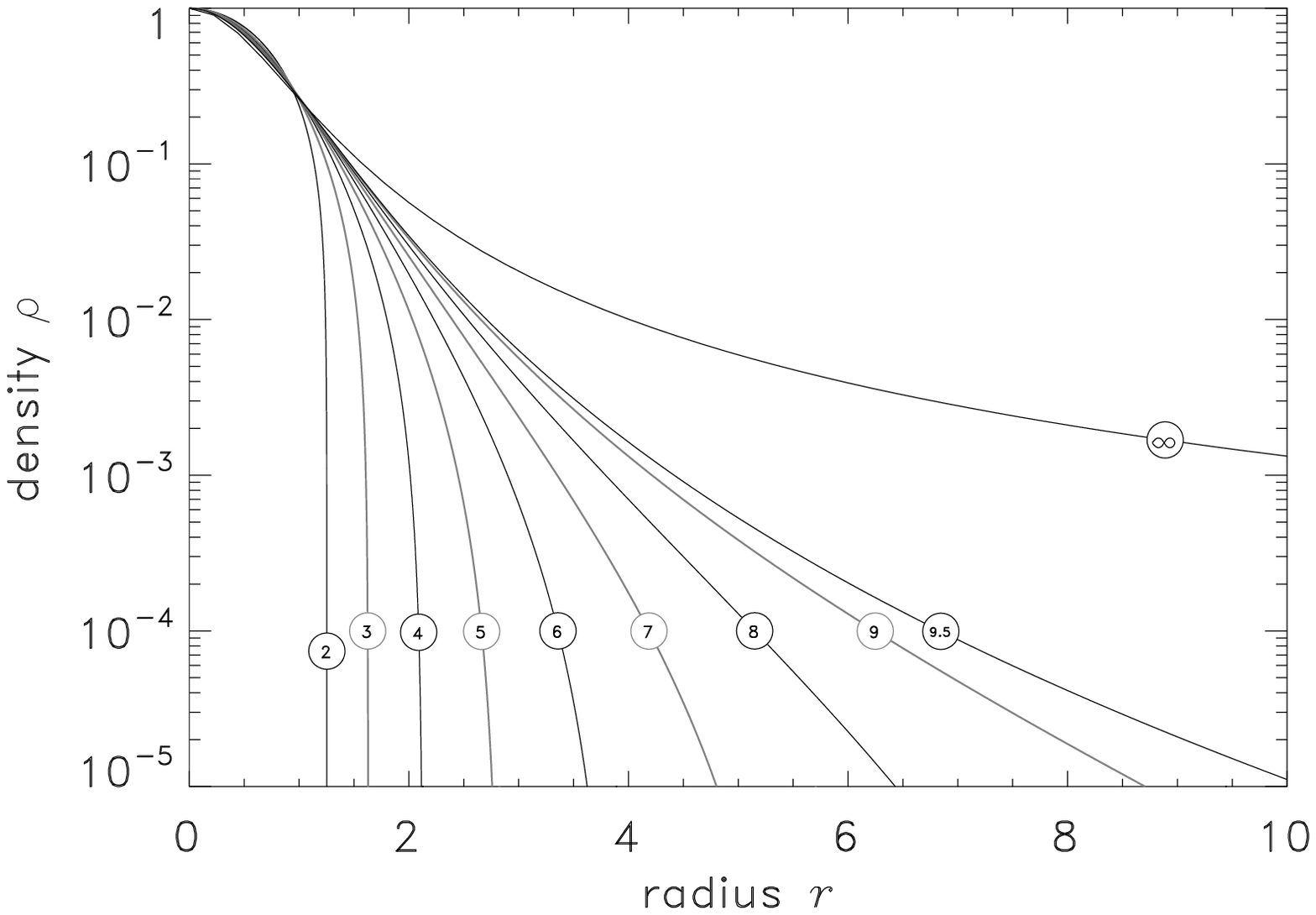}  
\end{center} 
\caption{ 
	Density profiles of polytropic dark halos
	with different adiabatic indices,
	$\gamma=1+2/F$,
	but no gas and no central mass.
	We set $s=1$ and $\rho=1$ at the origin.
	Labels denote the effective degrees of freedom $F$ in each case.
	For large $F$
	the flat core fills less of the total volume.
	The case $F=\infty$ is the non-singular isothermal sphere
	(``NIS'', \S\ref{s.isothermal}).
}
\label{fig.halo_gamma}
\end{figure}

The first subset lack a central mass ($m_*=0$),
   and share the same entropy and central density:
   $s=1$, $\rho(0)=1$.
   %(and thus $R_{\rm K}$ is fixed too).
Figure~\ref{fig.halo_gamma} shows their density profiles.
These solutions are classical Lane-Emden spheres of index $n=F/2$
   \citep{lane1870,emden1907,chandrasekhar1939,horedt1986a}.
We prefer $F$ as the more physically motivated notation.
Each sphere has a core of nearly uniform density,
   surrounded by declining outskirts.
For smaller $F$ the core is a larger fraction of the volume,
   and the fringe is steeper.
If $F<10$ then the halo possesses a zero-density outer surface at radius $R$.
These finite polytropes don't tend to any asymptotic outer slope;
   the density index steepens infinitely as $r\rightarrow R$.

\begin{figure}
\begin{center} 
\includegraphics[width=8cm]{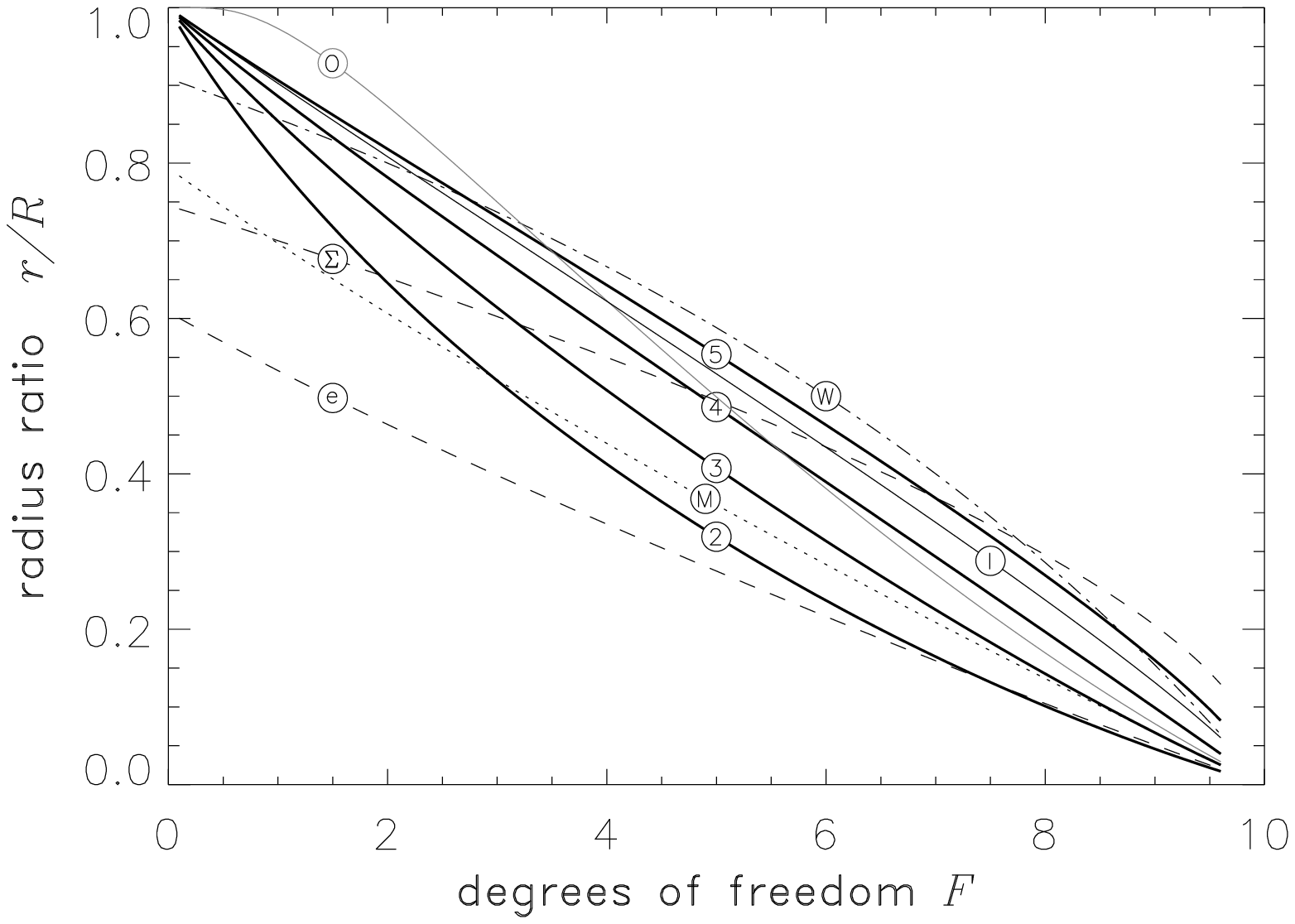}
\end{center} 
\caption{ 
        Signature radii (as fractions of the surface radius $R$)
	of non-singular ($m_*=0$) finite polytropic halo models.
Bold curves mark the density slope radii,
$R_2/R$, $R_3/R$, $R_4/R$ and $R_5/R$.
``I'' marks the lever radius, $R_I/R$.
``M'' marks the half-mass radius, $R_m/R$.
``W'' marks the gravitational radius, $R_w/R$.
``O'' traces peaks of the rotation velocity, $R_{\rm o}/R$.
Two-dimensional projected quantities 
are marked ``$\Sigma$'' ($R_\Sigma/R$ the mean-brightness radius)
and ``e'' ($R_{\rm e}/R$ the half-light radius).
        These radii stay bundled together
	but shrink in relation to the surface as $F$ rises.
}
\label{fig.poly_radii}
\end{figure}

With central conditions held constant,
   the radius and mass increase with $F$,
   and the binding energy increases both in absolute terms and per unit mass.
In absolute terms, $R_I$ increases slightly with $F$.
Proportionally, the core ($R_I/R$),
   the gravitational radius ($R_w/R$),
   the two-dimensional radii ($R_\Sigma/R, R_{\rm e}/R$)
   and the rotation peak ($R_{\rm o}/R$)
   all shrink with $F$.
The slope-radii ($R_2, R_3, R_4$)
also shrink with rising $F$.
They become multi-valued for large $F$
(the index of $\rho$ wobbles in some layers).
In such cases, we tabulate the steepening point
   farthest on the edge of the core.
This tends to be near the peak of the rotation curve.
We generally find that $R_{\rm o}>R_3$ and $R_I>R_4$.
Consecutive slope indices
   ($R_2, R_3, R_4 \ldots$)
   occur in roughly even steps,
   but the steps ultimately diminish near the true surface ($R$).
In non-singular halos, $R_w\approx R_I$.
As $F$ increases, all the signature radii shrink relative to $R$
   in a common manner (see Figure~\ref{fig.poly_radii}).

\begin{figure}
\begin{center} 
\includegraphics[width=8cm]{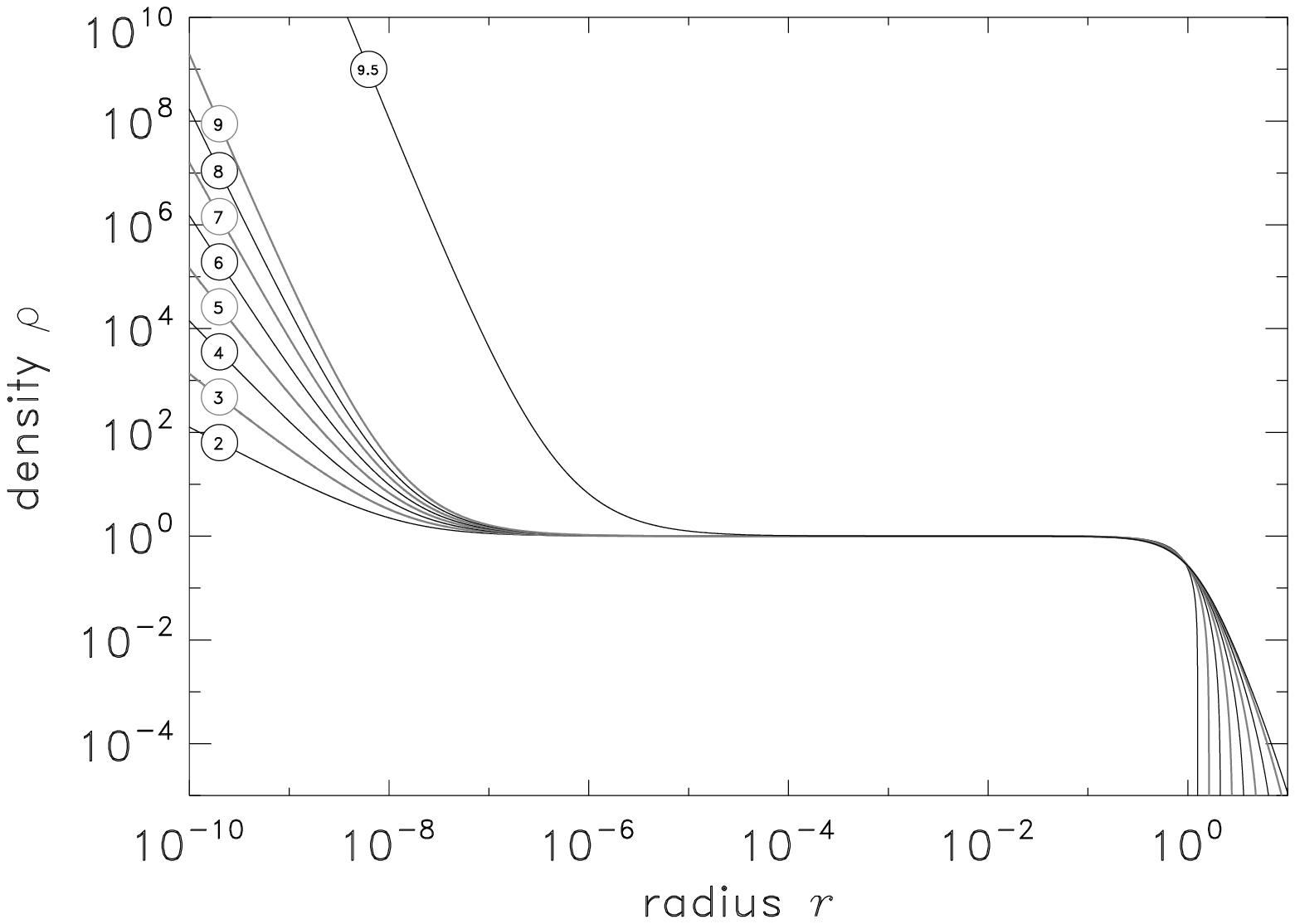}
\end{center} 
\caption{ 
	Density profiles of gasless polytropic halo models
	in the presence of a central gravitating mass
	($m_*=10^{-8} m$).
	Curves are annotated with their respective $F$ values.
}
\label{fig.halo_gamma_hole}
\end{figure}

However the non-singular model is a specially contrived condition.
Most galaxies are thought to contain a point-like central mass
   such as a nuclear star cluster or black hole.
At cluster scales, the analogous object
   is a cD galaxy or its black hole.
To clarify the effect of such a mass,
   we tabulate singular polytropic halos
   with central masses in astronomically realistic proportions:
   $m_*/m=10^{-8}, 10^{-6}, 10^{-5}$.
We fixed the total mass and $R$ of corresponding Lane-Emden spheres,
   but vary the entropy.
Figure~\ref{fig.halo_gamma_hole}
   shows density profiles for $m_*/m=10^{-8}$.
The central object draws a density spike about itself, of index $-F/2$.
Beyond this sharp sphere of influence,
  the halo flattens into a core, then steepens into outskirts
  and a surface like those of nonsingular models.
For small or medium $F$,
   the addition of $m*>0$ reduces
   the slope-radii ($R_2, R_3, R_4$) slightly.
However for sufficiently large $F$ and $m_*$,
   the spike steepens the entire core,
   to the point where $R_2$ and higher slope-radii vanish.

For $F>9$, the spike develops density undulations
  (at sub-parsec scales for a cluster).
In this tiny, deep core, the slope radii are multi-valued
  and the rotation curve peaks multiply.
Some undulations locally approach the brink of Jeans stability.
The dense spike 
   reduces $R_I$ and $R_w$ dramatically for $F\ge8$.

Table~\ref{table.halos} reveals several trends for as $m_*/m$ varies.
As $m_*$ increases, $R_w/R$ changes appreciably before
the slope-radii and rotation peak do.
The two-dimensional projected quantities
($R_\Sigma/R, R_{\rm e}/R$ are the least sensitive to the central mass.
Large $F$ enhances the sensitivity of all the of signature radii
   with respect to $m_*$.

If $F\ge10$ then the halo density attenuates indefinitely
   ($R=\infty$),
   regardless of $m_*$.
The borderline case of $F=10$ has an infinite radius but finite mass:
this is the well-known \cite{plummer1911} model,
where $\rho\propto r^{-5}$ at large radii.
In the isothermal limit, $F\rightarrow\infty$,
   the fringe declines like $\rho\propto r^{-2}$
   (Appendix~\ref{s.isothermal}).

\subsection{Isothermal spheres}
\label{s.isothermal}

If the particle velocities are isotropic
and $\sigma^2$ is constant everywhere,
then the halo is an ``isothermal sphere.''
This is essentially an extreme polytrope
in the limit $F\rightarrow\infty$.
Isothermality is plausible 
when some strong mechanism asserts global thermal equilibrium:
e.g. thorough and violent relaxation
\citep{lyndenbell1967},
or efficient thermal conduction.
Isothermal spheres are popular toy models
in gravitational lensing studies.
Such halos also predict flat rotation curves,
resembling the observed outer parts of disc galaxies.

The density profile depends on the central boundary condition.
The ``non-singular isothermal sphere'' (NIS)
   has $d\rho/dr=0$ at the origin,
   exhibiting a shallow density core,
   but at large radii it tends to a decline $\propto r^{-2}$
   \citep[see e.g.][and the $F=\infty$ curve in Figure~\ref{fig.halo_gamma}]{binney1987}.
The ``singular isothermal sphere'' (SIS) has a self-similar profile,
\begin{equation}
	\rho = {{\sigma^2}/{2\pi G r^2}}
	\ .
\end{equation}
For NIS and SIS, $R_3$ and higher slope radii never occur.
The density fails to vanish at any finite radius,
   so the halo lacks a distinct surface.
Within some ad~hoc truncation radius, the mass, moment of inertia
   and gravitational potential energy of the SIS are
\begin{eqnarray}
	m&\hspace{-0.25cm}=&\hspace{-0.25cm}{{2\sigma^2 r}/{G}}
	\ ,
\nonumber \\
	I&\hspace{-0.25cm}=&\hspace{-0.25cm}{{4\sigma^2 r^3}/{9G}}
	\ ,\mbox{~~~~~~~and}
\nonumber \\
	W&\hspace{-0.25cm}=&\hspace{-0.25cm}-{{4\sigma^4 r}/{G}}
	\ .
\end{eqnarray}
It follows that the gravitational radius $R_w=r$
   and the effective lever radius $R_I=\sqrt{5}r/3$.
For the SIS, the circular velocity is radially constant.
A SIS gravitational lens bends light rays
   by a constant angle at all projected radii.
The SIS virial radius is
\begin{equation}
R_{\rm v}=\left[{
	{3\sigma^2}\over{
			2\pi G \delta_{\rm c}\rho_{\rm c}
		}
	}\right]^{1/2}
	\approx 4.598 \sigma
\end{equation}
which is $\approx1.130$~Mpc for a halo at 1~keV temperature.

\subsection{pseudo-isothermal sphere}
\label{halo.pis}

In some observational studies
the exact non-singular isothermal sphere
is approximated by an empirical cored profile
that also has $\propto r^{-2}$ outskirts.
The ``pseudo-isothermal sphere'' (PIS) has a density profile
\begin{equation}
	\rho = {{\rho_{\rm s}}\over{1+x^2}}
	\ ,
\end{equation}
where $x=r/r_{\rm s}$
and $(\rho_{\rm s},r_{\rm s})$ are some density and radial scales.
The PIS has infinite radius, mass and moment of inertia.
At large radii, $R_I/r=\sqrt{5}/3$
and $R_w/r=1$.
The slope-2 radius is at infinity,
as is the peak circular velocity
($R_2=R_{\rm o}=\infty$).
Figure~\ref{fig.rvirial} shows
   numerical solutions for the virial radius depending on
   $(\rho_{\rm s},r_{\rm s})$,
   comparing PIS to some other radially infinite models.

\subsection{Hubble profile and $\beta$-model atmospheres}
\label{halo.hubble}
\label{halo.beta}

The modifield Hubble profile
   \citep{hubble1930,rood1972,king1972}
   approximates the projected mass or brightness
   of any non-singular cored sphere as
\begin{equation}
	\Sigma\approx{{\Sigma_0}\over{1+x^2}}
	\ ,
\end{equation}
   defined such that the intensity drops to half its central value ($\Sigma_0$)
   at the projected radius scale $r_{\rm s}$.
This empirical fit has been applied to the stellar light profiles of
   elliptical galaxies, globular clusters and galaxy clusters.
The corresponding spatial density profile is
\begin{equation}
	\rho \approx {{\rho_{\rm s}}\over{(1+x^2)^{3/2}}}
	\ .
\end{equation}
This distribution yields more realistic rotation curves
   than the isothermal models,
   as there is a peak at finite radius,
   $R_{\rm o}\approx 2.920 r_{\rm s}$
   \citep[e.g.][pp.39-41]{binney1987}.
However the model is only applicable in and near the core,
   otherwise it implies infinite radius, mass, potential energy
   and moment of inertia.
As $r\rightarrow\infty$ we have
   $R_I\rightarrow\infty$,
   $R_w\rightarrow\infty$,
   $R_I/r\rightarrow 0$ and
   $R_w/r\rightarrow 0$.
The density slope-radii are $R_2=\sqrt{2}r_{\rm s}$ and $R_3=\infty$.

The model was extended to describe the X-ray emitting gas 
   by assuming isothermality and local hydrostasis in cluster cores,
   e.g. \cite{lea1973}
   who assigned identical temperatures to gas and galaxies.
\cite{cavaliere1976}
   assumed a galaxy/gas temperature ratio of
   $\beta = \sigma_\star^2 / \sigma_{\rm g}^2$,
   deriving a {\em gas} density profile
\begin{equation}
	\rho = \rho_0\ [1+(r/r_{\rm c})^2]^{-3\beta/2}
	\ ,
\label{eq.beta.rho}
\end{equation}
   and X-ray surface brightness profile
\begin{equation}
	S = S_0 \left[{ 1 + (b/b_{\rm c})^2 }\right]^{-3\beta+{\frac12}}
\label{eq.beta.image}
\end{equation}
   where $b$ is the projected radius.
This became a commonplace fitting forumula for X-ray imaging observations,
   \citep[e.g.][]{bahcall1977,gorenstein1978,gbr1981,jones1984,neumann1999}.
Shorn of the original isothermal assumption,
   more recent studies adopt
   either (\ref{eq.beta.rho}) or (\ref{eq.beta.image})
   as a conventional template,
   and seek to infer actual radial variations of temperature from the data.
It is worth emphasising that the $\beta$-model is a parameterisation
   or an idealisation of the innermost observable regions.
The ultimate outer density index is $-3\beta$,
   and the implied asymptotic gas mass is infinite
   unless $\beta>1$
   (according to (\ref{eq.beta.rho})).
However (\ref{eq.beta.image}) implies that the luminosity is infinite if
   $1<\beta<2$.
This breakdown implies that the density and/or temperature
   must attenudate even more steeply in the outskirts of real clusters.
Indeed, some observations show that $\beta$ steepens 
   \citep[e.g][]{vikhlinin1999,neumann2005}.

\subsection{Hernquist profile}
\label{halo.hernquist}

\begin{figure}
\begin{center} 
\includegraphics[width=8cm]{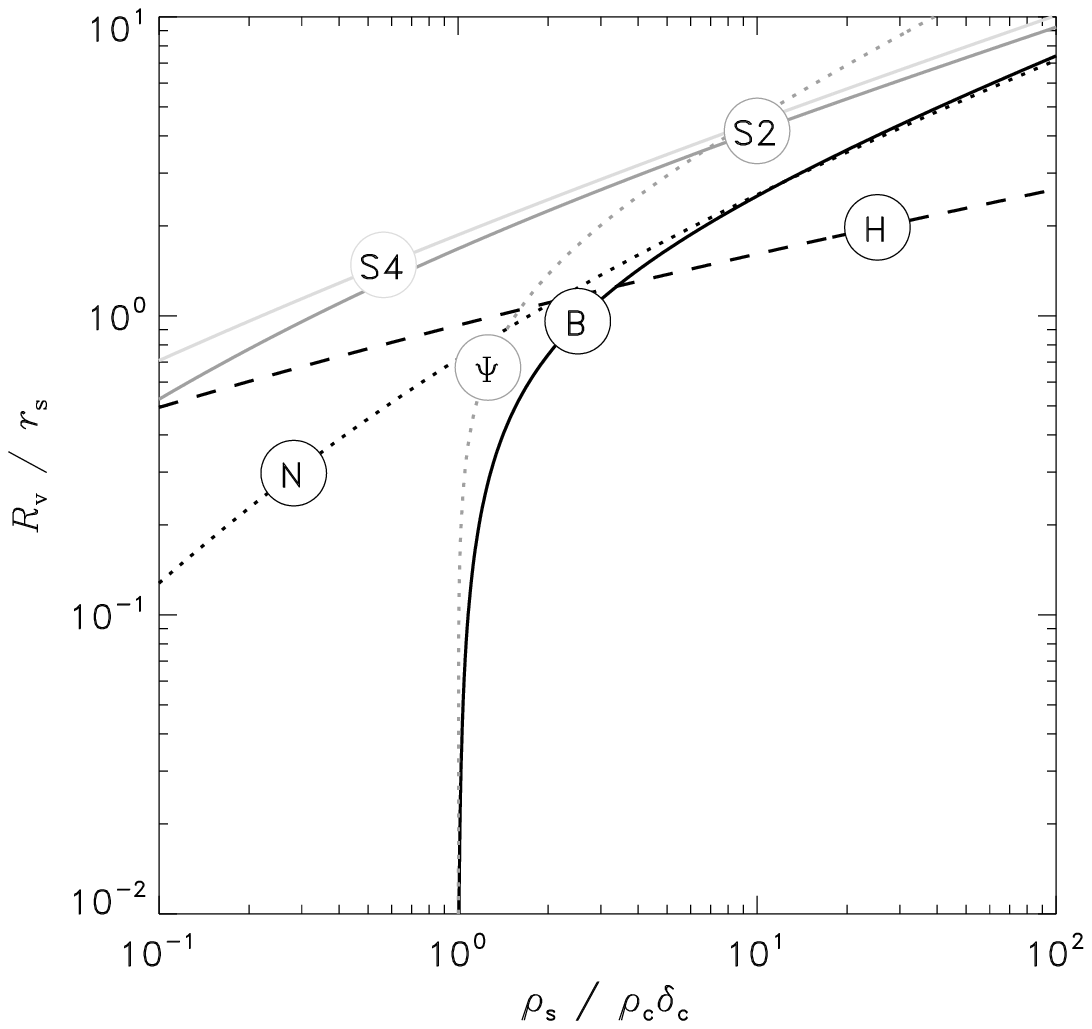}
\end{center} 
\caption{ 
The dependency of the virial radius upon
the halo's density at its scale radius,
for:
pseudo-isothermal ($\Psi$);
NFW (N);
Burkert (B);
Hernquist (H);
\sersic $n=2,4$ (S2,S4) models.
}
\label{fig.rvirial}
\end{figure}

\cite{hernquist1990}
proposed an analytic model for galaxy spheroids,
with density profile that attenuates infinitely
\begin{equation}
	\rho = {{m_{_\infty}}\over{2\pi x(1+x)^3} r_{\rm s}^3}
\end{equation}
but nonetheless yielding a finite mass profile
\begin{equation}
	m(r) = m_{_\infty} {{x^2}\over{(1+x)^2}}
	\ .
\end{equation}
The moment of inertia is infinite.
The density slope passes integer values at
$x_2={\frac12}$, $x_3=2$ and $x_4=\infty$.
The rotation curve peaks at the scale radius, $x_{\rm o}=1$.
The projected central brightness is infinite,
   but the outskirts decline fast enough that $R_{\rm e}$ is finite.
Figure~\ref{fig.rvirial} shows
   the relation between virial radius and $(\rho_{\rm s},r_{\rm s})$.

\subsection{N-body simulacra}
\label{halo.nfw}

\cite{dubinski1991}
found that collisionless structures emerging in N-body cosmological simulations
develop a cuspy, power-law central density profile.
The ``NFW profile''
\citep{nfw1996,nfw1997}
is a popular empirical fit to such halos,
with a density that follows
\begin{equation}
	\rho = {{\rho_{\rm s}}\over{x(1+x)^2}}
\label{eq.nfw.density}
\end{equation}
where $x=r/r_{\rm s}$ and $(\rho_{\rm s},r_{\rm s})$
are fitting parameters of a particular halo.
These parameters follow trends in relation to the halo mass,
which depend on cosmology
\citep[e.g.][]{bullock2001,dolag2004,shaw2006,maccio2007}.
Physically, the scale radius $r_{\rm s}=R_2$,
the slope-2 radius.
This radius defines a concentration,
$c\equiv R_{\rm v}/r_{\rm s}$.
A halo truncated at some finite radius has
a mass and moment of inertia of:
\begin{equation}
	m = 4\pi\rho_{\rm s} r_{\rm s}^3
		\left[{
			\ln(1+x) - {x\over{1+x}}
		}\right]
	\ ,
\label{eq.nfw.mass}
\end{equation}
\begin{equation}
	I = {{8\pi}\over{3}}\rho_{\rm s}r_{\rm s}^5
		\left[{
			{3\over2} + {1\over{1+x}} +{{(x+1)(x-5)}\over{2}}
			+3\ln(1+x)
		}\right]
	\ .
\end{equation}
There isn't an outer surface,
and both $m$ and $I$ are infinite as
$x\rightarrow\infty$.
The inertial radius $R_I\rightarrow\infty$ at infinity,
while the its concentration ratio vanishes ($R_I/r\rightarrow 0$),
which means that the rotational properties depend
on an ad~hoc truncation radius.
The gravitational potential energy is finite,
\begin{equation}
	W = -8\pi^2 G \rho_{\rm s}^2 r_{\rm s}^5
	\ .
\end{equation}
Thus $R_w\rightarrow\infty$ and $R_w/r\rightarrow 0$
at large radii.
The rotation curve peaks at $R_{\rm o}\approx 2.163~r_{\rm s}$.
Using (\ref{eq.nfw.density}) and (\ref{eq.nfw.mass}),
the virial radius equation
(\ref{eq.virial})
is transcendental.
(See Figure~\ref{fig.rvirial} for numerical solutions.)

\cite{moore1999} and \cite{zhao1996}
   proposed variations and generalisations to the NFW formula,
   consisting of a broken radial power-law again,
   but with different indices.
Expressions for the global quantities differ slightly from those above,
  but the models are qualitatively similar:
  infinite in mass and radius, 
  and ill-defined rotational properties.
More recent work
  \citep{merritt2005,graham2006}
  suggests that \sersic profiles (Appendix~\ref{s.sersic})
  fit simulated collisionless halos better.

Density slopes of the simulacra are least certain on the outskirts
  and near the origin,
  both places where mass resolution degrades.
Cuspy profiles emerge consistently from cosmological N-body simulations,
  but the causes of this shape
  still lack a comprehensive analytic derivation.
Possible causes may involve:
  cosmic expansion, ongoing infall and accretion history;
  the simplifying assumption of a collisionless medium;
  the approximation of discretised mass;
  implicit low-pass filtration in numerical Poisson solvers;
  or perhaps other less obvious computational and physical factors.

\subsection{Burkert profile}
\label{halo.burkert}

\cite{burkert1995}
proposed an empirical halo model
based on observed rotation curves of halo-dominated galaxies,
\begin{equation}
	\rho = {{\rho_{\rm s}}\over{
		(1+x)(1+x^2)
	}}
	\ ,
\end{equation}
where we denote a normalised radius again, $x=r/r_{\rm s}$.
The density index is $-2$ at $R_2\approx 1.521~r_{\rm s}$.
%1.5213797
The index ultimately approaches $-3$;
$R_3=\infty$.
There is a flat density core,
like in the nonsingular polytropic and King models
(and unlike the cuspy NFW and \sersic profiles).
The mass and inertial moment enclosed at $x$ are
\begin{equation}
	m = \pi\rho_{\rm s}r_{\rm s}^3 \left[{
		2\ln(1+x) +\ln\left({1+x^2}\right) - 2\arctan x
	}\right]
	\ ,
\end{equation}
\begin{equation}
	I = {{2\pi}\over{3}} \rho_{\rm s} r_{\rm s}^5
		\left\{{
			2x^2 -4x +2\arctan x 
			+\ln\left[{{(1+x)^2}\over{1+x^2}}\right]
		}\right\}
	\ ,
\end{equation}
and both are infinite as $r\rightarrow\infty$.
As with NFW,
the Burkert halo mass is intensely centrally concentrated:
$R_I/r\rightarrow 0$ as $r\rightarrow\infty$.
The gravitational potential energy is finite,
\begin{equation}
	W = -4 \pi^3 \ln~2 G\ \rho_{\rm s}^2 r_{\rm s}^5
	\ ,
\end{equation}
and $R_w/r\rightarrow 0$ at large radii.
The rotation curve peaks at $R_{\rm o}\approx 3.245 r_{\rm s}$.
%3.2446257
Like the NFW halo,
the virial radius equation is transcendental
(Figure~\ref{fig.rvirial} shows numerical solutions).

\subsection{King model}
\label{halo.king}

\cite{king1966}
   presented a cored stellar-dynamical model,
   derived from first principles.
Its basis is a phase-space density function,
\begin{equation}
	{\mathcal F}(\rvec,\vvec) = A \left[{
		e^{-a(\Phi+v^2/2)} - e^{-a \Phi_{\rm t}}
	}\right]
	\ ,
\end{equation}
   assuming locally isotropic particle velocities,
   truncating at some escape energy
   corresponding to the equipotential
   ($\Phi_{\rm t}$)
   of a zero-density outer surface at ``tidal radius'', $R=r_{\rm t}$.
The model is a self-consistent description
   of a non-isolated, self-bound, collisionless sphere.
It was originally applied to globular clusters with escaping stars.
\cite{firmani2001}
   reapplied it to cored, thermal, self-interacting dark halos.
The mass, moment of inertia and gravitational potential energy are all finite.
The local density,
\begin{equation}
	\rho = {{8\pi}\over{3}}\sqrt{{2}\over{a^3}} A e^{-a\Phi_{\rm t}}
		\ e^\psi\ \Gamma\left({ {\frac52},\psi }\right)
	\ ,
\end{equation}
depends on the dimensionless potential offset,
\begin{equation}
	\psi \equiv a \left[{
		\Phi_{\rm t} - \Phi(r)
	}\right]
	\ ,
\end{equation}
and $\Gamma$ is the lower incomplete gamma function.
At the outer boundary, $\rho=\psi=0$.
Radial coordinates for the equipotentials
are obtained by solving the Poisson equation,
subject to the inner boundary conditions
$\psi>0$ and $\nabla\psi=0$.
The latter condition precludes a central mass, $m_*=0$.

\begin{figure}
\begin{center} 
\includegraphics[width=8cm]{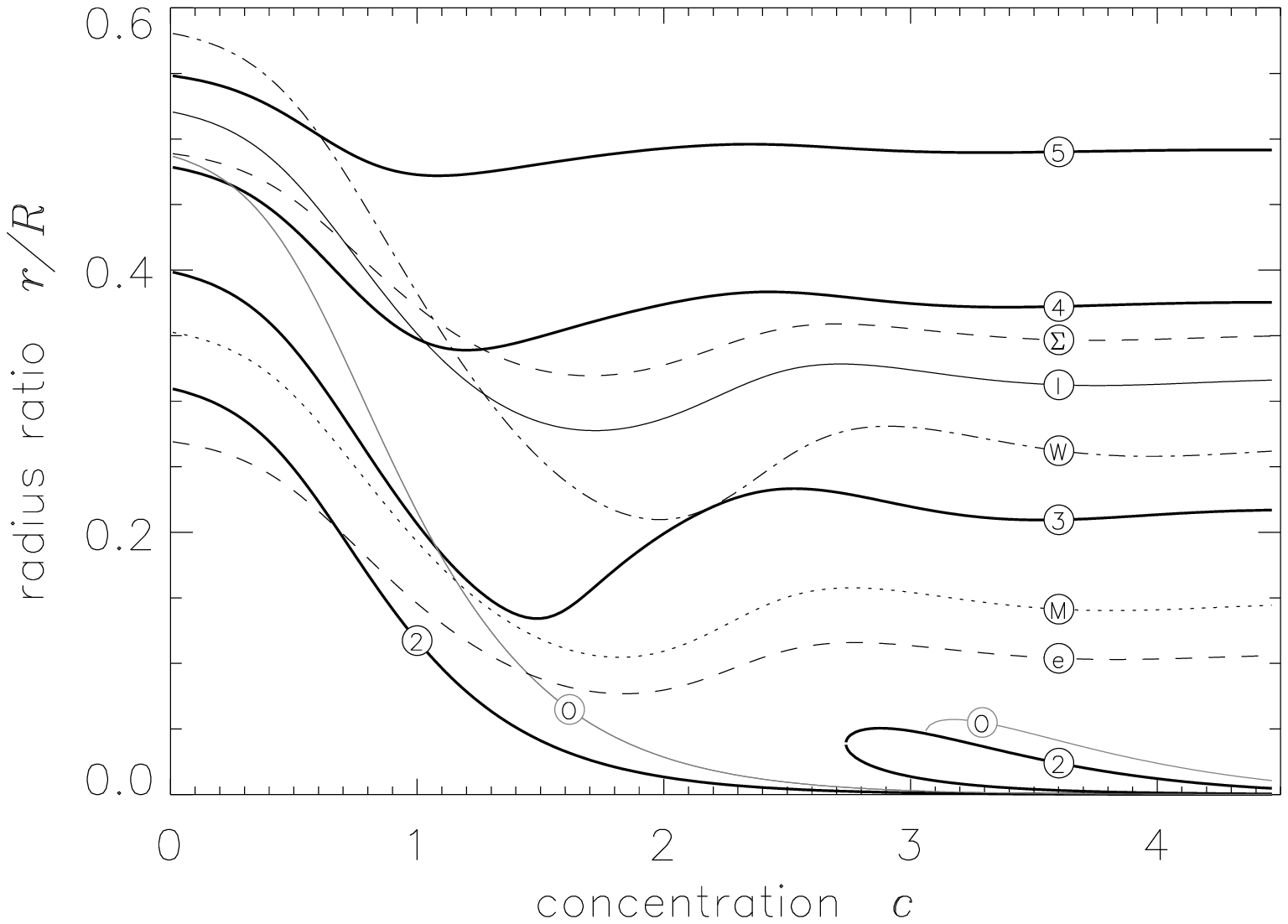}
\end{center} 
\caption{ 
Signature radii relative to the tidal surface,
for King models with various concentrations,
$c=\log_{10}(R/R_{\rm K})$.
Lines and annotations are the same as in Figure~\ref{fig.poly_radii}.
The signature radii span much of the halo volume
(unlike high-$F$ polytropic halos).
Their ratios stay roughly steady as $c$ increases.
}
\label{fig.king}
\end{figure}

Like the $F<10$ polytropes,
   the King model is radially finite.
Thus the slope radii $(R_2, R_3, R_4)$ are also finite,
   and the density index attains large negative values
   near the edge.
Concentrations are conventionally denoted by
   $c\equiv\log_{10}(R/R_{\rm K})$.
Figure~\ref{fig.king}
   shows the variation of signature radii with $c$.
Both $R_2/R$ and $R_{\rm o}/R$ drop rapidly with increasing $c$,
   and they are multi-valued for highly concentrated models,
   $c\ga2.7$.
The other main signature radii vary only within factors of a few
   in the shown domain ($0\le c\le 4.5$).

\subsection{\sersic profile}
\label{s.sersic}
\label{halo.sersic}

The \cite{sersic1968} model is an empirical fit
to the two-dimensional projected starlight
of spheroids such as elliptical galaxies and spiral bulges.
\begin{equation}
	\Sigma = \Sigma_{\rm e}\ \exp\left[{
		-b\left({ x^{1/n}-1 }\right)
	}\right]
\end{equation}
The radial coordinate is scaled in terms of the half-light radius,
$x=r/R_{\rm e}$.
The shape parameter
$n\approx4$ for elliptical galaxies
\citep[the classic profile of][]{devaucouleurs1948}
or $n\sim2$ for galaxy clusters.
The parameter $b$ depends on $n$ implicitly,
via lower-incomplete and complete gamma functions,
\begin{equation}
	2 \Gamma(2n,b)=\Gamma(2n)
\end{equation}
\cite{cotti1999}
derived a series expansion,
$b \approx2n -1/3 +4/405n +46/25515n^2 +131/1148175n^3
                -2194697/30690717750n^4$.
% + O(n^{-5})$
\sersic profiles appear ubiquitous among stellar spheroids in nature,
but the principal causes
have not yet been shown analytically.

The cuspy density profile of \cite{prugniel1997},
\begin{equation}
	\rho = \rho_{\rm s}\ x^{-p} \exp\left[{-b\left({
			\ x^{1/n}-1}\right) }\right]
	\ ,
\end{equation}
where $x=r/r_{\rm s}$ and $r_{\rm s}\approx R_{\rm e}$,
fits the \sersic light profile approximately.
The index of the inner cusp, $p\approx 1.0 -0.6097/n+0.05463/n^2$
\citep{limaneto1999,marquez2000}.
Expressed in terms of incomplete gamma functions,
the mass and moment of inertia
within some radius are
\begin{equation}
	m = 4\pi n b^{n(p-3)} e^b
		\rho_{\rm s} r_{\rm s}^3
		\ \Gamma[n(3-p),b x^{1/n}]
	\mbox{~~~~~and}
\end{equation}
\begin{equation}
	I = {{8\pi n}\over{3}} 
		\,b^{n(p-5)} e^b
		\rho_{\rm s} r_{\rm s}^5
		\ \Gamma[n(5-p),b x^{1/n}]
	\ .
\end{equation}
Evaluated as $x\rightarrow\infty$,
the core lever radius is finite,
\begin{equation}
	R_I = \left\{{
		{{5}\over{3b^{2n}}}
		{{\Gamma[n(5-p)]}\over{\Gamma[n(3-p)]}}
	}\right\}^{1/2}
	r_{\rm s}
	\ .
\end{equation}
For $n=2$ we have $R_I\approx3.421~r_{\rm s}$,
and for $n=4$ we have $R_I\approx7.418~r_{\rm s}$.
The gravitational radius is finite but needs some numerical integration;
$R_w/r_{\rm s}$ decreases with increasing $n$.
The $\sigma^2$ profile emerges from integration of
   a hydrostatic or Jeans equation,
   and it can be shown that the King radius vanishes,
   $R_K\rightarrow 0$ as $r\rightarrow 0$.
The dependence of the virial radius on $(\rho_{\rm s},r_{\rm s})$
is shown by grey curves in
Figure~\ref{fig.rvirial}.

\begin{figure}
\begin{center} 
\includegraphics[width=8cm]{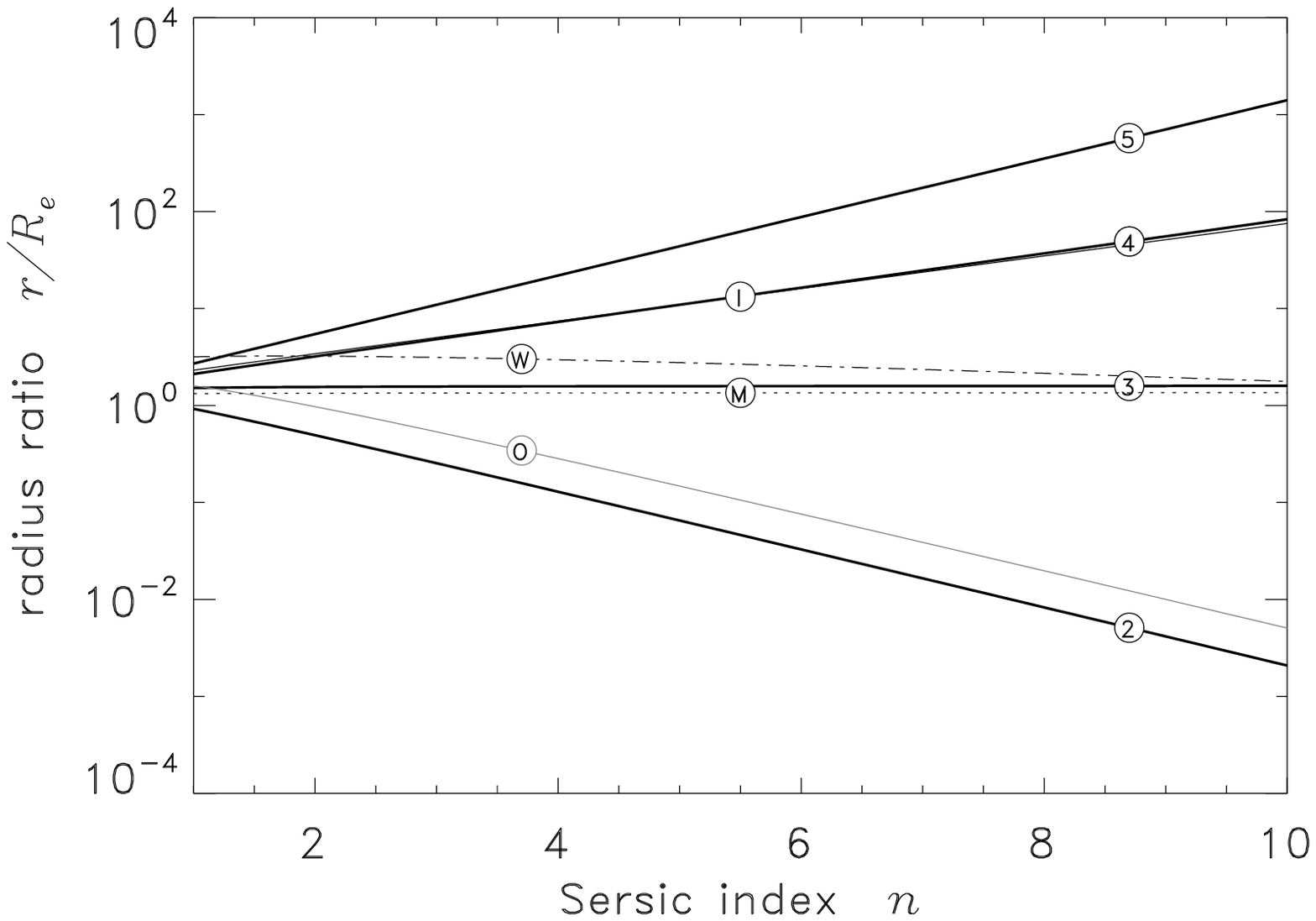}
\end{center} 
\caption{ 
Signature radii (relative to $R_{\rm e}$)
for \sersic models with indices $n$.
The curves are marked as in Figure~\ref{fig.poly_radii}.
In these halo models,
the signature radii splay out enormously as $n$ rises.
}
\label{fig.sersic}
\end{figure}

The density index drops with radius.
For all realistic $n$, $R_2$ exists uniquely,
along with all higher slope-radii.
In general, for index $-j$, we have
$R_j/R_{\rm e}=[n(j-p)/b]^n$.
For $n>1$ the consecutive integer-slope radii spread apart
   (vertical distribution of bold lines in Figure~\ref{fig.sersic}).
In contrast, finite polytropic halos have slope-radii
   spaced at shrinking intervals, converging at the true surface $R$.
For $n\ga 2$,
   the rotation curve peaks inside the effective radius,
   $R_{\rm o}<R_{\rm e}$.
For $n\ga 1.2$ we have $R_{\rm o}<R_3$.
The lever radius and slope-4 radius are of similar magnitude,
   $R_I\approx R_4$, with $R_I>R_4$ for $n\la 5$.
For $n\ge 1$ we have $R_I\ge R_3$.
These signature inequalities are potentially testable
   by gravitational lensing and kinematic studies 
   in halo outskirts.

\subsection{comparisons}
\label{s.comparisons}

Theoretical, numerical and observational halo models
   are in principle testable by measuring enough of their signature radii,
   derived from three-dimensional and projected quantities.
Table~\ref{table.halos}
   characterises gasless polytropic halo models with various $F$ values,
   and compares them to other models from the literature.

The density slope radii are the major discriminants between halo models.
Isothermal and pseudo-isothermal models have no $R_3$ radius,
   while the Hubble, NFW and Burkert models have $R_3=\infty$.
Galaxy densities in clusters appear to drop
   at least as steeply as $r^{-3}$ in the outskirts
   \citep{carlberg1997,adami2001}.
If this trend persists infinitely
   then it would fit NFW or Burkert interpretations.
However, the detection of steeper slopes
   (by any technique)
   would call for more sophisticated models.
All of the polytropic, King and \sersic models
   have finite $R_2$ and $R_3$,
   set in ratios depending on $F$, $c$ and $n$ respectively.
In \sersic models, the consecutive slope radii spread widely apart,
   but in the polytropes and King models these radii converge.
Low-concentration King models have somewhat evenly spread values of
   $\{R_2, R_3, R_4, \ldots\}$
   but for high concentrations $R_2\ll R_3 \sim R_{\rm e}$,
   and so the halo could be mistaken for an isothermal or NFW shape
   if it were probed at intermediate radii only.
For finite polytropic halos,
   $R_2/R, R_3/R, R_4/R, R_w/R$ and $R_{\rm o}/R$
   remain similar to each other in order of magnitude
   (even as $F\rightarrow 10$),
   while the projection scales $R_\Sigma/R$ and $R_{\rm e}/R$
   shrink slower with increasing $F$.

As their $I$ and $R_I/R$ values show,
   the concentrations and rotational properties
   of Hubble, NFW, Burkert and Hernquist models are hard to define.
In one sense the mass is centrally concentrated.
On the other hand, the outskirts dominate $I$.
Therefore the ability to spin such a halo up or down
   (e.g. in tidal interactions between unbound neighbours)
   depends on an ad hoc truncation.
The unphysical inability of simulacra to self-truncate
   may be part of the ``angular momentum problem''
   of simulated galaxy formation.
The polytropic, King and \sersic models are finite
   and consistent with respect to $I$.

Any real, isolated halo detaches from the Hubble flow
   possessing finite mass and energy.
Its radius may also become finite,
either due to intrinsic self-truncation (e.g. of a polytrope)
or extrinsic harrassment and evaporation (e.g. King models).
Infinite models cannot be a final or comprehensive description of any real halo.
They must be regarded as provisional approximations only.
The best physically motivated, consistent and plausible models
are the King, \sersic and general polytropic descriptions.
It is unclear how a \sersic model should adapt in a potential
   shared with other components,
   so it was unsuitable as the basis for our present study.
We chose the polytropic scenario,
   although a generalised King model has scope to represent
   a halo suffering surface evaporation or tidal truncation.

\begin{table*}
\caption{
Signature radii and masses of some representative gasless halo models.
From left to right, the columns are:
dark degrees of freedom,
surface radius;
inertial concentration;
gravitational concentration;
the density slope radii 
with indices $-2,, -3$ and $-4$;
the rotation-curve peak radius;
the projected mean-light radius and effective radius;
total mass.
Models marked ``$\star$'', ``$\bullet$'' and ``$\CIRCLE$''
   contain a central point mass,
$m_*=10^{-8}m, 10^{-6}m$ and $10^{-5}m$ respectively.
}
\begin{center}
$\begin{array}{lr@{.}lr@{.}lr@{.}lr@{.}lr@{.}lr@{.}lr@{.}lr@{.}lr@{.}lr@{.}lr@{.}lr@{.}lr@{.}lcccccccccccccc}
F
&\multicolumn{2}{c}{R}
&\multicolumn{2}{c}{R_I/R}
&\multicolumn{2}{c}{R_w/R}
&\multicolumn{2}{c}{R_{2}/R}
&\multicolumn{2}{c}{R_{3}/R}
&\multicolumn{2}{c}{R_{4}/R}
&\multicolumn{2}{c}{R_{\rm o}/R}
&\multicolumn{2}{c}{R_{\Sigma}/R}
&\multicolumn{2}{c}{R_{\rm e}/R}
&\multicolumn{2}{c}{m/\rho_{\rm s}r_{\rm s}^3}
\\
\hline
\\
2
&1&253
&0&8084
&0&8000
&0&6458
&0&7286
&0&7817
&0&8733
&0&6524
&0&4636
&2&507
\\
3
&1&630
&0&7152
&0&7368
&0&5200
&0&6141
&0&6813
&0&7500
&0&6025
&0&3980
&3&026
\\
4
&2&127
&0&6222
&0&6667
&0&4126
&0&5076
&0&5830
&0&6229
&0&5500
&0&3353
&3&534
\\
5
&2&826
&0&5287
&0&5882
&0&3191
&0&4076
&0&4860
&0&4989
&0&4943
&0&2748
&4&040
\\
6
&3&891
&0&4340
&0&5000
&0&2369
&0&3135
&0&3895
&0&3810
&0&4345
&0&2162
&4&555
\\
7
&5&706
&0&3374
&0&4000
&0&1645
&0&2251
&0&2932
&0&2708
&0&3693
&0&1594
&5&091
\\
8
&9&444
&0&2376
&0&2857
&0&1010
&0&1426
&0&1965
&0&1695
&0&2955
&0&1043
&5&668
\\
9
&21&06
&0&1313
&0&1540
&0&04587
&0&06678
&0&09880
&0&07829
&0&2050
&0&05094
&6&323
\\
9.5
&44&91
&0&07286
&0&08027
&0&02164
&0&03198
&0&04947
&0&03722
&0&1444
&0&02509
&6&707
\\
\\
9.5~\star
&44&91
&0&07286
&0&01019
&0&02164
&0&03198
&0&04947
&0&03722
&0&1444
&0&02506
&6&707
\\
\\
7~\bullet
&5&706
&0&3374
&0&3999
&0&1645
&0&2251
&0&2932
&0&2708
&0&3693
&0&1594
&5&091
\\
8~\bullet
&9&444
&0&2376
&0&04766
&0&1010
&0&1426
&0&1965
&0&1695
&0&2955
&0&1040
&5&668
\\
9~\bullet
&21&06
&0&003432
&1&429(^-5)
&1&702(^-4)
&3&674(^-4)
&9&730(^-4)
&2&451(^-4)
&0&02658
&2&754(^-4)
&6&323
\\
9.5~\bullet
&44&91
&1&766(^-4)
&5&217(^-6)
&1&089(^-5)
&1&871(^-5)
&3&656(^-5)
&1&827(^-5)
&0&005347
&1&474(^-5)
&6&707
\\
\\
6~\CIRCLE
&3&891
&0&4340
&0&4999
&0&2369
&0&3135
&0&3895
&0&3810
&0&4345
&0&2161
&4&555
\\
7~\CIRCLE
&5&706
&0&3374
&0&05895
&0&1645
&0&2251
&0&2932
&0&2708
&0&3693
&0&1594
&5&091
\\
8~\CIRCLE
&9&444
&0&04559
&1&061(^-6)
%&\multicolumn{2}{c}{\nexists}
&\multicolumn{2}{c}{-}
&0&008330
&0&04773
&\multicolumn{2}{c}{-}
&0&1138
&0&004818
&5&668
\\
9~\CIRCLE
&21&06
&5&554(^-5)
&1&252(^-7)
&6&603(^-7)
&1&432(^-6)
&3&829(^-6)
&9&464(^-7)
&0&002563
&1&072(^-6)
&6&323
\\
9.5~\CIRCLE
&44&91
&2&688(^-6)
&5&486(^-8)
&8&617(^-8)
&1&481(^-7)
&2&894(^-7)
&1&445(^-7)
&5&631(-4)
&1&167(^-7)
&6&707
\\
\\
\multicolumn{1}{l}{\mbox{SIS}}
&\multicolumn{2}{l}{\infty}
&\multicolumn{2}{l}{\sqrt{5}/3}
&\multicolumn{2}{l}{1}
&\multicolumn{2}{l}{0}
&\multicolumn{2}{l}{-}
&\multicolumn{2}{l}{-}
&\multicolumn{2}{l}{-}
&\multicolumn{2}{l}{-}
&\multicolumn{2}{l}{-}
&\multicolumn{2}{l}{\infty}
\\
\multicolumn{1}{l}{\mbox{NIS}}
&\multicolumn{2}{l}{\infty}
&\multicolumn{2}{l}{\sqrt{5}/3}
&\multicolumn{2}{l}{1}
&1&357R_{\rm K}
&\multicolumn{2}{l}{-}
&\multicolumn{2}{l}{-}
&2&998R_{\rm K}
&\multicolumn{2}{l}{-}
&\multicolumn{2}{l}{-}
&\multicolumn{2}{l}{\infty}
\\
\multicolumn{1}{l}{\mbox{PIS}}
&\multicolumn{2}{l}{\infty}
&\multicolumn{2}{l}{\sqrt{5}/3}
&\multicolumn{2}{l}{1}
&\multicolumn{2}{l}{\infty}
&\multicolumn{2}{l}{-}
&\multicolumn{2}{l}{-}
%&\multicolumn{2}{c}{\nexists}
&\multicolumn{2}{l}{1r_{\rm s}}
&\multicolumn{2}{l}{-}
&\multicolumn{2}{l}{-}
&\multicolumn{2}{l}{\infty}
\\
\\
\multicolumn{1}{l}{\mbox{Hubble}}
&\multicolumn{2}{l}{\infty}
&\multicolumn{2}{l}{0}
&\multicolumn{2}{l}{0}
&\multicolumn{2}{l}{\sqrt{2}r_{\rm s}}
&\multicolumn{2}{l}{\infty}
&\multicolumn{2}{l}{-}
&2&920r_{\rm s}
&\multicolumn{2}{l}{-}
&\multicolumn{2}{l}{-}
&\multicolumn{2}{l}{\infty}
\\
\multicolumn{1}{l}{\mbox{NFW}}
&\multicolumn{2}{l}{\infty}
&\multicolumn{2}{l}{0}
&\multicolumn{2}{l}{0}
&\multicolumn{2}{l}{1r_{\rm s}}
&\multicolumn{2}{l}{\infty}
&\multicolumn{2}{l}{-}
&2&163r_{\rm s}
&\multicolumn{2}{l}{-}
&\multicolumn{2}{l}{-}
&\multicolumn{2}{l}{\infty}
\\
\multicolumn{1}{l}{\mbox{Burkert}}
&\multicolumn{2}{l}{\infty}
&\multicolumn{2}{l}{0}
&\multicolumn{2}{l}{0}
&\multicolumn{2}{l}{1.521r_{\rm s}}
&\multicolumn{2}{l}{\infty}
&\multicolumn{2}{l}{-}
&\multicolumn{2}{l}{3.245r_{\rm s}}
&\multicolumn{2}{l}{-}
&\multicolumn{2}{l}{-}
&\multicolumn{2}{l}{\infty}
\\
\multicolumn{1}{l}{\mbox{Hernquist}}
&\multicolumn{2}{l}{\infty}
&\multicolumn{2}{l}{0}
&\multicolumn{2}{l}{6r_{\rm s}}
&\multicolumn{2}{l}{0.5r_{\rm s}}
&\multicolumn{2}{l}{2r_{\rm s}}
&\multicolumn{2}{l}{\infty}
&\multicolumn{2}{l}{1r_{\rm s}}
&\multicolumn{2}{l}{-}
&1&815r_{\rm s}
&\multicolumn{2}{l}{2\pi}
\\
\\
\multicolumn{1}{l}{\mbox{King $c$=${\frac12}$}}
&\multicolumn{2}{c}{10^c~R_{\rm K}}
&0&4685
&0&5277
&0&2484
&0&3400
&0&4316
&0&4075
&0&4535
&0&2318
&0&8142
\\
\multicolumn{1}{l}{\mbox{\hspace{\kinglen}$c$=$1$}}
&\multicolumn{2}{c}{10^c~R_{\rm K}}
&0&3514
&0&3823
&0&1174
&0&2064
&0&3476
&0&2156
&0&3721
&0&1457
&0&5436
\\
\multicolumn{1}{l}{\mbox{\hspace{\kinglen}$c$=${\frac32}$}}
&\multicolumn{2}{c}{10^c~R_{\rm K}}
&0&2557
&0&2840
&0&04112
&0&1342
&0&3493
&0&08297
&0&3236
&0&08841
&0&4024
\\
\multicolumn{1}{l}{\mbox{\hspace{\kinglen}$c$=$2$}}
&\multicolumn{2}{c}{10^c~R_{\rm K}}
&0&2098
&0&2868
&0&01340
&0&1990
&0&3738
&0&02852
&0&3276
&0&07970
&0&3951
\\
\multicolumn{1}{l}{\mbox{\hspace{\kinglen}$c$=${\frac52}$}}
&\multicolumn{2}{c}{10^c~R_{\rm K}}
&0&3238
&0&2575
&0&004282
&0&2332
&0&3831
&0&009384
&0&3559
&0&1100
&0&4669
\\
\\
\multicolumn{1}{l}{\mbox{\sersic} n$=2$}
&\multicolumn{2}{c}{\infty}
&3&421r_{\rm s}
&3&235r_{\rm s}
&0&4946r_{\rm s}
&1&557r_{\rm s}
&3&213r_{\rm s}
&0&9748r_{\rm s}
&\multicolumn{2}{c}{-}
&1&002r_{\rm s}
&33&27
\\
\multicolumn{1}{l}{\hspace{\sersiclen}n$=3$}
&\multicolumn{2}{c}{\infty}
&5&039r_{\rm s}
&3&132r_{\rm s}
&0&2541r_{\rm s}
&1&571r_{\rm s}
&4&840r_{\rm s}
&0&5355r_{\rm s}
&\multicolumn{2}{c}{-}
&1&001r_{\rm s}
&40&02
\\
\multicolumn{1}{l}{\hspace{\sersiclen}n$=4$}
&\multicolumn{2}{c}{\infty}
&7&418r_{\rm s}
&2&965r_{\rm s}
&0&1290r_{\rm s}
&1&578r_{\rm s}
&7&277r_{\rm s}
&0&2835r_{\rm s}
&\multicolumn{2}{c}{-}
&0&9999r_{\rm s}
&45&79
\\
\\
\hline
\end{array}$
\end{center}
\label{table.halos}
\end{table*}

\end{document}